\documentclass[structabstract]{aa} 
\usepackage{longtable,booktabs}
\usepackage{latexsym}
\usepackage{amssymb}
\usepackage{lscape}
\usepackage[]{natbib}
\usepackage{subfigure}
\usepackage{xcolor}
\usepackage{multicol}
\usepackage{multirow}
\usepackage{amsmath}
\usepackage{booktabs}
\usepackage{caption}
\usepackage{graphicx}
\usepackage{txfonts}
\def\lsim{\mathrel{\rlap{\lower 3pt \hbox{$\sim$}} \raise 2.0pt \hbox{$<$}}}
\def\gsim{\mathrel{\rlap{\lower 3pt \hbox{$\sim$}} \raise 2.0pt \hbox{$>$}}}

\title{Nuclear versus integrated spectroscopy of galaxies in the $Herschel$ Reference Survey.\thanks{based on observations taken at the 
1.5m Loiano telescope belonging to the Bologna Observatory.}
\thanks{Table A4 is only available in electronic form
at the CDS via anonymous ftp to cdsarc.u-strasbg.fr (130.79.128.5)
or via http://cdsweb.u-strasbg.fr/cgi-bin/qcat?J/A+A/}}

\subtitle{}

\author{G.Gavazzi \inst{1}                               
\and G. Consolandi \inst{1}                                                                  
\and S. Belladitta \inst{2,3}
\and A. Boselli \inst{4}  
\and M. Fossati \inst{5,6}                               
}
\authorrunning{G. Gavazzi et al.}
\titlerunning{Spectroscopic observations of HRS galaxies}
\institute{Universit\`a degli Studi di Milano-Bicocca, Piazza della Scienza 3, 20126 Milano, Italy\\
\email {giuseppe.gavazzi@mib.infn.it}
\and
DiSAT, Universit\`a degli Studi dell'Insubria, Via Valleggio 11, 22100 Como, Italy
\and
INAF - Osservatorio Astronomico di Brera, via Brera, 28, 20159 Milano, Italy\\
\email {silvia.belladitta@brera.inaf.it}
\and
Aix Marseille Universit\'e, CNRS, LAM (Laboratoire d'Astrophysique de Marseille) UMR 7326, F-13388, Marseille, France\\
\and
Max-Planck-Institut fur Extraterrestrische Physik, Giessenbachstrasse, D-85748 Garching, Germany\\
\and
Universitats-Sternwarte Munchen, Schenierstrasse 1, D-81679 Munchen, Germany}
\begin{document}

\date{Received; accepted}
\abstract
{The determination of the relative frequency of active galactic nuclei (AGN) versus other spectral classes, 
for example, HII region-like (HII), transition objects (TRAN), passive (PAS), and retired (RET), 
in a complete set of galaxies in the local Universe is of primary importance 
to discriminate the source of ionization in the nuclear region of galaxies.}
{Here we aim to provide a spectroscopic characterization of the nuclei of galaxies belonging to the $Herschel$ Reference Survey (HRS), 
a volume and magnitude limited sample representative of the local Universe, which has become a benchmark for local and high-$z$ studies, 
for semianalytical models and cosmological simulations.  
The comparison between the nuclear spectral classification and the one determined on the global galactic scale 
provides information about how galaxy properties change from the nuclear to the outer regions.
Moreover, the extrapolation of the global star formation (SF) properties from the SDSS fiber spectroscopy
compared to the one computed by H$\alpha$ photometry can be useful for testing the method based on aperture correction 
for determining the global star formation rate (SFR) for local galaxies.}
{By collecting the existing nuclear spectroscopy available from the literature, complemented with new observations obtained using the Loiano 1.52m telescope, 
we analyze the 322 nuclear spectra of HRS galaxies.} 
{Using two diagnostic diagrams (the BPT and the WHAN) we provide a nuclear and an integrated spectral classification for the HRS galaxies. 
The BPT and the WHAN methods for nuclei consistently give a frequency of 53-64\% HII, around 21-27\% AGNs (including TRAN), 
and 15-20\% of PAS (including RET), whereas for integrated spectra they give 69-84\% HII, 4-11\% of AGNs and 12-20\% PAS.
}
{We find that the fraction of AGNs increases 
with stellar mass, such that at $\rm M_{\ast}$ > $10^{10.0}$ $\rm M_{\odot}$ $\sim$66\% of the LTGs are AGNs or TRAN. 
}


\maketitle
%

\section{Introduction}
Characterizing the nuclear properties of galaxies in the local Universe, 
that is, establishing the mass dependence of the relative frequency of active galactic nuclei (AGN) 
ionized by supermassive black holes with respect to HII region-like nuclei excited by young stars, or 
with respect to galaxies ionized by old stars, 
is an urgent task of today's research in astrophysics (see e.g., Kauffmann et al. 2003b, hereafter K03).
There is a clear need to establish the frequency of AGNs of various types, in different environments, 
locally and from a cosmological perspective, to improve our understanding of galaxy assembly.
As Boselli et al. (2010) pointed out, observing the local Universe is relevant because it represents the endpoint of galaxy evolution, 
providing important boundary conditions to models and simulations. 
Galaxies can only be completely characterized  at low-redshift  or by multifrequency observations.
Moreover, dwarf galaxies can only be observed locally.
In recent years, considerable effort has been made to select representative samples of galaxies, like SINGS (Kennicutt et al. 2003), 
the recent KINGFISH (Kennicutt et al. 2011), and VNGS (Bendo et al. 2012), to study the properties of the local Universe. 
In recent years, the $Herschel$ Reference Survey (HRS) (Boselli et al. 2010a) has become a benchmark for the representation of the properties of local galaxies 
(Hughes, T. M. et al. 2013, Boquien et al. 2014, Ciesla et al. 2016), 
and has therefore become a reference for comparing observations of local galaxies with those of increasing redshift (Bassett et al. 2017, Fossati et al. 2017, 
Schreiber et al. 2016) or resulting from simulations (Cattaneo et al. 2017, Fontanot et al. 2017, Lagos et al. 2016). 
The HRS has been observed with SPIRE (250, 350, 500 $\mu$m) (Ciesla et al. 2012, 2013) and with PACS (100, 160 $\mu$m) (Cortese et al. 2014) on board $Herschel$: 
At D $\leq$ 30 Mpc the angular resolution of SPIRE ($4$ arcmin) allows us to resolve the different galaxy components, such as nuclei, bulges, discs and spiral arms.
The HRS is a volume-limited sample and is $k-band$ selected, therefore, by design, 
it is suitable for statistical studies of the multifrequency properties of local galaxies. 
Many surveys have been done to cover the whole sample in different bands: FUV and NUV photometry obtained from GALEX is reported by Cortese et al. (2012); 
24-160 $\mu$m photometry from MIPS is given by Bendo et al. (2012); Spitzer/IRAC photometry is reported by Ciesla et al. (2014); 
Boselli et al. (2014) report the gas properties (HI and CO) of the HRS, whereas Boselli et al. (2015, hereafter B15) describe an H$\alpha$ imaging survey of the full HRS. 
Finally Boselli et al. (2013) give the integrated (drift-mode) spectroscopy of the HRS.
Conversely no systematic information is available from the literature on the nuclear spectroscopic classification of the HRS, a task that we tackle in this paper.  
Not all the 322 galaxies constituting the HRS have been observed spectroscopically so far (only 277/322 objects). 
In this paper we complete the HRS by presenting new nuclear long-slit spectroscopy for 45 galaxies, 
25 of which in the whole optical range (from H$\beta$ to [SII]) and 20 only in the red band near H$\alpha$.
Generally, the classification of galaxies based on their optical nuclear spectra is performed using the BPT (Baldwin, Phillips \& Terlevich, 1981) 
diagnostic diagram, which requires the measurement of different spectral lines: 
H$\beta$, [OIII]$\lambda$5007, H$\alpha$, [NII]$\lambda$6583.4 and [SII]$\lambda$6717,6731. 
In this diagram AGNs are distinguished from HII-region-like nuclei using the ratio [NII]/H$\alpha$, while strong AGNs (sAGN) can be separated from the weaker 
(weak AGNs or wAGN) low-ionization nuclear emission-line region (LINERs, Heckman et al. 1980) using the ratio [OIII]/H$\beta$. 
There is also a recent two-line diagnostic diagram, named WHAN, which is based only on the [NII]/H$\alpha$ ratio combined with the equivalent width (EW) of the H$\alpha$ line. 
It was introduced by Cid Fernandes et al. (2010, 2011) to divide both
strong and weak AGNs from $fake$ $AGNs$, 
namely the retired galaxies (RET), whose ionization mechanism is probably provided by an old stellar population (Capetti \& Baldi 2011). 
Using the BPT and the WHAN diagrams, we obtain a robust determination of the frequency of AGNs in a complete sample of local late-type galaxies (LTGs) 
and compare it with the frequency of HII-region-like nuclei as a function of stellar mass.\\
Moreover the comparison between the nuclear spectral classification and the one determined on the global galactic scale 
provides information about how galaxy properties change from the nuclear to the outer regions. 
The issue is hotly debated and only spectroscopy obtained with Integral Field Units (IFU) such as MaNGA (Belfiore et al.  2016, 2017; Sanchez et al. 2017), 
CALIFA (S{\'a}nchez et al. 2016) and SAMI (Richards et al. 2016) will bring the issue to an end. 
These surveys have confirmed previous claims that regions of low ionization conditions
can in fact extend to the bulges of galaxies and even further, mimicking the presence of AGNs, being in fact due to old post-AGB stars.\\
We compare the global with the nuclear properties of HRS galaxies  adopting a method similar
to Moustakas et al. (2010) who used long-slit spectroscopy, 
while  Iglesias-P{\'a}ramo et al. (2013, 2016) adopted the two-dimensional (2D) spectroscopy.
The availability of the global properties of HRS galaxies allows us to test the program of Brinchmann et al. (2004, hereafter B04) 
who claimed that a satisfactory estimate of the global star formation rate (SFR) of galaxies in the local Universe ($z<0.2$) can be achieved using SDSS fiber spectroscopy, 
once corrected for aperture effects using SDSS photometry (e.g., cmodel - fiber colors), 
both quantities being available  for hundreds of thousands of local galaxies thanks to the SDSS. \\
The paper is structured as follows: in section 2 we describe the sample, in section 3 we outline the data reduction, 
in section 4 we present the results, and in section 5 the discussion and conclusions. 
A description of the tables provided is given in the appendix. 
Standard cosmology is assumed, with H$_{0}$ = 73 km/s/Mpc.\\
The spectroscopic data presented in this work, as well as those collected at other frequencies, 
are available to the community through the HeDaM database (http://hedam.lam.fr/HRS/).

\section{The sample}
\label{sample}
The sample of galaxies analyzed in this work is the $Herschel$ Reference Survey (HRS) 
(Boselli et al. 2010a; the reader should refer to this paper to find the catalog names of HRS galaxies).
It consists of a volume limited (15 $<$ $Dist$ $<$ 25 Mpc) sample of 322 galaxies, 
located in the sky region in the ranges 10h17m < R.A.(2000) < 14h43m and -6$^o$ < dec < 60$^o$ (see Fig. \ref{atlas}), 
of which 64 are early-type galaxies (ETGs: E, S0 and S0a) 
and 258 are late-type galaxies (LTGs: Sa-Sd-Im-BCD).\footnote{With respect to the original sample given in Boselli et al. (2010a), 
we revised the morphological classification for some HRS galaxies: HRS-032, HRS-090, HRS-202, HRS-229 and HRS-291 become ETGs, while HRS-256 becomes LTG.  
After this revision, the LTGs are 254 and the ETGs are 68. In this work we use this new number for our analysis.} 
All LTGs with a 2MASS K band total magnitude K$_{Stot}$ $\leq$12 mag have been selected 
and all ETGs with K$_{Stot}$ $\leq$8.7 mag have been included. 
The stellar mass range of HRS galaxies is: 5x10$^{8}$ $\leq$ $\rm M_{\ast}$ $\leq$ 10$^{11.3}$ $\rm M_{\odot}$.
The sample spans a large range of environments: it includes the Virgo cluster, many galaxy groups, as well as isolated objects. 
In the Virgo cluster region ($12$h<R.A.<$13$h and 0$^o$<dec<18$^o$), 
the sample includes all galaxies with vel < $3000$ km/s and those belonging to cluster A, the North (N) and East (E) clouds, 
the southern extension (S; at $17$ Mpc) and Virgo B ($23$ Mpc), where the subgroup membership has been taken from Gavazzi et al. ($1999$a). 
The W and M clouds, at a distance of $32$ Mpc, have been excluded. 
In the sky projection map of Fig. \ref{atlas}, the Virgo cluster region, which includes 148 HRS objects from Virgo A, Virgo B, the N and E clouds and the S extension, 
is easily recognizable as the density enhancement at RA$\simeq$190 and Dec$\simeq$10. 

\subsection{Data from the literature}
\begin{figure*}
\centering
\includegraphics[angle=0, scale=0.7]{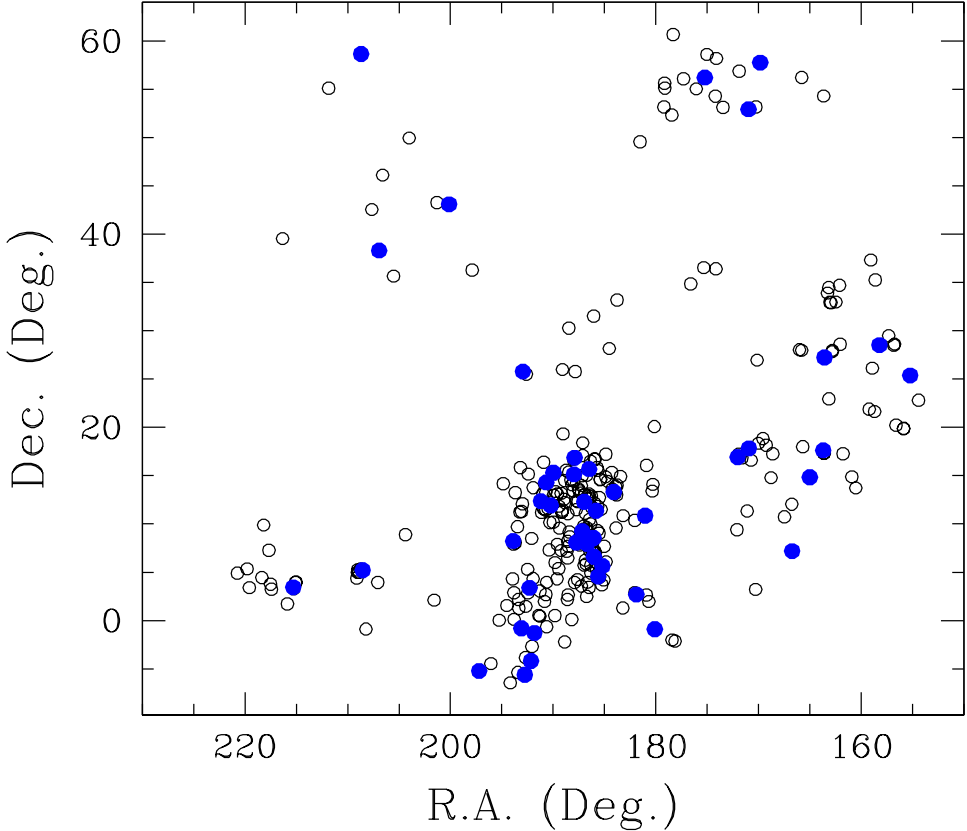}
\vspace{-0.7cm}
\caption{Sky projection of the full HRS survey (empty symbols) and the subsample of 45 targets whose nuclear spectra were observed with the Cassini Telescope (filled blue).}
\label{atlas}  
\end{figure*}

Most of HRS nuclear optical spectra were found in the literature. 
In particular, 163 spectra were downloaded in FITS format from the SDSS DR13 and DR12 database (Albareti et al. 2016);
114 spectra were found in the NASA/IPAC Extragalactic Database (NED) and 
104 of these are taken from Ho et. al (1997). 
For these galaxies, the double spectrograph on the Hale 5m telescope at Palomar Observatory 
yielded simultaneous spectral coverage of $\sim$4230-5110$\rm \dot{A}$ and $\sim$6210-6860$\rm \dot{A}$, 
with a spectral resolution of $\sim$4$\rm \dot{A}$ in the blue and $\sim$2.5$\rm \dot{A}$ in the red.
We joined the blue and red spectra by projecting them on a common wavelength solution and 
by performing a linear interpolation between the average flux of the last ten pixels 
of the blue spectrum and the first ten of the red spectrum.
HRS-070, 072, 209, 265, 275, 288 and 284 have been observed with the 3.9m Anglo Australian Telescope (AAT) by Jones et al. (2009). 
HRS-193 was observed by Falco et al. 2000 with the Z-Machine at Mt. Hopkins 1.5m telescope (Arizona). 
HRS-251 was observed by Kim \& Veilleux (1995) with the 2.2m telescope at the University of Hawaii.
HRS-271 is part of the work of Jansen et al. (2000) who observed 196 nearby galaxies with the FAST instrument at Tillinghast 1.5m telescope on Mt. Hopkins. 
Finally the remaining 45 spectra were observed by us with the 1.52m Cassini telescope at Loiano (Bologna) 
and their spectrum is published for the first time in this paper. 
Filled symbols in Fig. \ref{atlas} represent these 45 HRS targets. 

\section{Long-slit nuclear spectroscopy: observations and data reduction}
Nuclear spectroscopic observations of 45 HRS galaxies that were found neither in the SDSS spectroscopic catalog
nor within the NED database, were obtained by us during several observing campaigns from 2000 to 2017
using the Bologna Faint Object Spectrograph and Camera (BFOSC, Gualandi \& Merighi 2001) mounted on the 152cm F/8 Cassini Telescope located
in Loiano, belonging to the Observatory of Bologna. Following the work of Gavazzi et al. 2011, 2013,
these are long-slit spectra taken through a slit of 2 arcsec width and 12.6 arcmin length, combined with an intermediate-resolution red-channel grism (G8; $R$ $\sim$ 2200)
covering the 6100 - 8200 \AA ~portion of the spectrum containing the $\rm H\alpha$, [NII], and [SII] lines.
Twenty-five of these galaxies were also observed with the same instrument during six nights in April and May, 2017, (see Table \ref{Tobs} for details),
using a blue-channel grism (G7) providing ($R$ $\sim$ 1400) cover from 4200 to 6600 \AA~.
The observing log of these 45 galaxies is given in Table \ref{Tobs}.\\
BFOSC is equipped with a back illuminated EEV LN/1300-EB/1 CCD detector of 1300x1340 pixels, reaching 90\% QE near 5500 \AA ~and  70\% QE near 4000 \AA.
For the spatial scale of 0.58 arcsec/pixel, 
and a dispersion of 8.8 nm/mm, the resulting spectra have a resolution of 1.6 ~\AA/pix for G8 
and a dispersion of 11 nm/mm, with a resolution of 1.9 ~\AA/pix for G7.\\
Exposures of 2-7 minutes (G8) and  7-15 min (G7) were repeated typically three times to remove cosmic ray hits. 
The slit was generally set in the E-W direction and the typical seeing conditions at Loiano ranged from 1.5 to 2.5 arcsec. 
The wavelength calibration was secured by means of frequent exposures of a He-Ar hollow-cathode lamp and was further refined using bright OH sky lines.
The spectrograph response (sensitivity function) was obtained by daily exposures of the stars Feige34 and Hz44 (Massey 
et al. 1988). 
The spectra were not flux calibrated, and measurements of the lines EW were derived, along with flux ratios of adjacent lines, 
namely [OIII]/H$\beta$, [NII]/H$\alpha$ and [SII]/H$\alpha$.\\ 
The spectra were reduced using standard IRAF procedures. 
To derive the wavelength calibration we used the tasks $identify$, $reidentify$, and $fitcoord$ on the hollow-cathode lamp exposures, 
and the 2D wavelength solution was transferred to the galaxy exposures by means of the task $transform$.
After verification of  the wavelength calibration on several known sky lines, the sky was subtracted using $background$.
One-dimensional (1D) nuclear spectra were extracted from the 2D images using $apsum$, adopting an aperture of 10 pixel width
(5.8 arcsec). 
We adopt this value for the aperture to be consistent with Ho et al. (1997), who extracted (red) spectra from an aperture of 4.1 arcsec.
The 1D spectra were response-calibrated using the median sensitivity function of each night.
After normalization to the flux in the interval 6400-6500 $\mbox{\AA}$ the spectra were Doppler shifted to $\lambda_0$:
each redshift was taken from NED and then applied to the galaxy spectrum using $dopcor$. 
For the galaxies observed with both the G7 and G8 grisms, the blue and red spectra were joined using the task $scombine$.\\
Plots of the nuclear spectra obtained at Loiano are divided among Figures \ref{spectra} and \ref{spectra1} in the Appendix.
Out of 19 spectra taken with the G8 grism, 8 can be found in Gavazzi et al. (2011, 2013). 
The remaining 11 are given in Figure \ref{spectra1} and are published for the first time in this paper.
25 spectra observed with both the G7 and G8 grisms are shown in Figure \ref{spectra} of the appendix.\footnote{Actually 26 spectra  are shown. 
The additional, HRS-110, is a combination of a red spectrum taken at Loiano and a spectrum from the SDSS (DR7), whose red part near H$\alpha$  was ruined.}\\

\begin{figure}
\centering
\includegraphics[angle=0, scale=0.4]{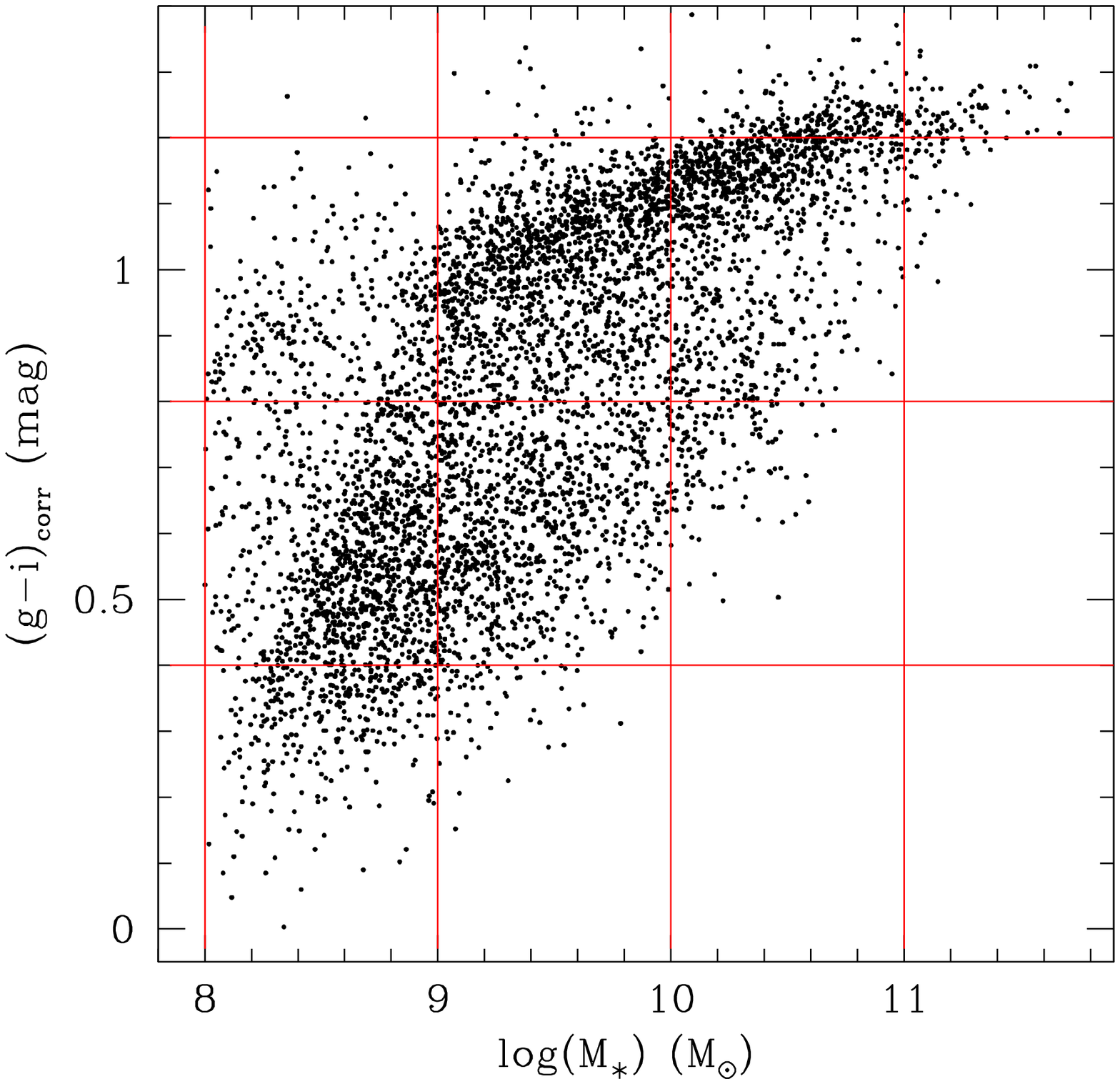}
\vspace{-0.7cm}
\caption{The color-stellar mass relation for a complete sample of
galaxies in the Coma and Local supercluster from Consolandi et al. (2016), subdivided into nine bins of stellar mass and color.
In these bins we compute the average SDSS continuum correction to H$\alpha$ (see Table \ref{Thalpha}).}
\label{colmag}  
\end{figure}

The Balmer hydrogen lines are affected by stellar absorption, thus they have been corrected for this effect for proper use in the diagnostic diagrams.\\ 
We homogeneously corrected the 26 spectra that we took at Loiano with the G7+G8 grisms, the spectra downloaded from SDSS, the spectra 
observed by Ho et al. (1997) and the spectra downloaded from the NED database 
using the GANDALF fitting code (Gas AND Absorption Line Fitting, Sarzi et al. 2006; Falcon-Barroso et al. 2006) complemented by the Penalize 
Pixel-Fitting code (Cappellari \& Emsellem, 2004) 
to simultaneously model the stellar continuum and the emission lines in individual pixels. 
Consistently with B15, we used the $MILES$ stellar library (Vazdekis et al. 2010). 
GANDALF is a simultaneous emission and absorption line fitting algorithm designed to separate the relative contribution of the stellar continuum 
from that of the nebular emission in the spectra of nearby galaxies, while measuring the gas emission and kinematics.
This code implements the pPXF method, which combines and adjusts the observed spectra of several stars of all spectral type 
to the stellar continuum to quantify and remove the underlying absorption.
GANDALF was run several times, adjusting the algorithm input parameters,
especially on spectra of Ho et al. (1997), because of the presence of the 1000\AA~ gap in these spectra (see section \ref{sample} for details). \\
For the remaining 19 galaxies observed at Loiano with the G8 grism, the insufficient spectral coverage prevents GANDALF from converging.
For these galaxies, a statistical absorption correction was applied according to the following procedure.
We took a sample of about 5000 galaxies from Consolandi et al. (2016) and constructed their color-stellar mass diagram in Figure \ref{colmag}. 
For these 5000 galaxies, we downloaded the absorption correction to H$\alpha $  from SDSS ($GalSpecLine$) and constructed a mean correction in 
nine bins of stellar mass and color (see Table \ref{Thalpha}) 
that we applied to the spectra according to each bin of mass and color. 
The $g$ and $i$ magnitudes for these HRS galaxies were taken from Cortese et al. (2012) and were subsequently corrected for Galactic extinction using 
the Cardelli et al. (1989) reddening curve; 
the value of stellar mass is from B15.
The resulting H$\alpha$ correction is consistent with the correction obtained using the GANDALF analysis for the individual spectra.\\

\begin{table}
\centering
\caption{Continuum correction to H$\alpha$ EW as derived from the color-mass relation of Figure \ref{colmag}.} 
\begin{tabular}{|c||c|c|c|}
\hline
                              & 8 $\leq$ $\rm M_{\ast}$ < 9  & 9 $\leq$ $\rm M_{\ast}$ $\leq$ 10  & 10 < $\rm M_{\ast}$ $\leq$ 11 \\    
\hline                        
0.8< g-i $\leq$ 1.2           & 1.94  $\pm$  0.41    &  1.91  $\pm$  0.29          &  1.81  $\pm$  0.24      \\
\hline                        
0.4 $\leq$ g-i $\leq$ 0.8     & 2.43  $\pm$  0.63    &  2.25  $\pm$  0.45          &  1.98  $\pm$  0.32      \\
\hline                        
g-i < 0.4                     & 2.57  $\pm$  0.63    &  2.48  $\pm$  0.51          &  -                      \\
\hline
\end{tabular}
\label{Thalpha}
\end{table}

\section{Spectral classification}
\label{res}
\begin{figure*}
\centering
  {\includegraphics[width=4.4cm]{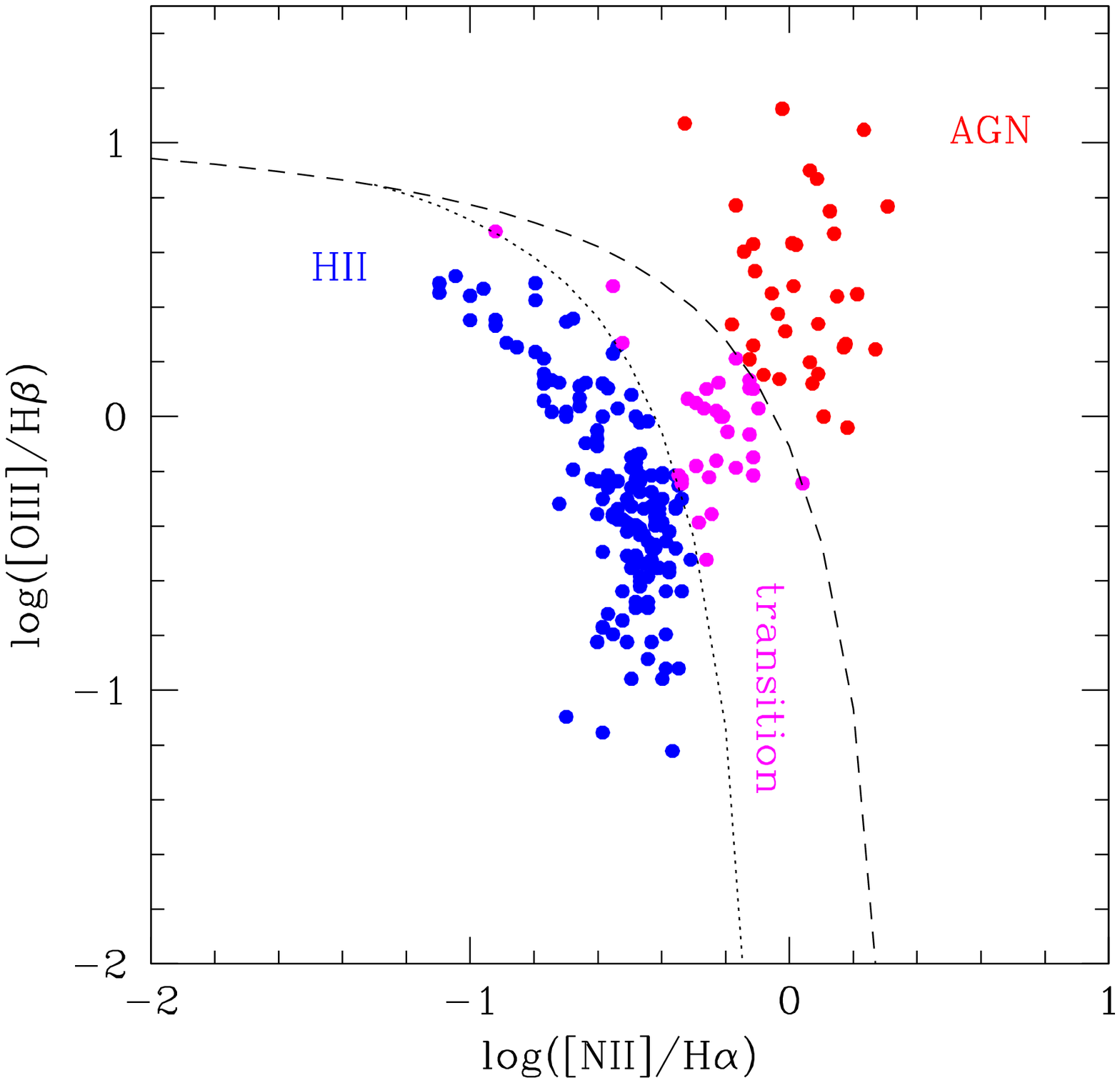}}{\includegraphics[width=4.4cm]{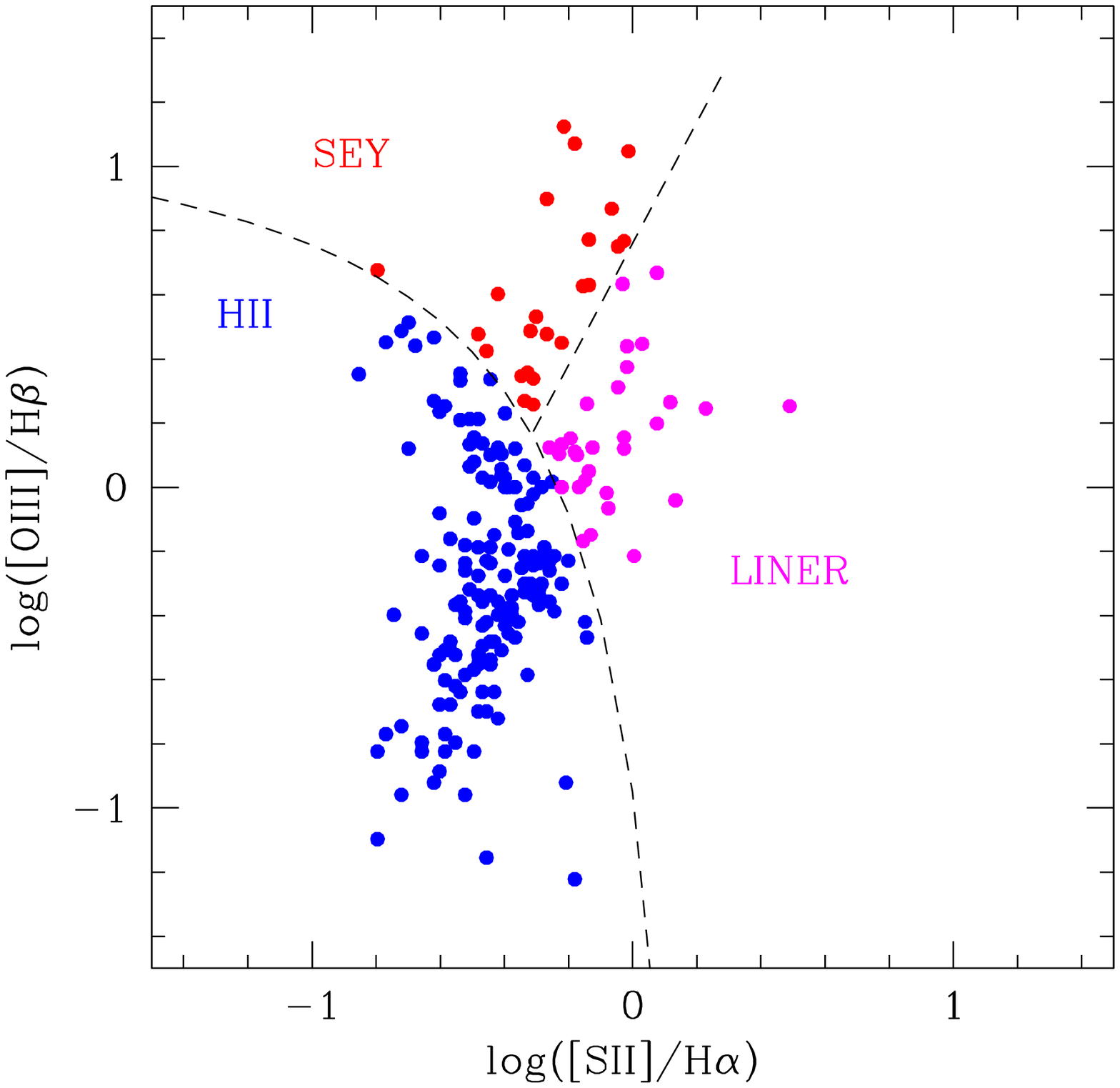}}{\includegraphics[width=4.4cm]{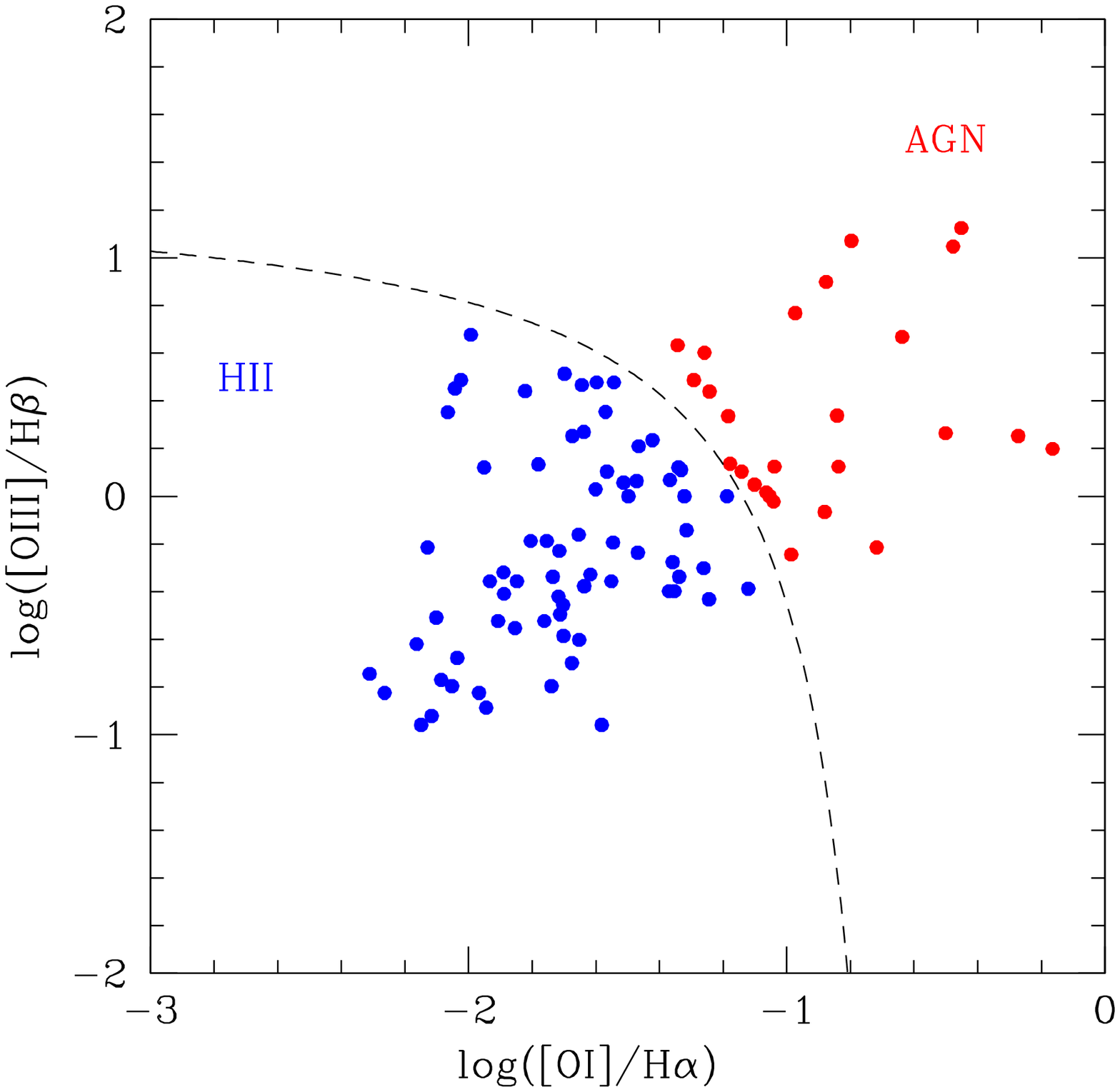}} {\includegraphics[width=4.4cm]{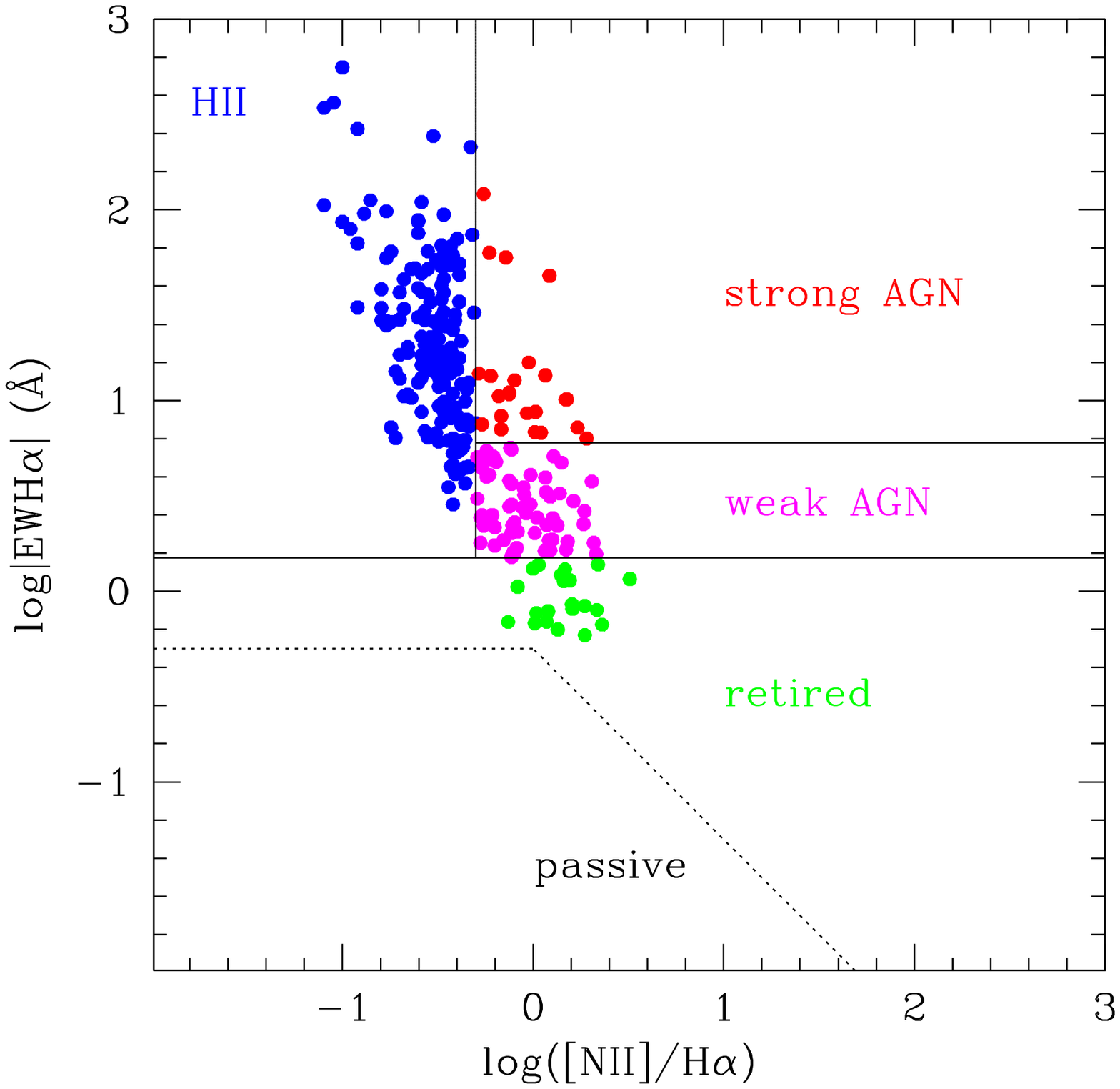}} 
\vspace{-0.5cm}
\caption{Nuclear diagnostic diagrams for the whole HRS sample limited to emission-line objects with S/N ratio > 3. 
In the left BPT diagram the broken separation line between AGNs (red) and TRAN (pink) galaxies is from Kewley et al. (2001), while
the dotted separation line between TRAN (pink) and HII region-like nuclei (blue) is from K03. 
In the second BPT diagram the extreme starburst classification line is from Kewley et al. (2001), while the separation between LINER (pink) and SEY (red) is from Kewley et al. (2006).  
We note that the pink filled symbols refer to different types of objects in the three diagrams.
The third panel reports the BPT diagram for the minority of galaxies (101) with [OI] in emission.
In the WHAN diagram the separations between the different classes are from Gavazzi et al. (2011).}
\label{nuc}
\end{figure*}

\begin{figure*}
\centering
  {\includegraphics[width=5.9cm]{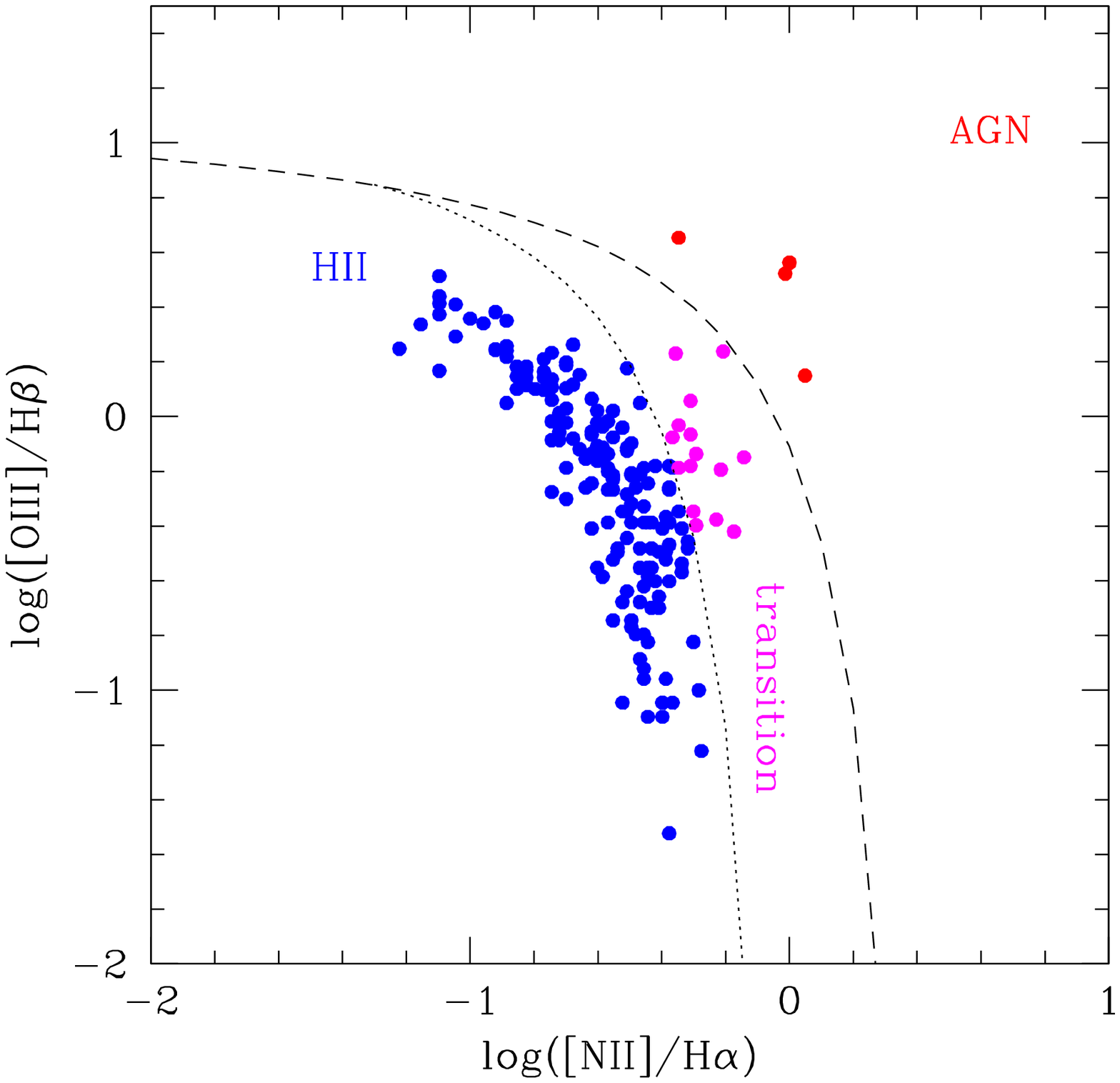}}{\includegraphics[width=5.9cm]{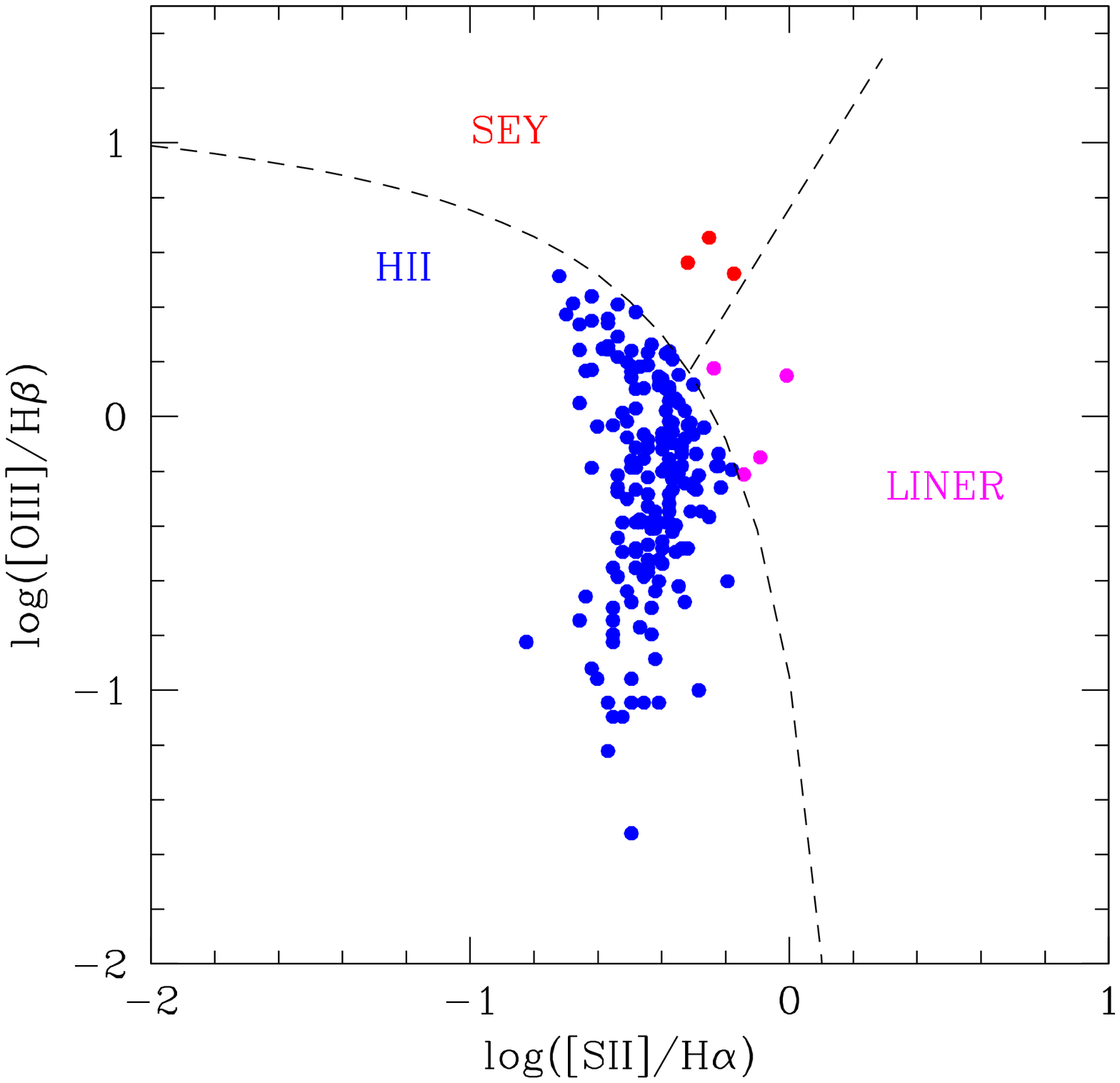}}{\includegraphics[width=5.9cm]{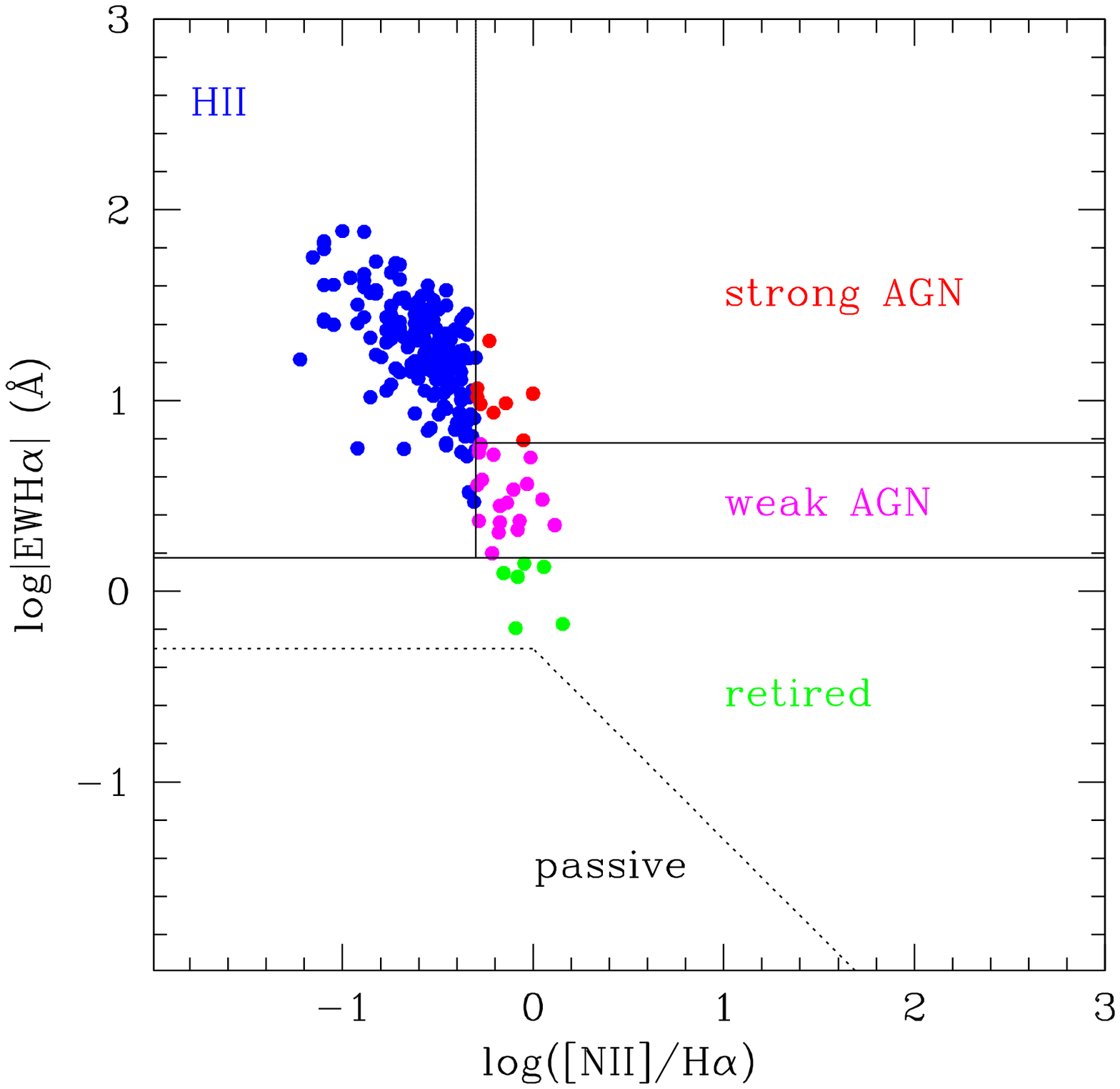}}
\vspace{-0.5cm}
\caption{Integrated diagnostic diagrams for emission-line objects with S/N ratio > 3. Broken separations in the diagrams are the same as in Fig. \ref{nuc}.}
\label{int}
\end{figure*}

The determination of the spectral properties of HRS galaxies in this paper is based  on four indicators (three versions of the BPT and the WHAN diagram)
which exploit the ratio of [OIII]/H$\beta$ on 
[NII]/H$\alpha$ or [SII]/H$\alpha$  or occasionally  on the [OI]/H$\alpha$   
using the BPT diagnostic diagram, and on the ratio [NII]/H$\alpha$ versus the H$\alpha$ EW, using the WHAN 
diagram, which holds when the blue emission lines are not available.
These diagnostics only classify galaxies with emission lines.
Therefore galaxies classified as passive (PAS, showing only absorption lines) or post-starburst (PSB,   showing the Balmer series in absorption)
have been added a posteriori, by inspecting their spectra by eye.\\
According to Kobulnicky et al. (2003), the ratio of the two adjacent lines can be derived using the equivalent widths instead of the line fluxes.\\
In order to perform a robust spectral classification, we do not classify integrated and nuclear spectra with signal-to-noise ratio (S/N) of H$\alpha$ and H$\beta$ <3.
With such a threshold, 48 nuclear spectra (marked with the letter "d" in Table \ref{Tclass}) 
and 36 integrated spectra (letter "b" in Table \ref{Tclass}) were not classified according to the BPT diagrams. 
Instead, all nuclear spectra could be classified according to the WHAN diagram and only 8 integrated spectra (marked with letters "b,c" in Table \ref{Tclass}) 
were excluded from this classification.\\ 
In the following spectral analysis we refer always to the LTGs subsample, because their statistic is complete (both nuclear and integrated).
As far as concerns the ETGs subsample, 50 integrated spectra were not observed by Boselli et al. (2013): 
this subsample was not considered in our statistical analysis when we compare the nuclear to the integrated classification. 
Therefore we refer the reader to a sample of 360 ETGs from the ATLAS3D whose nuclear spectra are given in Gavazzi et al. (2018) for the properties of ETGs.\\
Figure \ref{nuc} gives the nuclear classification of HRS galaxies according to the BPT diagnostic diagram based on the [NII]/H$\alpha$ ratio (217 objects), 
on the [SII]/H$\alpha$ ratio (217 objects),  on the [OI]/H$\alpha$ ratio (101 objects) and according to the WHAN diagram (288 objects). 
Approximately 30 additional galaxies with no emission lines, not included  among the 217 nor 288 classified objects, 
were classified as PAS or PSB. \\
The region labeled HII (blue) lies to the left of the dotted line defined by K03, while the
region occupied by AGNs (red) is to the right of the broken line of Kewley et al. (2001) in both the NII]/H$\alpha$ (first panel) and in the
[OI]/H$\alpha$  diagrams (third panel).
In between the two lines an intermediate region defines the position of the Transition objects (TRAN) (pink).
The  second panel gives the BPT diagnostic using the ratio [SII]/H$\alpha$.
Here the straight line (from Kewley et al. 2006) separates Seyfert (SEY, in red) from $LINERs$ (LIN, in pink).
The  fourth panel of Fig. \ref{nuc} shows the WHAN diagnostic: the different spectral classification 
and separations between them are from Gavazzi et al. (2011).
In this diagram galaxies are divided into five spectral types: HII-regions (blue), sAGN (red), wAGN (pink), RET (green) and PAS. \\
Using the same diagnostic criteria, Figure \ref{int} gives the BPT and WHAN classifications for integrated spectra with available emission lines, 
according to the S/N threshold.\\
In our analysis we refer always to AGNs as TRAN+AGN, LIN+SEY, and sAGN+wAGN. 
We adopted this definition for the AGNs spectral class to be able to compare our results with those of K03, who defined as AGNs all galaxies 
beyond the left line in the BPT ([NII]/H$\alpha$) diagram. 
In specific cases we refer only to "pure" AGNs, namely galaxies above the curve defined by Kewley et al. (2001) in the BPT ([NII]/H$\alpha$) diagram.
Also, galaxies classified as PAS or PSB (and RET in the WHAN diagnostic) are always held together.\\

\begin{figure*}
\centering
  {\includegraphics[width=7.5cm]{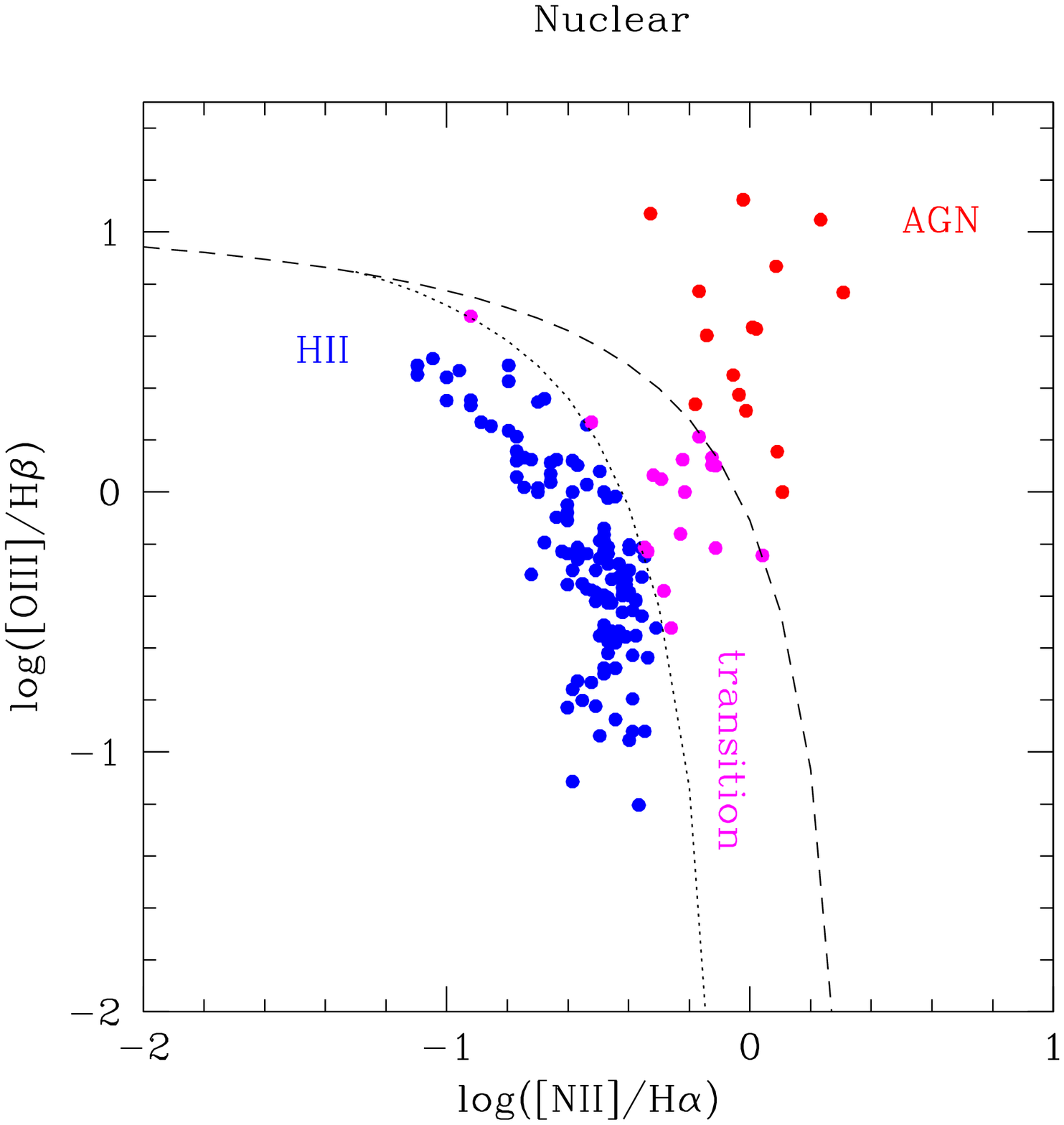}}{\includegraphics[width=7.5cm]{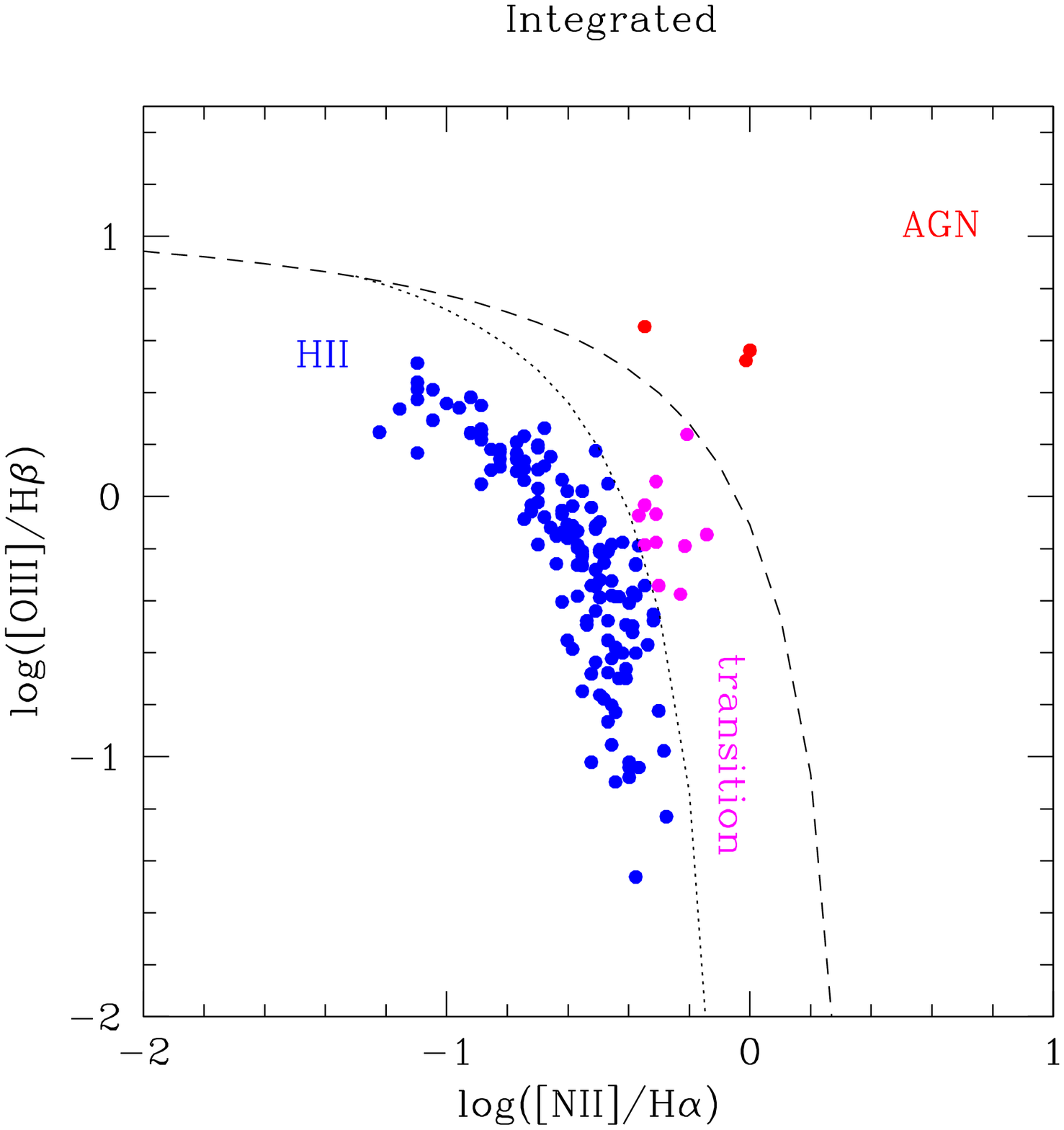}}
\vspace{-0.5cm}
\caption{BPT diagrams for 164 HRS objects that have both the nuclear and integrated classification available in the [OIII]/H$\beta$ vs.
[NII]/H$\alpha$ diagram. The separation lines are the same as in Figure \ref{nuc}.}
\label{BPT}
\end{figure*}

\begin{figure*}
 \centering
 \subfigure[AGNs]
   {\includegraphics[width=5.9cm]{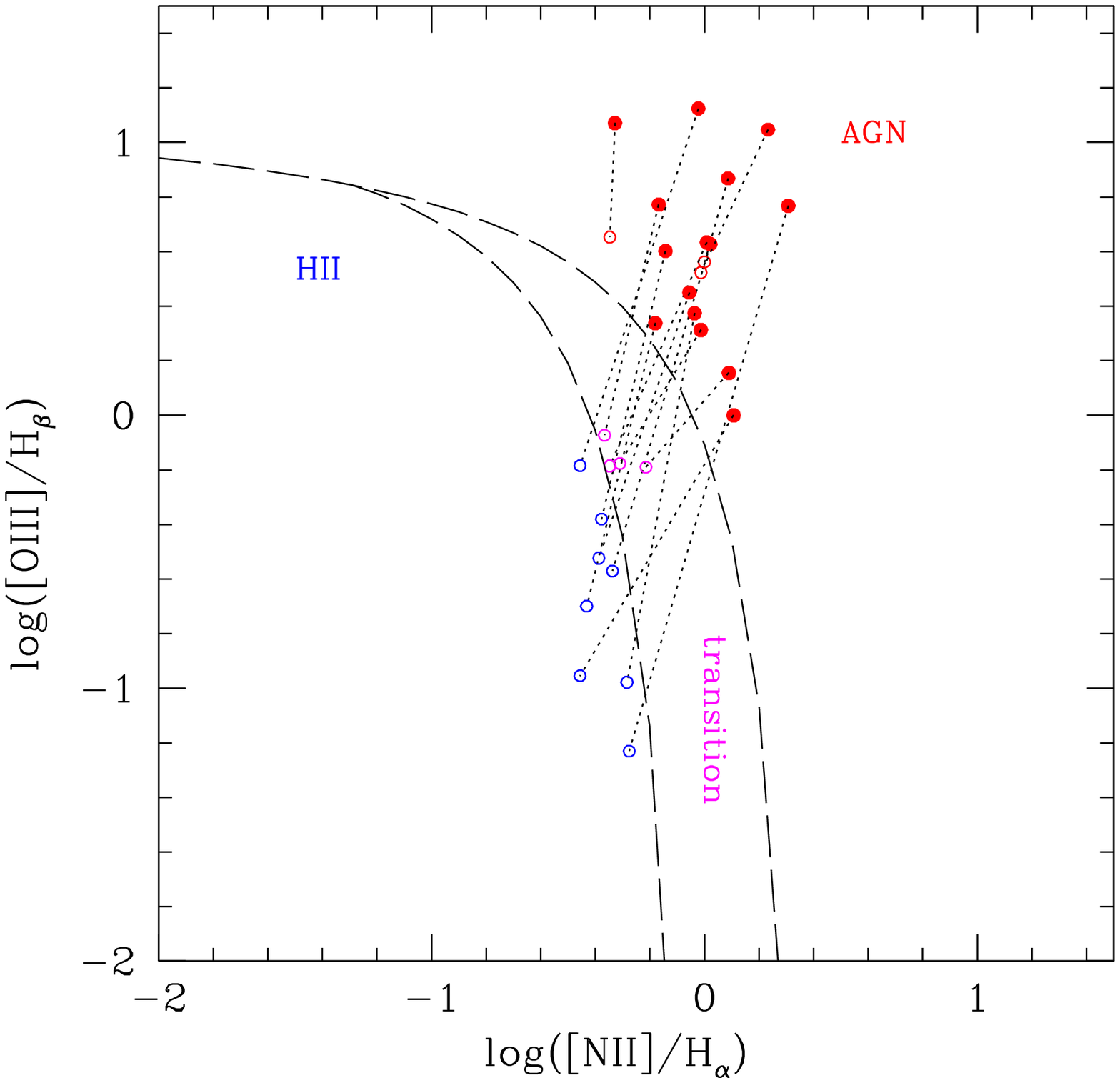}}
 \hspace{0.5mm}  
 \subfigure[Transition]
   {\includegraphics[width=5.9cm]{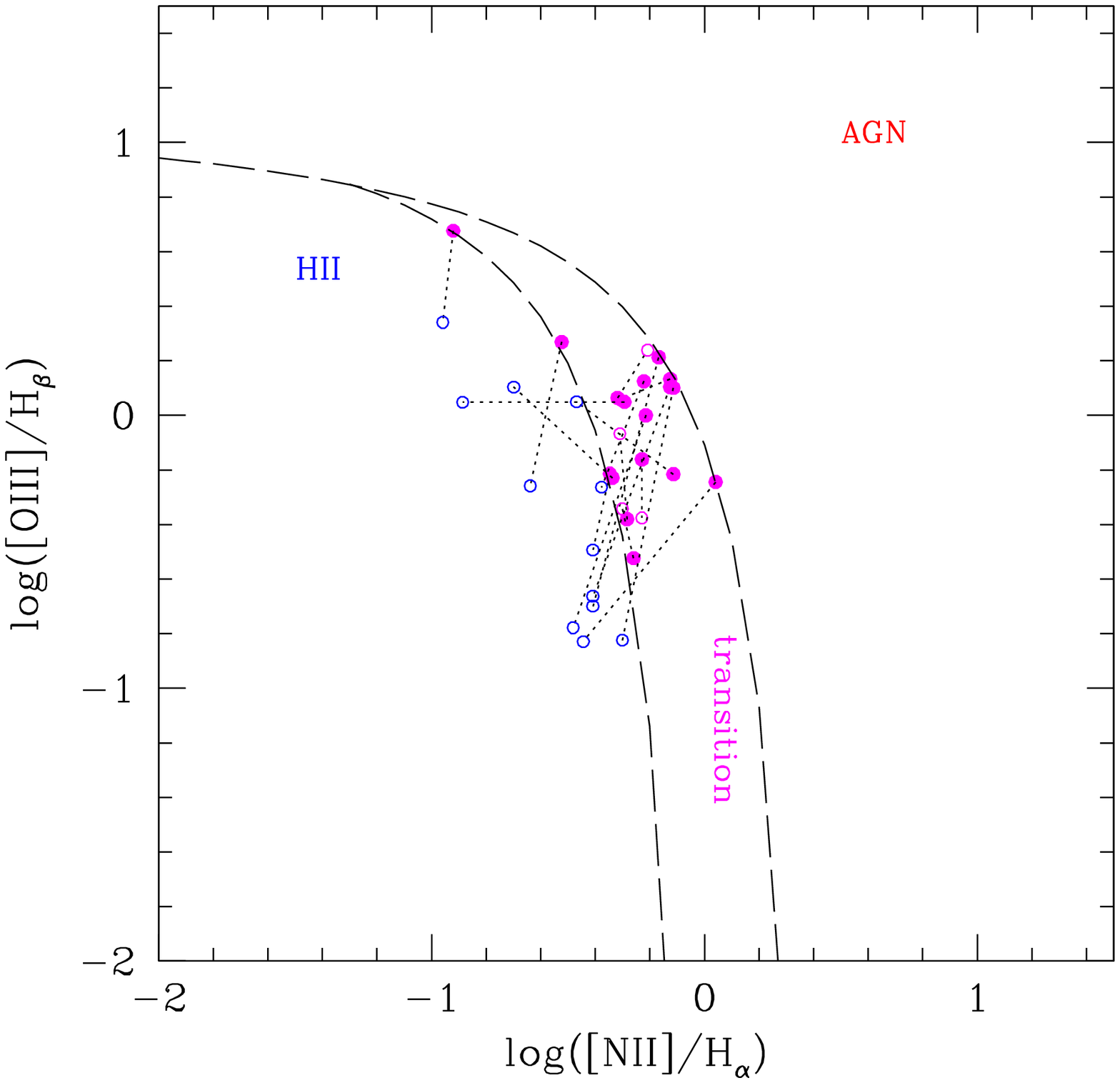}}
 \hspace{0.5mm}
 \subfigure[HII regions]
   {\includegraphics[width=5.9cm]{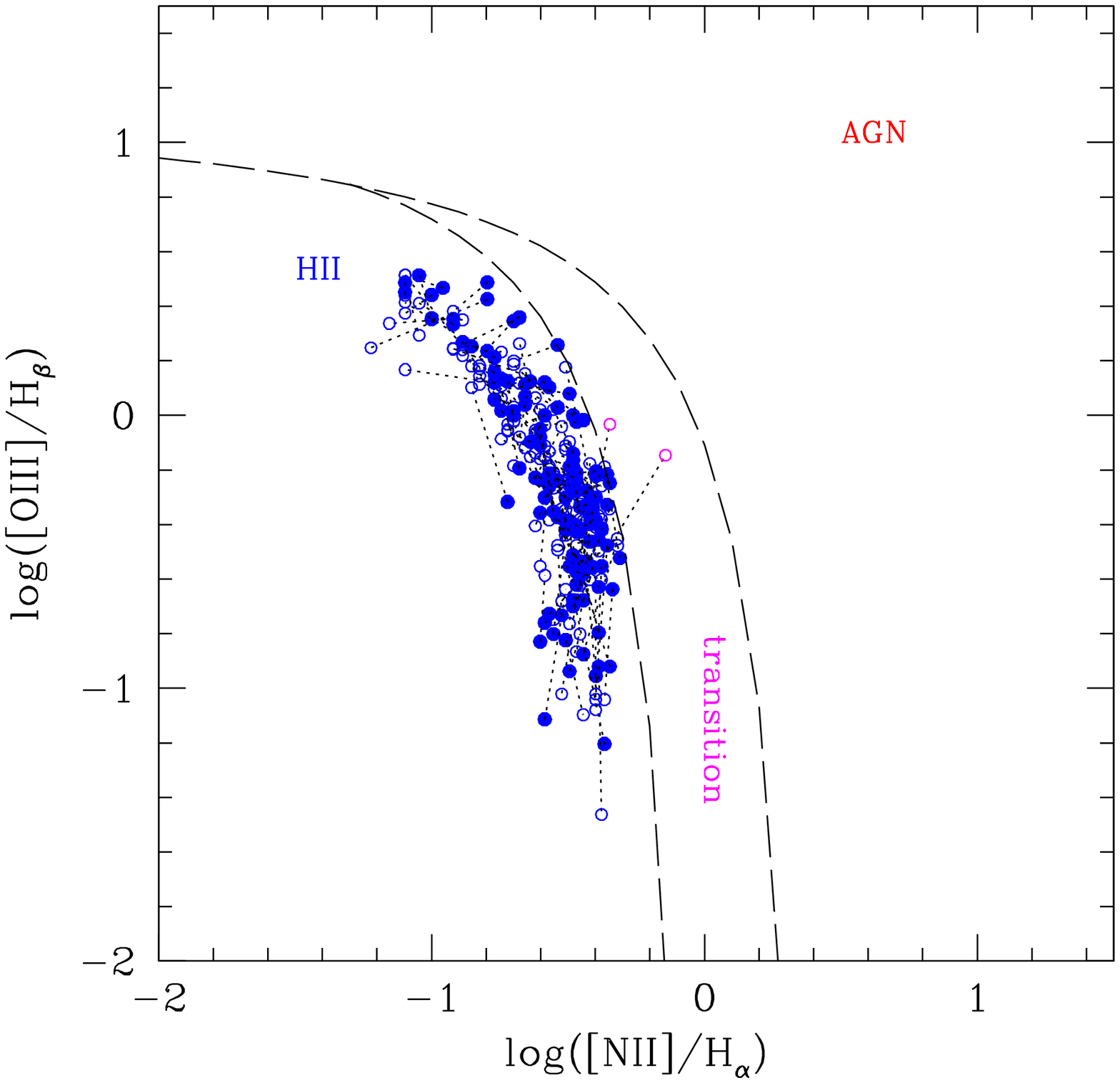}}
\vspace{-0.3cm} 
 \caption{Migration from nuclear (filled symbols) to integrated (empty symbols) in the $BPT$ diagram.}
 \label{transition}
\end{figure*}

\begin{figure*}
\centering
  {\includegraphics[width=7.5cm]{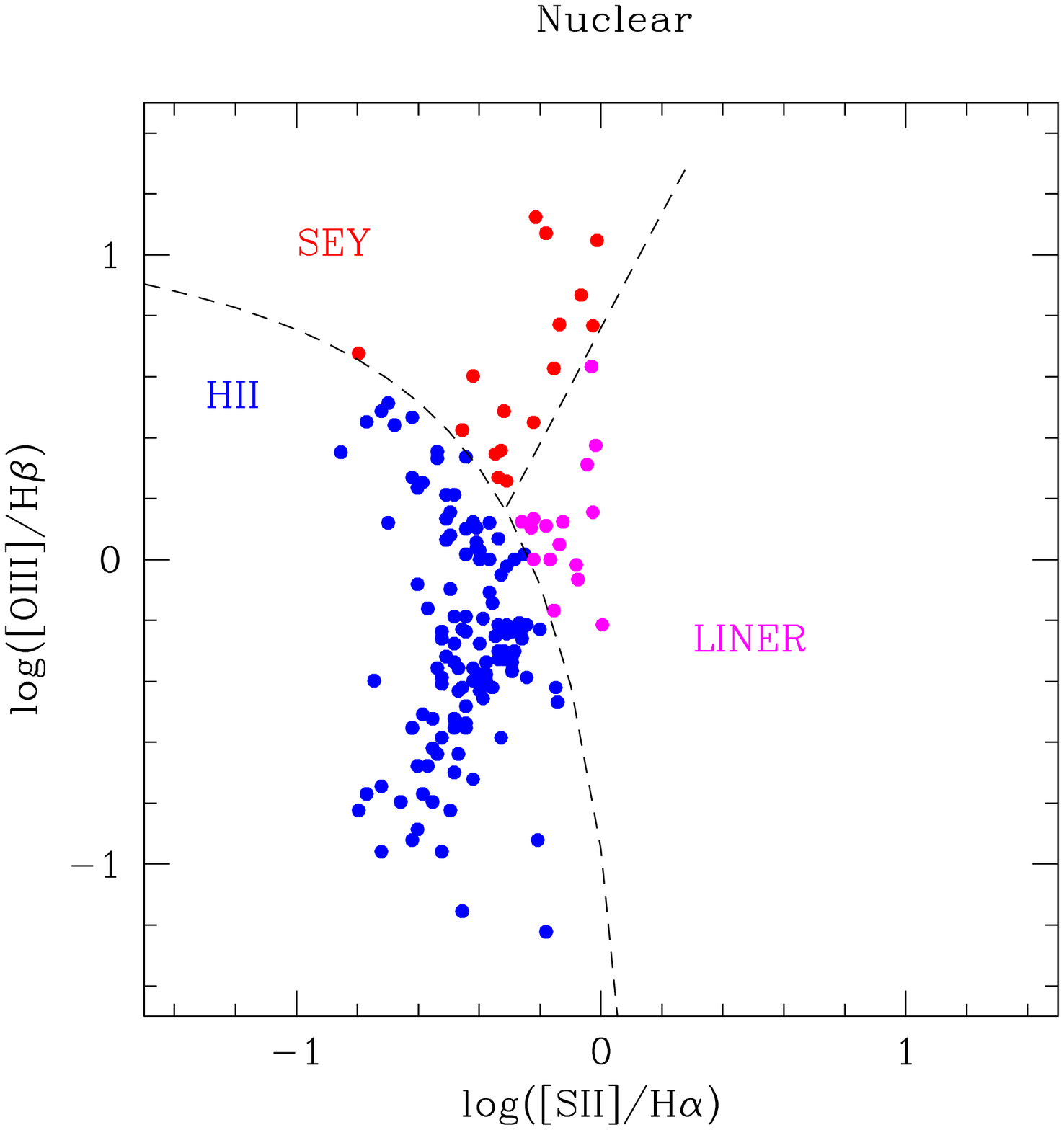}}{\includegraphics[width=7.5cm]{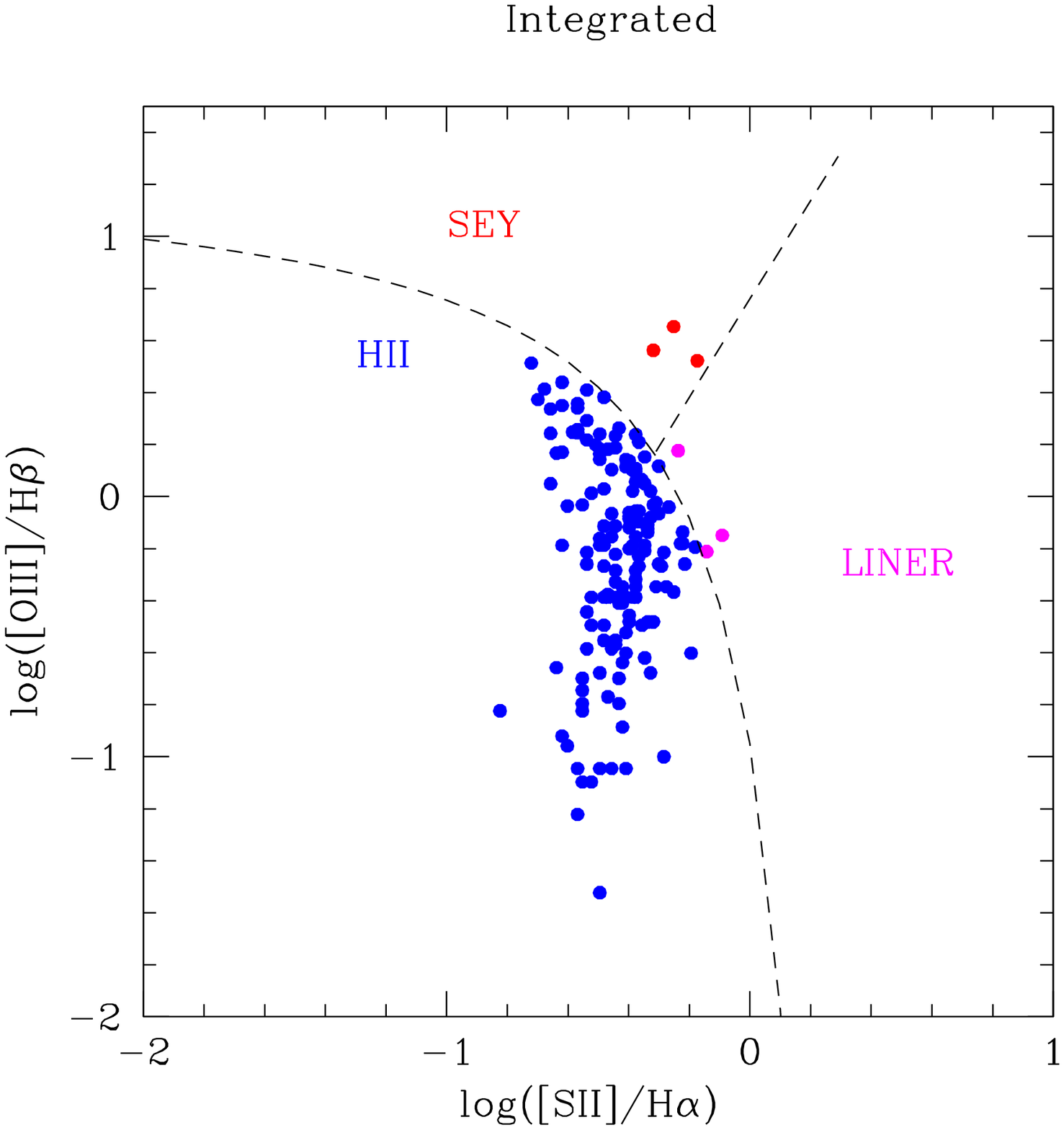}}
\vspace{-0.5cm}
\caption{BPT diagrams for 164 galaxies that have the ratios [OIII]/H$\beta$ and [SII]/H$\alpha$ available both from the nuclear and integrated spectra.
The dotted separation lines are the same as in Figure \ref{nuc}.}
\label{bptSII}
\end{figure*}

\begin{figure*}
\centering
  {\includegraphics[width=7.5cm]{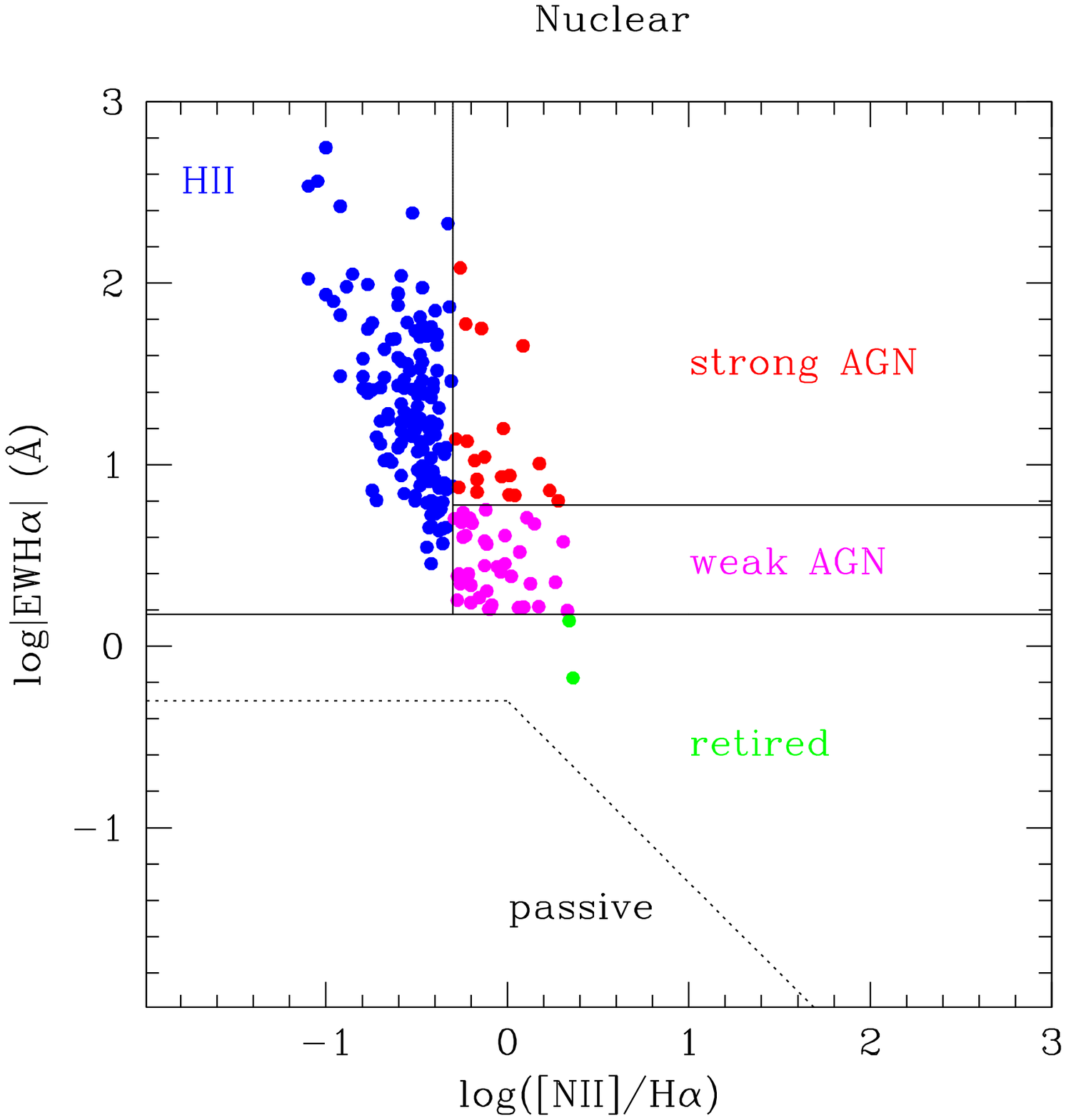}}{\includegraphics[width=7.5cm]{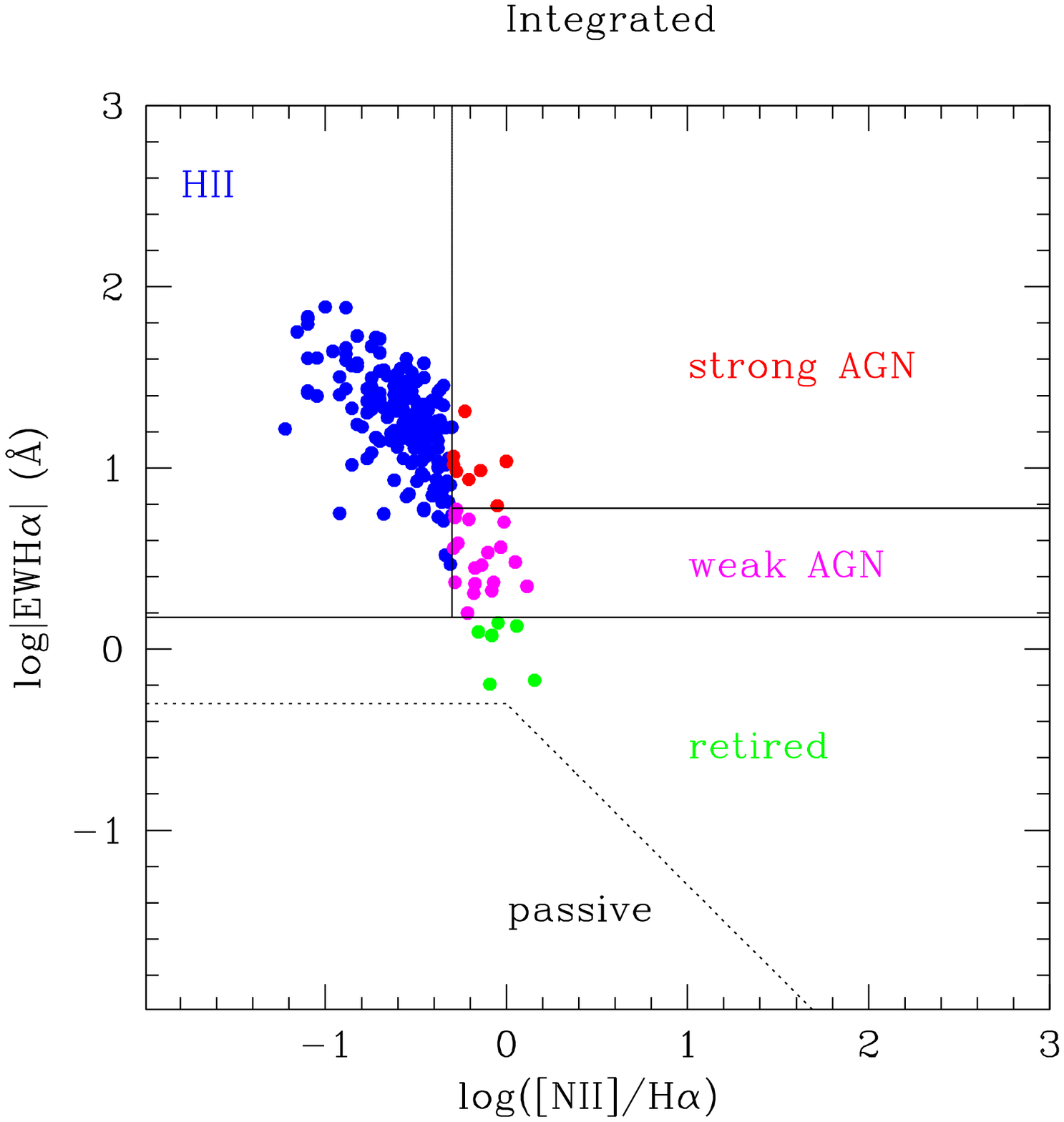}}
  \vspace{-0.5cm}
\caption{WHAN diagrams for 244 galaxies with the ratios [NII]/H$\alpha$ and H$\alpha$ EW available both from nuclear and integrated spectra. 
The separations are the same as in Figure \ref{nuc}.}
\label{whan}
\end{figure*}

\noindent
Despite the different populations and classification criteria, the WHAN and the BPT nuclear diagnostics give consistent results (see Table \ref{statnuc}).
 Hereafter two different percentages are always reported, one is the value resulting from one of the [NII]/H$\alpha$ or [SII]/H$\alpha$ BPT diagrams 
and the other is from the WHAN diagram.
Between 53 and 64\% of the nuclear spectra are HII region-like nuclei.
The remaining percentage is divided between AGNs (21-27\%) and PAS (15-21\%).
The "pure" AGNs are half of the AGNs percentage: 7-13\%. 
These are galaxies ionized by a SMBH.\\
Restricting to the 254 LTGs alone, the statistics is altered (see right columns of Table \ref{statnuc}): 
the HII-regions increase to 67-77\%; the AGNs become 21-27\%; and the PAS spectra decrease to 2-7\%.
The latter class includes two PSB galaxies: HRS-116 and HRS-195.
These two objects show a late-type morphology, with a central bulge and spiral arms, but the SF activity has been stopped suddenly.
Also HRS-164 is a LTG with a PAS spectrum: its morphology shows a strong barred structure in the center, but it is an old and red object.\\
A comparison between the nuclear and integrated spectral classifications can be performed among the subsample 
of HRS galaxies with available nuclear and integrated classification (see Fig. \ref{BPT}, Fig. \ref{bptSII}, and Fig. \ref{whan}); the subsample is reported in Table \ref{statcomb}. 
This comparison provides information about how the excitation properties of galaxies change from the center to the outskirts.
As can be seen in the three figures, AGNs represent 18-26 \% in the nuclear spectra; while in the integrated they decrease to 4-11\%. 
HII-regions are 59-71\% in the nuclear classification, becoming 69-84\% among the integrated spectra.
PAS (including PSB) and RET are 11-15\% in the nuclear spectra, becoming 12-20\% in the integrated spectra. 
Altogether, the two diagnostics give consistently that the integrated spectra contain $\sim$ 10\% more HII-region-like spectra than the nuclear ones, 
together with 10\% less AGNs.
PAS spectra have an equal frequency: they are 11-15\% in the nuclear classification, becoming 12-20\% in the integrated one.
Figure \ref{transition} gives a graphical representation of the migration from nuclear to integrated classification, separately for the various spectral classes.
Filled symbols refer to the position of nuclear spectra in the BPT ([NII]/H$\alpha$) diagram, 
while empty symbols give the position of integrated spectra in the same diagram. 
Unsurprisingly, only three nuclear AGNs remain AGNs in the integrated classification; all the remaining ones end up as TRAN or HII-regions.
All but four nuclear TRAN become HII-regions in the integrated spectra.
All but two nuclear HII-regions remain integrated HII-regions.
We must highlight that the integrated statistics is biased toward the LTGs, 
because not all the integrated spectra were taken for the ETGs by Boselli et al. (2013) (only 18/68).\\

\begin{table*}
\centering
\caption{The frequency of the different types of nuclear spectra in the whole HRS sample (left) and among the 254 LTGs (right), 
including galaxies a posteriori classified as passive or post-starburst.}
\begin{tabular}{|c|c|}
\hline
                                 & AGN  $34$ $(13\%)$ \\
                                 & TRAN $31$ $(12\%)$ \\
BPT ([NII]/H$\alpha$) $255/322$  & HII $152$ $(60\%)$ \\
                                 & PAS $36$ $(14\%)$ \\
                                 & PSB $2$ $(1\%)$ \\
                                 
\cline{1-2}
\hline
                                  & SEY  $23$ $(9\%)$ \\
                                  & LIN $30$ $(12\%)$ \\
BPT ([SII]/H$\alpha$)  $255/322$  & HII $164$ $(64\%)$ \\
                                  & PAS $36$ $(14\%)$ \\
                                  & PSB $2$ $(1\%)$ \\  
                                     
\cline{1-2}
\hline
               & sAGN  $24$ $(7\%)$ \\
               & wAGN $63$ $(20\%)$ \\
WHAN $322/322$ & HII $171$ $(53\%)$ \\
               & RET $28$ $(9\%)$ \\
               & PAS $34$ $(11\%)$ \\
               & PSB $2$ $(1\%)$ \\
\cline{1-2}
\end{tabular}\qquad
\begin{tabular}{|c|c|}
\hline
                                & AGN  $26$ $(13\%)$ \\
                                & TRAN $27$ $(13\%)$ \\
BPT ([NII]/H$\alpha$) $206/254$ & HII $149$ $(72\%)$ \\
                                & PAS $2$ $(1\%)$ \\    
                                & PSB $2$ $(1\%)$ \\  
\cline{1-2}
\hline
                                 & SEY  $20$ $(10\%)$ \\
                                 & LIN $24$ $(11\%)$ \\
BPT ([SII]/H$\alpha$) $206/254$  & HII $158$ $(77\%)$ \\
                                 & PAS $2$ $(1\%)$ \\
                                 & PSB $2$ $(1\%)$ \\  
\cline{1-2}
\hline
                & sAGN  $21$ $(8\%)$ \\
                & wAGN $47$ $(19\%)$ \\
WHAN $254/254$  & HII $168$ $(67\%)$ \\
                & RET $14$ $(6\%)$ \\
                & PAS $2$ $(0.5\%)$ \\
                & PSB $2$ $(0.5\%)$ \\
\cline{1-2}
\end{tabular}
\label{statnuc}
\end{table*}

\begin{table*}
\centering
\caption{Statistics of integrated and nuclear spectra for HRS galaxies with line ratios available both from the nuclear and integrated spectra, 
including galaxies a posteriori classified as passive or post-starburst.}
\begin{tabular}{|c|c|c|}
\hline
\multicolumn{1}{|c|}{} & \multicolumn{1}{|c|}{Nuclear} & \multicolumn{1}{|c|}{Integrated} \\ \hline
                                & AGN  $15$ $(8\%)$   &  AGN  $3$ $(2\%)$    \\
BPT  ([NII]/H$\alpha$)          & TRAN $20$ $(11\%)$  &  TRAN $11$ $(6\%)$   \\
$185/322$                       & HII $129$ $(70\%)$  &  HII $148$ $(79\%)$  \\
                                & PAS $19$ $(10\%)$   &  PAS $23$ $(13\%)$   \\
                                & PSB $2$ $(1\%)$     &                       \\
\hline
\hline
                         & SEY $16$ $(9\%)$   &  SEY $3$ $(2\%)$     \\
BPT ([SII]/H$\alpha$)    & LIN $16$ $(9\%)$   &  LIN $3$ $(2\%)$     \\
$185/322$                & HII $132$ $(71\%)$ &  HII $156$ $(84\%)$  \\
                         & PAS $19$ $(10\%)$  &  PAS $23$ $(12\%)$   \\
                         & PSB $2$ $(1\%)$    &                      \\
               
\hline
\hline
               & sAGN  $22$ $(8\%)$  & sAGN  $8$ $(3\%)$   \\
WHAN           & wAGN $46$ $(18\%)$  & wAGN $20$ $(8\%)$   \\
$264/322$      & HII $155$ $(59\%)$  & HII $182$ $(69\%)$  \\
               & RET $20$ $(7\%)$    & RET $21$ $(8\%)$   \\
               & PAS $18$ $(7\%)$    & PAS $32$ $(12\%)$   \\
               & PSB $2$ $(1\%)$     &                     \\
\hline
\end{tabular}
\label{statcomb}
\end{table*}

\subsection{Mass and environment dependence}

\begin{figure*}
\centering
{\includegraphics[width=8.5cm]{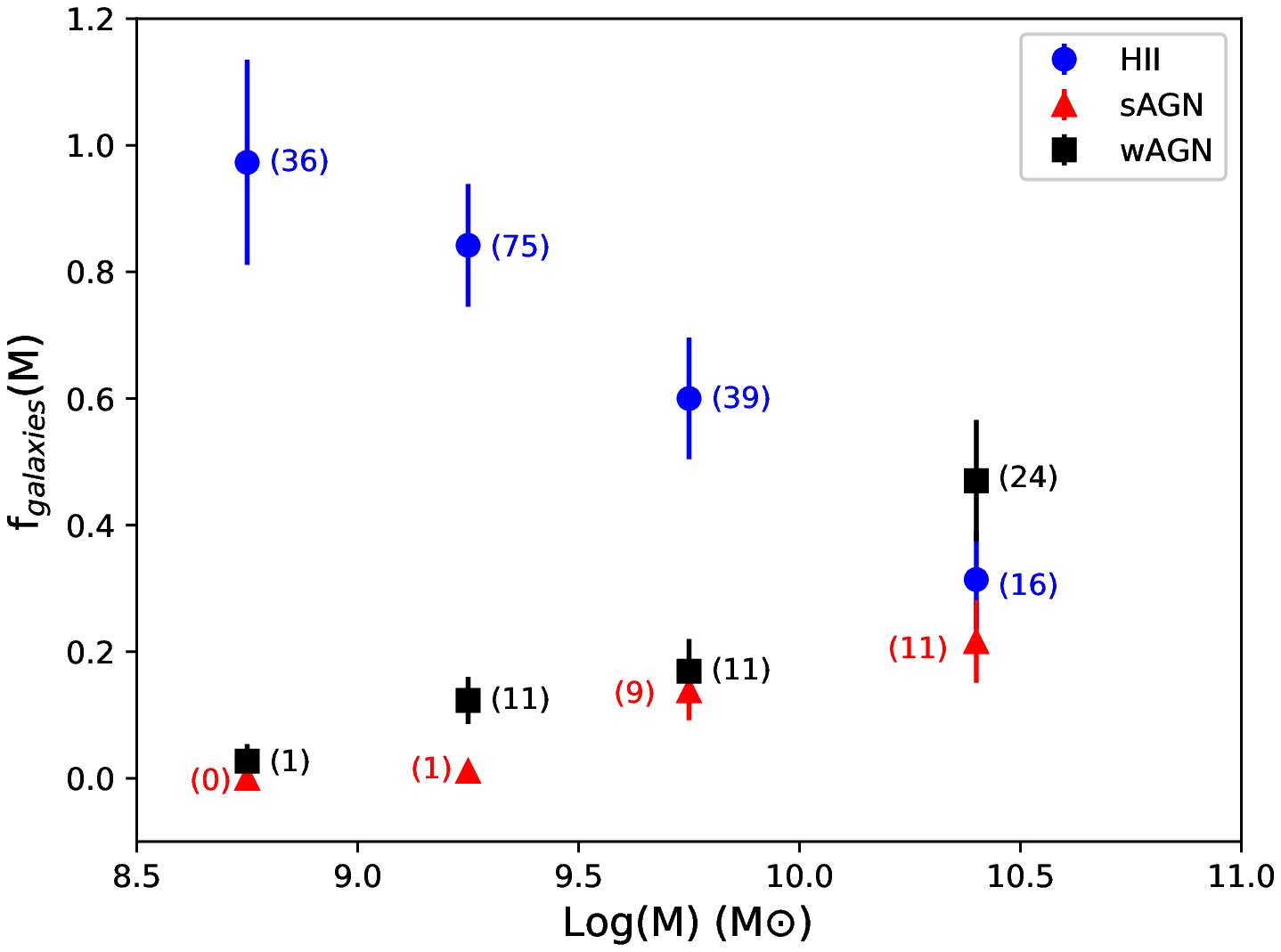}}{\includegraphics[width=8.5cm]{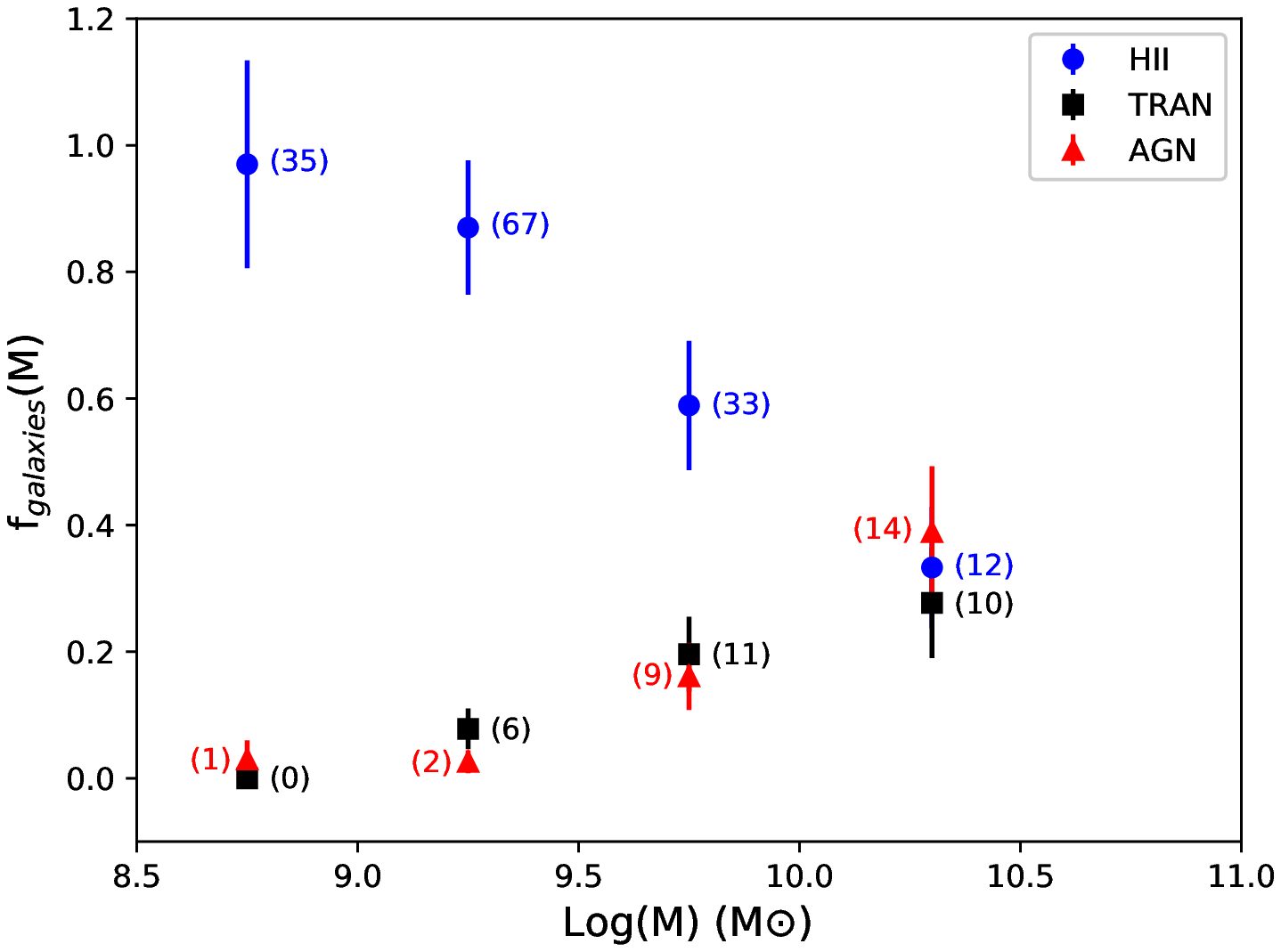}}
\caption{Distribution of nuclear spectral type fractions for LTGs as a function of the stellar mass, using the classification provided by the WHAN (left) 
and by the BPT ([NII]/H$\alpha$) (right) diagram.}
\label{histo}
\end{figure*}

Figure \ref{histo} shows the distribution of the different nuclear spectral type fractions for HRS late-type galaxies 
as a function of the stellar mass (from B15) with the respective error bars, computed using the WHAN diagram (left) 
and the BPT ([NII]/H$\alpha$) diagnostic (right). 
The number of galaxies in each specific mass bin is reported in parentheses.
In spite of the different classification criteria, WHAN and BPT diagrams give consistent dependence of the nuclear spectral 
properties from stellar mass, excluding PAS spectra because their statistics in the LTG subsample is very small (only 4 objects).
As shown in Fig. \ref{histo}, nuclear HII regions (blue) are exclusively found among galaxies with $\rm M_{\ast}$ < $10^{9.5}$ $\rm M_{\odot}$ (90$\pm$9\%), 
and then their fraction decreases with increasing stellar mass.
On the contrary, the frequency of AGNs (red) (including TRAN, black) is low, at $\rm M_{\ast}$ < $10^{9.5}$ $\rm M_{\odot}$ 
(8$\pm$2\%) and increases significantly with stellar mass, reaching 66$\pm$14\% above $10^{10.0}$ $\rm M_{\odot}$, 
suggesting that approximately two thirds of the late-type galaxies in the local Universe, at that stellar mass value, contain an AGN or at least a TRAN object.\\
This also suggests that there is a clear dependency of the nuclear ionization source from galaxy stellar mass: 
less-massive galaxies are more likely to contain a nucleus ionized by young stellar population, compared to more-massive galaxies, which contain an AGN or a TRAN object.\\  
Moreover previous works, such as K03, demonstrate that the AGNs fraction continues to grow at higher stellar mass.
 We conclude that even on nuclear scales, LTG galaxies suffer from decreasing SFR with increasing
mass, a "downsizing" behavior, reinforcing the early claim by Gavazzi et al. (1996).  \\


\subsection{Star formation: the Brinchmann 04 program}
\label{Br04}

\begin{figure}
\centering
 \includegraphics[width=8.5cm]{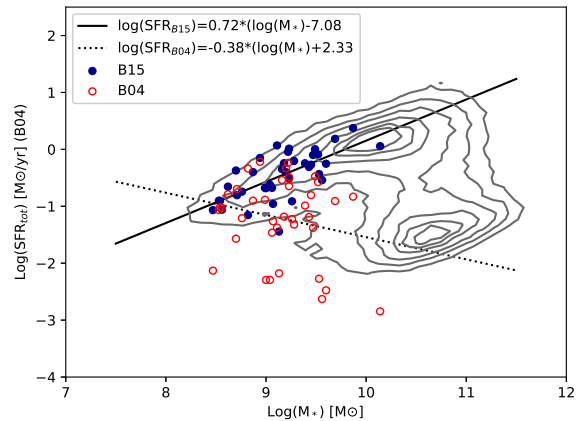}
\caption{The relation between the stellar mass and the global SFR as computed by B04 (contours). 
Open symbols represent 38 HRS galaxies computed by B04 (dotted line represents the best fit to these data),  
while filled symbols are the integrated SFR derived for them by B15 using H$\alpha$ integrated photometry 
(solid line is the best fit to these data).
The filled symbols precisely overlap the SFMS of galaxies in the local Universe; 
instead the value of SFR estimated by B04 is underestimated and does not follow the SFMS trend.}
\label{sfr}
\end{figure}

Understanding the physical processes that drive SF and the rate at which galaxies form stars is crucial for the study of galaxy evolution. 
In recent decades, the available multiwavelength observations, from X-rays to radio bands, have been used to define a set of SF  indicators.
The most frequently used are: the UV continuum, nebular recombination lines (primarily H$\alpha$), far-IR dust emission, 
and the synchrotron radio continuum at 21 cm (Kennicutt 1983).
Only the comparison of two or more SF indicators provides reliable measure of the SFR in galaxies.\\
Many works in the literature proposed to compute the SFR directly from nuclear spectroscopy.
Whether or not the global SF properties of nearby galaxies can be extrapolated from the available SDSS fiber spectroscopy 
($aperture$ $correction$ problem) is still debated. \\
\noindent
In this section, we test for the HRS galaxies the global SFR derived by B04 from SDSS nuclear spectra with the H$\alpha$-derived SFR of B15. 
B04 proposed a method to extrapolate the global SFR of local galaxies from SDSS nuclear spectra using aperture correction based on photometric optical colors.
Summarizing, B04 computed the SFR of AGNs and TRAN objects from the 4000 \AA~ break and for HII-regions from a global 
emission-line estimate obtained by fitting Charlot \& Longhetti (2001) models to the nuclear spectra for galaxies belonging to the SDSS database. 
These values were corrected using an estimate of the external color of galaxies derived from the difference in color between 
the $cModel$ magnitude (the magnitude based on the best-fitting of the exponential and de Vaucouleurs models in the $r$ band, Abazajian et al. 2004) 
and the $fiber$ magnitude (the magnitude in 3-arcsec-diameter SDSS fiber radius).
The global SFR is obtained by adding the external to the nuclear SFR.
Both the stellar mass and SFR in B04 are computed assuming a Kroupa (2001) IMF. B04 data are available through the SDSS database ($galSpecExtra$ table)
for 205186 galaxies with $z$<0.1 in the sky region corresponding to the one covered by the HRS. 
Figure \ref{sfr} represents the relation between the stellar mass and the log(SFR$_{total}$) for the aforementioned SDSS galaxies (see contours). 
The appearance of the galaxy main sequence is evident, separated by a cloud of quenched galaxies, a factor of one hundred less star forming. \\
Among galaxies listed in galSpecExtra we found the data for 38 HRS galaxies. For these we plot  the value derived by B04 with open symbols. 
Conversely the filled symbols represent the SFR obtained by B15 directly from the H$\alpha$ integrated photometry 
(corrected to transform from Salpeter 1955 to Kroupa 2001 IMF).
The filled symbols precisely overlap the galaxies SF main sequence (SFMS). 
On the contrary, most B04 SFR estimates are at least a factor of one hundred less than the actual integrated values.
In particular, for three galaxies, the B04 value is smaller by a factor of one hundred; 
for eight galaxies, the B04 value is smaller by a factor of between ten and one hundred.
The logarithmic mean deviation between B04 and B15 SFR estimations is 0.7.
However we must highlight that the B04 study includes SDSS galaxies  with $0.005 < z < 0.22$,
while the HRS comprises galaxies with $cz<3000 ~km ~s^{-1}$, therefore suffering from 
more severe aperture correction than B04.

\section{Discussion and conclusions}
\label{summary}
In this paper we have used the BPT  and the WHAN diagnostic diagrams
to derive the nuclear spectral classification of galaxies belonging to the HRS, a statistically representative 
magnitude and volume limited sample of 322 local galaxies, spanning a wide range in morphology and stellar mass, 
and belonging to different environments.  
The determination of the relative frequency of AGNs versus other spectral classes, for example, HII region-like, PAS, and RET,
in a statistically complete sample of local galaxies is important to discriminate 
the source of ionization in the nuclear region of galaxies (e.g., black holes vs. young or old stars). \\
We present new nuclear long-slit spectroscopy of 45 HRS galaxies, which, added to the ones available form the literature, gives a complete 
nuclear spectroscopic census of all HRS galaxies more massive than $10^{8.5}$$\rm M_\odot$. 
The completeness of the observations allows us to make a statistical analysis of the spectroscopic nuclear properties of this sample.\\
We note that, after checking that using different diagnostics does not change the global statistics, 
we separated the different spectral classes by means of only the BPT(NII/Ha). 
We separated AGNs from HII and TRAN objects following the prescriptions of Kewley et al. (2001) and K03.
Our main results can be summarized as follow:
\\
\begin{figure*}
\centering
{\includegraphics[width=8.5cm]{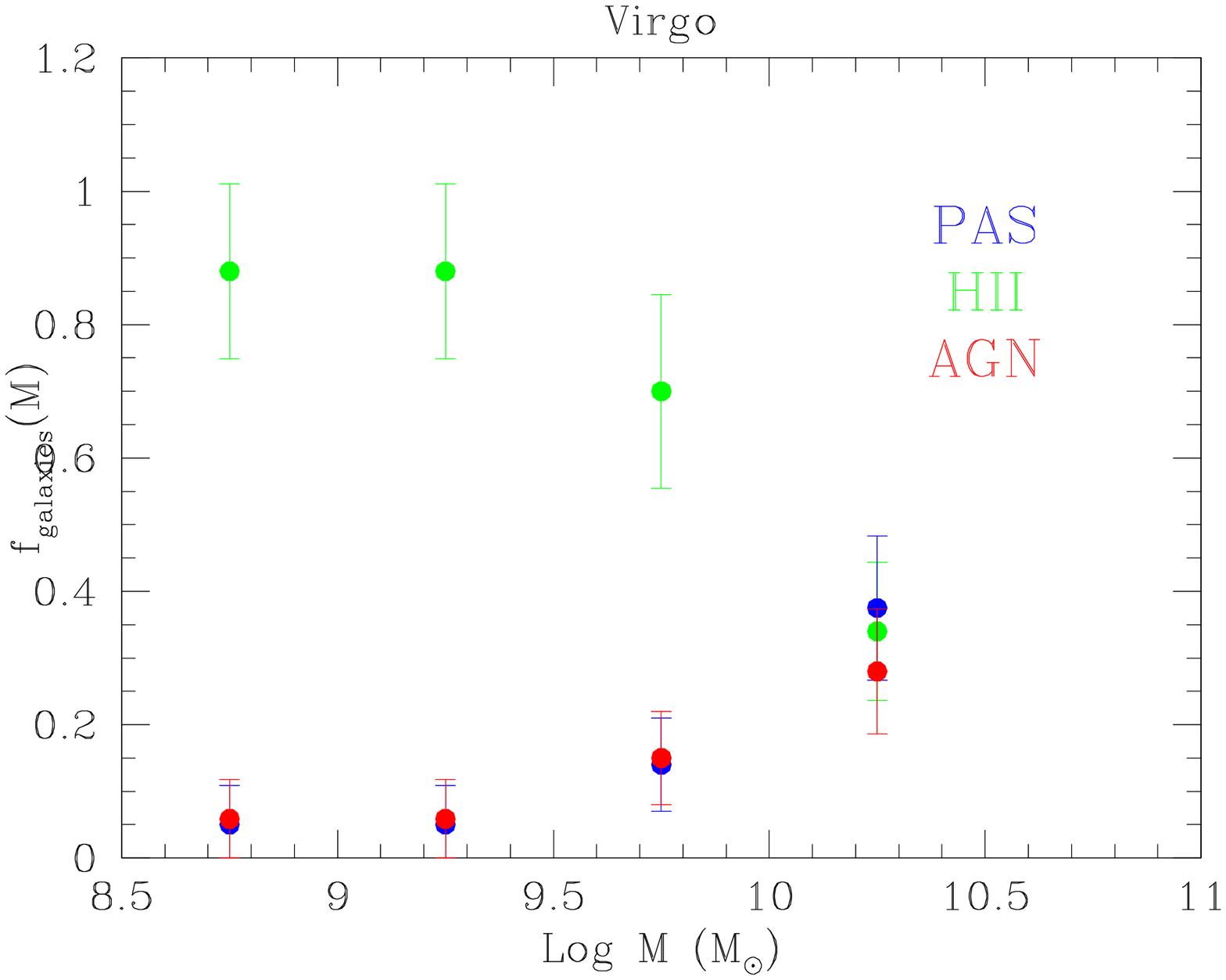}}{\includegraphics[width=8.5cm]{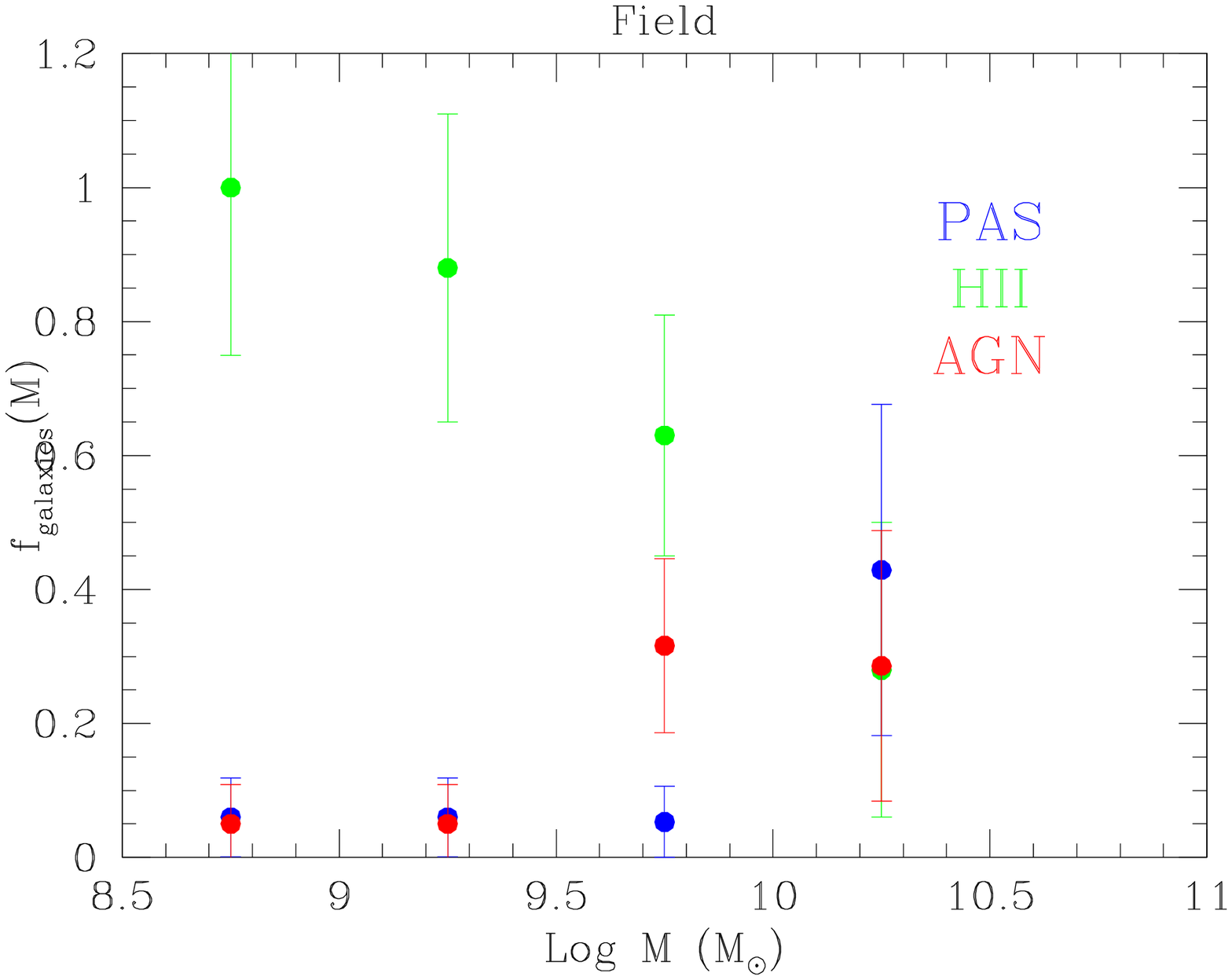}}
\caption{Mass distributions of the  $mode$ of four different nuclear spectral classifications  for the late-type HRS galaxies
 inside the Virgo cluster (left) and outside it (right).}
\label{frac}
\end{figure*}

1) As for the whole HRS sample about half of the nuclear spectra (53-64\%) are HII region-like nuclei, while 
AGNs (including TRAN) represent 21-27\% and PAS (including RET) spectra make up 15-21\%. Restricting to the 254 LTGs, the percentage of AGNs+TRAN varies between 22\% and 27\%, 
the fraction of HII-regions increases to 67-77\%, and PAS (including RET) spectra represent a mere 1-7\%.
However, the classification schemes provided by the various diagnostics  cannot be fully reconciled: for example, the class "Transition" exists only in the classic BPT diagram, and the division between strong and weak AGNs is provided only by the WHAN diagram; the  same applies to the classification of retired galaxies 
(that were identified by Cid Fernandez et al. (2011) as presently inactive nuclei, previously ionized by post-AGB stars).
 Moreover, the scatter between the various classifications appears 
large, as it was also concluded, in particular for LINERs, by Balmaverde \& Capetti (2014) who
compared HST high spatial- and spectral-resolution observations of low-luminosity AGNs with ground-based lower-resolution spectra.  
To reduce as much as possible the inconsistencies of the individual spectral classification schemes, we evaluate the $mode$ between the  four classifications in Figure 
\ref {nuc}.
In order to find a criterion for reliable classification, we select all galaxies whose $mode$ is consistently provided by at least three methods
(we note that spectra in which emission lines are absent, no matter if they were considered RET in the WHAN, PSB (poststarbust) or purely passive (PAS),
were reclassified by hand as PAS, and therefore their $mode$ classification is PAS with four votes).  
In the end, 195 targets out of 264 LTG galaxies have their $mode$ composed of three coherent classifications. 
In order to study a possible environmental dependence of these frequencies we plot the mass distributions of the spectral fractions 
in Fig. \ref{frac}, separately inside (left panel) and outside (right panel) the Virgo cluster 
(projected distance from M87 equal to two virial radii, assuming that the virial radius of the Virgo cluster is 1.7 Mpc from Boselli \& Gavazzi 2006).
From Fig. \ref{frac} it is clear that the dependency of the nuclear spectral types on galaxy stellar mass remain visible also observing galaxies in different environments.
However, no significant environmental dependence is found for the AGN frequency: among galaxies with $\rm M_{\ast}$ $\geq$ $10^{10.0}$ $\rm M_{\odot}$ 
they are consistent with 30$\pm$7\% inside or outside the Virgo cluster.
In both environments the decrease of the frequency of HII region-like nuclei with increasing stellar mass is compensated by a combined increase 
of the fraction of AGNs and of passive nuclei (including those previously classified as RET that are known to derive from post AGB stars).

2) Regarding the comparison between the nuclear and the integrated classification, as expected, 
the fraction of AGNs+TRAN integrated spectra is significantly lower than in the nuclei; they are only 4-11\% instead of 18-26\% ($\sim$10\% lower). 
This decrease is compensated by an equal increase of HII region-like spectra (from 59-71\% in the nuclear to 69-84\% in the integrated classification). 
There is no significant change from the nuclear to the integrated classification in the percentage of PAS (including RET) spectra. This confirms previous results by Moustakas et al. (2006, 2010) and Iglesias-P{\'a}ramo et al. (2013, 2016) who found that the relative fraction of 
star-forming galaxies versus AGNs is a strong function of the integrated light enclosed by the spectroscopic aperture.
Iglesias-P{\'a}ramo et al. (2013, 2016) observed 104 spiral galaxies in the CALIFA survey with an IFS, finding that the H$\alpha$ flux and 
the H$\alpha$ EW increase from the nucleus to the outer regions of galaxies with no dependence on galaxy inclination and stellar mass.
Furthermore, Moustakas et al. (2010) provided optical nuclear, circumnuclear, and semi-integrated spectra of 65 galaxies in SINGS (Kennicutt et al. 2003), 
finding that the fraction of the sample classified as AGNs decreases by $\sim$30\% as the light fraction increases by $\sim$50\%, 
with a corresponding increase in the fraction of galaxies classified as HII-regions.
Finally, Belfiore et al. (2016), who studied the spatially resolved excitation properties of the ionized gas for 646 galaxies belonging to SDSS-IV MaNGA, 
concluded that the excitation properties of galaxies change from the center to the outskirts in a very complex way.
 As the coverage of ETGs in the HRS is incomplete, we refer the reader to Gavazzi et al. (2018) who analyze the nuclear properties of a sample of 360 ETGs from the ATLAS3D.

3) Restricting to the 254 LTGs, we computed the spectral classes distribution as a function of stellar mass.
The percentage of HII regions is inversely proportional to the stellar mass (90\% for galaxies with $\rm M_{\ast}$ < $10^{9.5}$ $\rm M_{\odot}$) 
while that of AGNs (including TRAN) is directly proportional. We find that approximately two thirds (66\%) of spiral galaxies in the local Universe ($z$<0.03) 
with $\leq$ $\rm M_{\ast}$ > $10^{10.0}$ $\rm M_{\odot}$ 
contain an AGN or a TRAN object, that is, a nucleus not ionized by a young stellar population (O and B stars), but ionized by a supermassive black 
hole or old post-AGB stars.
Our result is consistent with K03 who examined the properties of the host galaxies of 22623 AGNs with 0.02<$z$<0.3 selected from the SDSS.
First of all, we note that K03 defines AGNs as all galaxies that lie to the right of the line defined by K03 in the BPT diagram, 
so we can compare our results on AGNs+TRAN to these.
K03 found that AGNs reside almost exclusively in massive galaxies: between 3x$10^{10}$ $\rm M_{\odot}$ and $10^{11}$ $\rm M_{\odot}$ 
, they represent 50\% at very low redshift ($z\sim$0.02). 
More recently, Sanchez et al. (2017) remarks that AGNs are hosted in the most massive galaxies, namely mostly ETGs or early-type spirals, 
with an important bulge.\\

4) Restricting to the 254 LTGs we also computed the spectral type distribution as a function of environment, considering that 148 HRS galaxies belong to the Virgo cluster.
No significant environmental dependence is found for AGNs (including TRAN): for galaxies with $\rm M_{\ast}$ $\geq$ $10^{9.0}$ 
they are consistent with $\sim$33\% inside the Virgo cluster and with $\sim$30\% outside it. 
Similar results are obtained restricting to the "pure" AGNs, at the same stellar mass; they reach $\sim$13\% inside the Virgo cluster and $\sim$14\% outside it. 
These results suggest that AGNs, including or excluding TRAN objects, exist with similar frequencies in clusters and in the field. 
Both inside and outside Virgo, the frequency of AGNs found in this work is higher than the percentage found by Marziani et al. (2017) in the WINGS survey (Fasano et al. 2006)
and by B04 in the SDSS.
However, after inspection of the individual AGN spectra in this work (prior and after the GANDALF correction) we found that 
the classification of 4 out of 26 AGN spectra is highly uncertain, making the above difference insignificant.

\noindent
5) B15 computed the SFR for the HRS galaxies. Thirty-eight of these are also part of B04.
The SFR derived from the global measurements (both spectroscopic and imaging) for this set of 38 HRS galaxies is inconsistent 
with the value obtained by B04 with his aperture-corrected method: 
the former being in most cases two orders of magnitude larger than the values of B04.
This confirms previous results by Richards et al. (2015) who tested the B04 method for 1212 galaxies belonging to the SAMI Galaxy survey (Croom et al. 2012).
Richards et al. (2015) conclude that the nuclear spectrum is not representative of the properties of the entire galaxy,
mainly because of the different dust content: H$\alpha$/H$\beta$ ratio (extinction) decreases from the center towards larger radii; therefore,
galaxies with different morphology require different dust attenuation corrections. 
Similar results have been obtained by Iglesias-P{\'a}ramo et al. (2013) who confirmed that local star-forming galaxies observed through a small aperture 
are misclassified as quiescent if the aperture-correction method is based only on the nuclear properties, 
because the fraction of a galaxy covered by a fixed aperture varies with redshift. 
We maintain that the B04 aperture-correction method does not provide a robust extrapolation of the global SFR for local galaxies ($z$<0.03).

\begin{acknowledgements}
This research has made use of the GOLDmine database (Gavazzi et
al. 2003, 2014b) and of the NASA/IPAC Extragalactic Database (NED) which is
operated by the Jet Propulsion Laboratory, California Institute of Technology, 
under contract with the National Aeronautics and Space Administration.  
Funding for the Sloan Digital Sky Survey (SDSS) and SDSS-II has been provided  
by  the  Alfred  P.  Sloan  Foundation,  the  Participating  Institutions,  
the National  Science  Foundation,  the  U.S.  Department  of  Energy,  
the  National  Aeronautics and Space Administration, the Japanese Monbuk
agakusho, and the Max Planck Society, and the Higher Education  Funding Councill  
for England. The  SDSS  Web  site  is http://www.sdss.org/.
The  SDSS  is  managed  by  the Astrophysical  Research  Consortium  (ARC)  for  the  Participating  
Institutions.The  Participating  Institutions  are  the  American  Museum  of  Natural  History, 
Astrophysical Institute Potsdam, University of Basel, University of Cambridge, 
Case Western Reserve University, The University of Chicago, Drexel University, 
Fermilab,  the  Institute  for  Advanced  Study,  the  Japan  Participation  Group, 
The  Johns  Hopkins  University,  the  Joint  Institute  for  Nuclear  Astrophysics, 
the  Kavli   Institute   for  Particle   Astrophysics   and   Cosmology,  
the  Korean Scientist  Group,  the  Chinese  Academy  of  Sciences  (LAMOST),
Los  Alamos National  Laboratory,  the  Max-Planck-Institute   for  Astronomy  (MPIA),  
the Max-Planck-Institute  for  Astrophysics  (MPA),  New  Mexico  State  University, 
Ohio  State  University,  University  of  Pittsburgh,  University  of  Portsmouth, 
Princeton  University,  the United States  Naval Observatory,  and the University of Washington.
\end{acknowledgements}

\appendix
\section{Tables description}
Here we report the description of the observations at the Cassini telescope and the three tables which include our results.
The observing log of 45 galaxies whose nuclear spectra were taken at Loiano is given in Table \ref{Tobs}.1 

The spectroscopic parameters of HRS galaxies observed at Loiano only with the red-channel grism are given in Table \ref{TloiG8}  


The spectroscopic parameters of 25 HRS galaxies observed at Loiano with the red and blue-channel grism are given in Table \ref{TloiGandalf} 

The spectroscopic parameters derived for all HRS galaxies are given in Table \ref{Tclass} 

\newpage
\begin{onecolumn}
\setlength{\oddsidemargin}{0cm}
\setlength{\evensidemargin}{0cm}
\setlength{\headsep}{0cm}
\tiny
\captionsetup{width=13cm}
\label{Tobs}
\begin{longtable}{|c c c c |c c c | c c c|}
\hline
\multicolumn{4}{|c|}{} & \multicolumn{3}{|c|}{Blue} & \multicolumn{3}{|c|}{Red} \\ \hline
   HRS             & Alpha      &   Delta   &  PA   &  Obs date   &Ncomb& expT  &    Obs date   &Ncomb& expT      \\  
                   & hhmmss.ss  &  ddppss.s &  deg  &  yyyy-mm-dd &    &   sec  &    yyyy-mm-dd &    &   sec      \\  
    (1)            &   (2)      &  (3)      &  (4) &     (5)      & (6)&   (7)  &      (8)      & (9)&   (10)     \\  
\hline
      2*           &  102057.13 & +252153.4 &  30   &  2017-04-20 & 3  &   480  &    2013-2017  & 6  &   300      \\  
      9            &  103255.45 & +283042.2 &  90   &             &    &        &    2015-03-20 & 3  &   180      \\  
     28*           &  105420.89 & +271423.1 &  90   &  2017-04-21 & 3  &   480  &    2009-2017  & 9  &   300      \\  
     32*           &  105448.63 & +173716.4 &  90   &             &    &        &    2009-2013  & 6  &   300      \\  
     33*           &  110002.38 & +145029.7 &  90   &  2017-04-21 & 3  &   480  &    2009-2017  & 15 &   300      \\  
     38*           &  110656.63 & +071026.1 &  90   &             &    &        &    2009-2013  & 9  &   240      \\  
     45            &  111921.60 & +574527.8 &  90   &             &    &        &    2015-03-23 & 4  &   180      \\  
     51*           &  112345.53 & +174907.2 &  90   &  2017-04-20 & 3  &   480  &    2009-2017  & 3  &   300      \\  
     52*           &  112355.58 & +525515.6 &  90   &             &    &        &    2009-2013  & 9  &   180      \\  
     58            &  112809.41 & +165513.7 &  90   &             &    &        &    2015-03-23 & 6  &   300      \\  
     65*           &  114053.42 & +561207.3 &  87   &  2017-05-25 & 3  &   600  &    2013-2017  & 6  &   300      \\  
     77*           &  120023.63 & -010600.3 &  90   &             &    &        &    2012-03-15 & 3  &   360      \\  
     84            &  120411.55 & +105115.8 &  59   &             &    &        &    2015-03-20 & 6  &   300      \\  
     86*           &  120737.15 & +024125.8 &  140  &  2017-04-20 & 3  &   480  &    2009-2017  & 6  &   300      \\  
     98*           &  121622.52 & +131825.4 &  57   &  2017-05-25 & 3  &   600  &    2012-2013  & 6  &   300      \\  
    108*           &  122048.49 & +053823.6 &  74   &  2017-05-26 & 3  &   600  &    2009-2013  & 10 &   300      \\  
    110*           &  122117.79 & +113040.0 &  145  &  SDSS DR7   &    &        &    2012-03-16 & 3  &   300      \\
    118            &  122227.25 & +043358.7 &  55   &  2017-04-20 & 3  &   600  &    2015-2017  & 6  &   360      \\  
    124            &  122317.25 & +112204.7 &  58   &             &    &        &    2015-03-23 & 3  &   300      \\  
    132            &  122407.44 & +063626.9 &  90   &  2017-04-21 & 3  &   600  &    2015-2017  & 6  &   300      \\  
    134*           &  122414.53 & +083209.1 &  90   &             &    &        &    2013-03-12 & 3  &   120      \\  
    148            &  122558.80 & +154017.3 &  134  &  2017-04-22 & 3  &   900  &    2015-2017  & 7  &   420      \\  
    153*           &  122646.72 & +075508.4 &  10   &  2017-04-22 & 3  &   600  &    2012-2017  & 6  &   300      \\  
    164*           &  122753.57 & +121735.8 &  90   &             &    &        &    2012-2009  & 6  &   180      \\  
    171            &  122840.55 & +091532.2 &  157  &             &    &        &    2015-04-14 & 3  &   300      \\  
    185*           &  123059.71 & +080440.3 &  90   &             &    &        &    2006-02-27 & 3  &   300      \\  
    188            &  123139.57 & +165110.1 &  122  &  2017-04-21 & 3  &   420  &    2015-2017  & 6  &   300      \\  
    189*           &  123154.76 & +150726.2 &  90   &  2017-05-27 & 3  &   600  &    2012-03-16 & 3  &   300      \\  
    230            &  123951.91 & +151752.1 &  90   &  2017-05-26 & 3  &   600  &    2015-03-20 & 4  &   300      \\  
    232*           &  124057.53 & +115444.0 &  31   &  2017-05-27 & 3  &   600  &    2012-03-15 & 3  &   300      \\  
    239            &  124240.96 & +141745.0 &  157  &  2017-05-27 & 3  &   600  &    2015-03-23 & 3  &   240      \\  
    249            &  124445.99 & +122105.2 &  72   &             &    &        &    2015-04-14 & 3  &   420      \\  
    253            &  124717.52 & -024338.6 &  22   &  2017-04-22 & 3  &   480  &    2015-2017  & 5  &   300      \\  
    258            &  124835.91 & -054803.1 &  90   &             &    &        &    2015-03-20 & 3  &   180      \\  
    259            &  124911.56 & +032319.4 &  90   &  2017-05-26 & 3  &   600  &    2015-03-20 & 3  &   240      \\  
    266            &  125101.09 & -062335.0 &  125  &  2017-05-27 & 3  &   600  &    2015-03-23 & 3  &   240      \\  
    267            &  125145.96 & +254638.3 &  33   &  2017-04-20 & 3  &   480  &    2015-2017  & 6  &   300      \\  
    270            &  125222.11 & -011158.9 &  90   &             &    &        &    2015-03-20 & 3  &   180      \\  
    282            &  125533.67 & +081425.8 &  90   &             &    &        &    2015-03-20 & 3  &   240      \\  
    289            &  130848.74 & -064639.1 &  90   &             &    &        &    2015-04-14 & 3  &   300      \\  
    291*           &  132030.08 & +430502.3 &  90   &             &    &        &    2009-02-19 & 4  &   300      \\  
    298*           &  134744.99 & +381816.6 &  90   &  2017-04-20 & 4  &   420  &    2012-2017  & 6  &   300      \\  
    302*           &  135411.26 & +051338.8 &  178  &  2017-05-26 & 3  &   600  &    2012-2017  & 11 &   300      \\  
    303*           &  135445.99 & +584000.7 &  90   &  2017-04-20 & 3  &   480  &    2009-2017  & 6  &   360      \\  
    313*           &  142113.11 & +032608.8 &  90   &  2017-04-21 & 3  &   600  &    2009-2017  & 14 &   300      \\  
\hline
\caption{Galaxies observed at the Cassini Telescope. An asterisk marks galaxies whose red spectrum was published in Gavazzi et al. (2011) or Gavazzi et al. (2013).
 The spectrum of HRS110 is a combination of a red spectrum taken at Loiano and a blue spectrum from the SDSS (DR7), whose red part near H$\alpha$  was damaged.
 }
 \footnote{Columns are: (1): HRS name; (2) and (3): J2000 Celestial coordinates; (4): Position Angle (P.A.) of the slit with respect to N (counterclockwise);
(5): Observing date (blue grism);  (6): Number of exposures (blue grism);  (7): Exposure time (in seconds) of the individual exposures (blue grism); 
(8): Observing date (red grism); (9): Number of exposures (red grism); 
(10): Exposure time (in seconds) of the individual exposures (red grism)}
\end{longtable}
\tiny

\newpage
\setlength{\oddsidemargin}{0cm}
\setlength{\evensidemargin}{0cm}
\setlength{\headsep}{0cm}
\tiny
\begin{longtable}{|c|c c c c c c c c|}
\hline
HRS          &  Mass      &    g-i    &  H$\alpha$ correction &  H$\alpha$EW &  H$\alpha$EW corr    &  [NII](6583)EW   &  rms        & WHAN      \\
             & $\rm M_{\odot}$&    mag    &     $\AA$         &   $\AA$      &       $\AA$           &     $\AA$        & count  &           \\
 (1)         &    (2)     &   (3)     &     (4)               &    (5)       &       (6)     &     (7)          &  (8)   &  (9)      \\
\hline                                                                                                                               
   9         &  10.08     &  1.081    &    1.82               &   1.69       &        3.51           &     3.12         &  0.04  &  wAGN     \\
  32         &  9.46      &  1.044    &    1.91               &  -1.28       &        0.63           &     0.85         &  0.02  &  RET      \\
  38         &  9.1       &  0.572    &    2.25               &   13.1       &        15.35          &     4.00         &  0.16  &  HII      \\
  45         &  10.25     &  1.248    &    1.82               &   0.23       &        2.05           &     1.70         &  0.05  &  wAGN     \\
  52         &  9.11      &  0.933    &    1.91               &   3.63       &        5.54           &     2.34         &  0.02  &  HII      \\
  58         &  8.94      &  0.661    &    2.44               &   3.89       &        6.33           &     1.99         &  0.10  &  HII      \\
  77         &  10.54     &  0.938    &    1.82               &   0.68       &        2.5            &     1.36         &  0.02  &  wAGN     \\
  84         &  9.39      &  0.851    &    1.91               &  -0.28       &        1.63           &     1.87         &  0.05  &  wAGN     \\
 124         &  9.52      &  0.732    &    2.25               &   4.96       &        7.21           &     -            &  0.18  &  HII      \\
 134         &  9.96      &  0.797    &    2.25               &  -1.12       &        1.13           &     -            &  0.11  &  HII      \\
 164         &  9.91      &  1.061    &    1.91               &  -1.48       &        0.43           &     -            &  0.02  &  PAS      \\
 171         &  9.69      &  0.75     &    2.25               &   14.92      &        17.17          &     5.07         &  0.07  &  HII      \\
 185         &  10.06     &  1.171    &    1.82               &  -2.61       &       -0.79           &     0.91         &  0.07  &  RET      \\
 249         &  9.01      &  0.688    &    2.25               &   1.86       &        4.11           &     1.63         &  0.11  &  HII      \\
 258         &  11.1      &  1.12     &    1.82               &  -0.31       &        1.51           &     1.16         &  0.06  &  wAGN     \\
 270         &  10.93     &  1.151    &    1.82               &   -          &        -      &     1.48         &  0.04  &  RET      \\
 282         &  9.3       &  1.055    &    1.91               &  -2.27       &       -0.36           &     -            &  0.05  &  PAS      \\
 289         &  10.31     &  0.281    &    0.49               &   5.322      &        5.81           &     2.26         &  0.07  &  HII      \\
 291         &  9.86      &  1.075    &    1.91               &  -1.22       &        0.69           &     0.51         &  0.02  &  RET      \\
\hline 
\caption{Derived spectroscopic parameters for galaxies observed at Loiano with red-channel grism.}
\footnote{Columns are:    (1): HRS name;  
   (2): log of stellar mass $\rm M_{\ast}$  ($\rm M_{\odot}$), from B15; 
   (3): $g-i$ color index;  
   (4): Absorption continuum correction to H$\alpha$ EW (see section \ref{cont});  
   (5): Observed H$\alpha$ EW (positive EW means emission);  
   (6): Corrected H$\alpha$ EW (positive EW means emission); 
   (7): Observed [NII] EW ($\lambda$6583 $\mbox{\AA}$) (positive EW means emission); 
   (8): rms near H$\alpha$; \\
   (9): Nuclear spectral classification (from WHAN) }
\label{TloiG8} 
\end{longtable}
\tiny

\setlength{\oddsidemargin}{0cm}
\setlength{\evensidemargin}{0cm}
\setlength{\headsep}{0cm}
\tiny
\begin{longtable}{|c| c c c c c c c|}
\hline
   HRS              & H$\alpha$EW  &   H$\alpha$ correction & H$\alpha$EW$_{C}$ &  [OIII]/H$\beta$  &  [NII]/H$\alpha$  &  BPT  & WHAN   \\   
                    & $\mbox{\AA}$ &   $\mbox{\AA}$         & $\mbox{\AA}$      &                   &                   &           &    \\    
    (1)             &    (2)       &         (3)            &    (4)            &        (5)        &     (6)           & (7)   &   (8)  \\    
\hline                                                                                                                                    
      2             &   25.78      &       1.56             &    27.34           &     0.40          &      0.33         &  HII  &   HII  \\    
     28             &   7.63       &       1.35             &    8.98            &     0.58          &      0.34         &  HII  &   HII  \\    
     33             &   9.93       &       1.55             &    11.49           &     0.61          &      0.45         &  TR   &   HII  \\    
     51             &   30.81      &       2.01             &    32.82          &     0.58            &      0.29         &  HII  &   HII  \\    
     65             &   10.98      &       1.86             &    12.84          &     0.89            &      0.25         &  HII  &   HII  \\
     86             &   88.48      &       2.34             &    90.83           &     0.58          &      0.25         &  HII  &   HII  \\    
     98             &   24.54      &       1.86             &    26.40          &     0.77            &      0.25         &  HII  &   HII  \\
    108             &   24.30      &       1.79             &    26.09          &     0.19            &      0.27         &  HII  &   HII  \\
    110             &   57.05      &       1.75             &    58.80          &     1.43            &      0.17         &  HII  &   HII  \\
    118             &   22.14      &       1.77             &    23.91          &     1.63            &      0.17         &  HII  &   HII  \\    
    132             &   30.36      &       2.76             &    33.12           &     2.66          &      0.16         &  HII  &   HII  \\    
    148             &   47.51      &       1.83             &    49.34          &     0.8     &      0.23         &  HII  &   HII  \\    
    153             &   17.03      &       1.52             &    18.55          &     0.29            &      0.33         &  HII  &   HII  \\    
    188             &   13.98      &       1.96             &    15.94           &     0.41          &      0.31         &  HII  &   HII  \\    
    189             &   18.17      &       1.74             &    19.91          &     0.61            &      0.27         &  HII  &   HII  \\
    230             &   12.77      &       1.71             &    14.48          &     1.86            &      0.3          &  TR   &   HII  \\
    232             &   16.90      &       1.89             &    18.79           &     0.23          &      0.3          &  HII  &   HII  \\
    239             &   26.97      &       1.44             &    28.42           &     0.41          &      0.39         &  HII  &   HII  \\
    253             &   48.64      &       1.24             &    49.88           &     3.0           &      0.28         &  TR   &   HII  \\    
    259             &   31.36      &       2.05             &    33.41          &     0.23            &      0.41         &  HII  &   HII  \\
    266             &   22.78      &       2.34             &    25.12          &     1.04            &      0.18         &  HII  &   HII  \\
    267             &   13.71      &       2.07             &    15.78          &     1.81            &      0.29         &  HII  &   HII  \\    
    298             &   10.63      &       1.67             &    12.31          &     0.95            &      0.34         &  HII  &   HII  \\    
    302             &   15.85      &       1.97             &    17.83           &     0.59          &      0.33         &  HII  &   HII  \\
    303             &   55.10      &       1.95             &    57.05           &     0.53          &      0.34         &  HII  &   HII  \\    
    313             &   6.36       &       1.41             &    7.76            &     0.12          &      0.45         &  HII  &   HII  \\    
\hline
\caption {Derived spectroscopic parameters for galaxies observed at Loiano with red and blue channel grisms.}
\footnote{Columns are:   (1): HRS name; 
   (2): Absorption continuum correction to H$\alpha$ EW; 
   (3): Observed H$\alpha$ EW (positive EW means emission);  
   (4): Corrected H$\alpha$ EW (positive EW means emission);  
   (5): [OIII]/H$\beta$;  
   (6): [NII]/H$\alpha$;  
   (7): Nuclear classification using the BPT ([NII]/H$\alpha$) diagram; 
   (8): Nuclear classification using the WHAN diagram} 
\label{TloiGandalf} 
\end{longtable}
\tiny

\newpage
\begin{landscape}
\vskip -2cm
\setlength{\oddsidemargin}{0cm}
\setlength{\evensidemargin}{0cm}
\setlength{\headsep}{0cm}
\tiny
\captionsetup{width=16cm}
\begin{longtable}{|c| c c c c c c c| c c c c c c c|} 
\hline
\multicolumn{1}{|c|}{} & \multicolumn{7}{|c|}{Integrated} & \multicolumn{7}{|c|}{Nuclear} \\ \hline
 HRS    &    Ref        &  EW H$\alpha$$_{C}$ &  [OIII]/H$\beta$   & [NII]/H$\alpha$ & [SII]/H$\alpha$ &      BPT     &   WHAN     & Ref     & EW H$\alpha$$_{C}$  &  [OIII]/H$\beta$  &[NII]/H$\alpha$  & [SII]/H$\alpha$ &    BPT        &   WHAN   \\                              
        &               &  $\mbox{\AA}$       &                    &                 &                 &                  &            &         &  $\mbox{\AA}$       &             &                 &                 &               &   \\     
 (1)    &     (2)       &        (3)          &         (4)        &      (5)        &      (6)        &      (7)         &   (8)      & (9)     &   (10)        &     (11)          &     (12)        &   (13)          &   (14)        &   (15)   \\
\hline  \endhead                                                                                                                                                                                                                                 
  1     &     1         &     18.25           &       0.41          & 0.37           &     0.42        &    HII / HII       &    HII     &  4      &   8.02                &      0.53         &   0.37          &   0.4           &  HII / HII    &    HII   \\
  2     &     1         &     30.05           &       0.41          & 0.32           &     0.34        &    HII / HII       &    HII     &  3      &   27.34               &      0.40         &   0.33          &   0.38          &  HII / HII    &    HII   \\
  3     &     ""        &       ""            &       ""            & ""             &      ""         &     "" / ""      &     ""     &  5      &   3.93                &      1.58         &   1.16          &   1.19          &  AGN / LIN    &   wAGN   \\
  4     &     1a        &     10.88           &       3.12          & 0.39           &     0.48        &    AGN / SEY       &   sAGN     &  5      &   43.15               &      7.38         &   1.22          &   0.86          &  AGN / SEY    &   sAGN   \\
  5     &     1         &     6.18            &       -             & 0.89           &     1.14        &  (PAS) / (PAS)  &   sAGN     &  4 &   10.55               &      0.33         &   0.38          &   0.37          &  HII / HII    &    HII   \\
  6     &     ""        &       ""            &       ""            & ""             &      ""         &     "" / ""      &    ""      &  4      &   6.16                &      0.71         &   0.32          &   0.74          &  HII / LIN    &    HII   \\
  7     &     2         &       -             &       -             & -              &      -          &   (PAS)/ (PAS)  &    PAS     &  5d      &   3.29                &      0.06         &   0.9           &   0.47          &   - / -       &   wAGN   \\
  8     &     ""        &       ""            &       ""            & ""             &      ""         &     "" / ""      &    ""      &  5d     &   2.11                &      5.36         &   1.02          &   0.86          &   - / -       &   wAGN   \\
  9     &     ""        &       ""            &       ""            & ""             &      ""         &     "" / ""      &    ""      &  3      &   3.51                &       -           &   0.89          &   0.83          &   - / -       &   wAGN   \\
  10    &     1         &     28.36           &       0.85          & 0.24           &     0.35        &    HII / HII       &    HII     &  4      &   18.86               &      0.42         &   0.29          &   0.41          &  HII / HII    &    HII   \\
  11    &     1         &     20.99           &       0.54          & 0.27           &     0.43        &    HII / HII       &    HII     &  4      &   24.13               &      0.28         &   0.32          &   0.33          &  HII / HII    &    HII   \\
  12    &     1         &     27.96           &       1.28          & 0.18           &     0.42        &    HII / HII       &    HII     &  4      &   26.63               &      1.14         &   0.17          &   0.39          &  HII / HII    &    HII   \\
  13    &     1         &     16.65           &       0.20          & 0.39           &     0.37        &    HII / HII       &    HII     &  4      &   6.94                &      1.63         &   0.68          &   0.31          &  TR / HII     &   sAGN   \\
  14    &     ""        &       ""            &       ""            & ""             &      ""         &     "" / ""      &    ""      &  5d     &   1.91                &        -          &   1.26          &   1.68          &   -  /  -     &   wAGN   \\
  15    &     ""        &       ""            &       ""            & ""             &      ""         &     "" / ""      &    ""      &  4      &   1.04                &      1.42         &   0.83          &   0.64          &  AGN / LIN    &    RET   \\
  16    &     1         &     12.73           &       0.33          & 0.34           &     0.4         &    HII / HII       &    HII     &  4      &   13.87               &      0.29         &   0.37          &   0.34          &  HII / HII    &    HII   \\
  17    &     1         &     30.04           &       0.84          & 0.28           &     0.31        &    HII / HII       &    HII     &  5d     &   5.36                &      0.11         &   0.38          &   0.2           &   - / -       &    HII   \\
  18    &     1         &     10.40           &       0.84          & 0.43           &     0.4         &    TR / HII        &    HII     &  4      &   6.98                &      5.91         &   0.68          &   0.73          &  AGN / SEY    &   sAGN   \\
  19    &     1         &     29.14           &       0.32          & 0.29           &     0.3         &    HII / HII       &    HII     &  4      &   246.83              &      0.18         &   0.30          &   0.19          &  HII / HII    &    HII   \\
  20    &     1         &     53.45           &       1.48          & 0.15           &     0.24        &    HII / HII       &    HII     &  5      &   28.47               &      1.0          &   0.33          &   0.4           &  HII / HII    &    HII   \\
  21    &     1         &     11.25           &       0.73          & 0.27           &     0.6         &    HII / HII       &    HII     &  4      &   7.67                &      0.68         &   0.33          &   0.7           &  HII / LIN    &    HII   \\
  22    &     ""        &       ""            &       ""            & ""             &      ""         &     "" / ""      &    ""      &  5      &   2.26                &      1.32         &   1.18          &   0.94          &  AGN / LIN    &   wAGN   \\
  23    &     1         &     10.97           &       0.85          & 0.49           &     0.5         &    TR / HII        &    HII     &  4      &   13.61               &      0.41         &   0.52          &   0.39          &  TR / HII     &   sAGN   \\
  24    &     1         &     18.07           &       0.32          & 0.41           &     0.33        &    HII / HII       &    HII     &  5      &   17.45               &      0.21         &   0.36          &   0.25          &  HII / HII    &    HII   \\
  25    &     1         &     22.48           &       0.47          & 0.35           &     0.36        &    HII / HII       &    HII     &  4      &   69.33               &      0.11         &   0.40          &   0.19          &  HII / HII    &    HII   \\
  26    &     1         &     22.10           &       1.05          & 0.25           &     0.47        &    HII / HII       &    HII     &  4      &   19.01               &      1.17         &   0.22          &   0.46          &  HII / HII    &    HII   \\
  27    &     1         &     40.10           &       0.61          & 0.28           &     0.29        &    HII / HII       &    HII     &  4      &   24.62               &      0.46         &   0.35          &   0.33          &  HII / HII    &    HII   \\
  28    &     1         &     24.24           &       0.54          & 0.28           &     0.33        &    HII / HII       &    HII     &  3      &   8.98                &      0.58         &   0.34          &   0.51          &  HII / HII    &    HII   \\
  29    &     1         &     20.69           &       0.78          & 0.25           &     0.46        &    HII / HII       &    HII     &  4      &   9.19                &      0.55         &   0.32          &   0.55          &  HII / HII    &    HII   \\
  30    &     1         &     18.99           &       0.52          & 0.31           &     0.42        &    HII / HII       &    HII     &  4      &   7.93                &      0.56         &   0.45          &   0.45          &  HII / HII    &    HII   \\
  31    &     1         &     44.07           &       2.19          & 0.11           &     0.27        &    HII / HII       &    HII     &  5      &   293.92              &      4.75         &   0.12          &   0.16          &  TR  / SEY    &    HII   \\
  32    &     1         &     -0.34           &       -             & -              &      -          &   (PAS) ,(PAS)  &    RET     &  3       &   0.63                &       -           &   1.35          &    -            &   - / -       &    RET   \\
  33    &     1         &     15.41           &       0.32          & 0.39           &     0.44        &    HII / HII       &    HII     &  3      &   11.49               &      0.61         &   0.45          &   0.49          &  TR  / HII    &    HII   \\
  34    &     1b        &     9.97            &       0.14          & 0.42           &     0.6         &     -  / -         &    HII     &  4      &   3.91                &      0.44         &   0.57          &   0.55          &  TR  / HII    &   wAGN   \\
  35    &     1         &     -0.18           &        -            & 0.16           &      -          &   (PAS) ,(PAS)  &    RET     &  5d        &   0.75                &       -           &   1.04          &   1.43          &   - / -       &    RET   \\
\hline
\caption{Derived spectroscopic parameters of a sample of 35 HRS galaxies. The full catalog is available at the CDS in the electronic form. 
Columns marked "" were not observed, while columns marked - were not detected; positive EW means emission.
A letter `a' marks galaxies with integrated spectra remeasured by us using the GANDALF code; letter `b' marks galaxies with H$\beta$ SN < 3.0 
instead letter `c' marks galaxies with H$\alpha$ SN < 3.0 in the integrated spectrum; letter  `d' marks galaxies with nuclear H$\beta$ SN < 3.0.
In columns 6 and 12 the classification "(PAS)" is added to galaxies whose spectra do not contain H$\alpha$, [NII], [OIII], H$\beta$, [SII](6716\AA) and [SII](6731\AA) in emission.
In columns 7 and 14, two BPT classifications are reported: the first spectral type is from the BPT with [NII]/H$\alpha$ ratio, the second is from the BPT with [SII]/H$\alpha$ ratio.}

\footnote{ Columns 2-8 refer to integrated spectra, columns 9-15 to nuclear spectra; columns marked "" were not observed, 
      while columns marked - were not detected; positive EW means emission. 
(1): HRS name;  
(2): Reference to the integrated spectrum: 1: B15; 2: Kennicutt et al. (1992). 
(3): (Integrated) H$\alpha$ EW corrected for continuum absorption according to GANDALF;  
(4): (Integrated) [OIII]/H$\beta$; 
(5): (Integrated) [NII]/H$\alpha$;
(6): (Integrated) [SII]/H$\alpha$;
(7): (Integrated) classification using the BPT ([NII]/H$\alpha$ and [SII]/H$\alpha$) diagrams;  
(8): (Integrated) classification using the WHAN diagram; 
(9): Reference to the nuclear spectroscopy: 3: this work or Gavazzi et al. (2013); 4: SDSS DR13 or DR12 (Albareti et al. 2016); 5: Ho et al. (1997); 6: Jones
     et al. (2009); 7: Veilleux et al. (1995); 8: Jansen et al. (2000); 9: Falco et al. (1999). 
(10): (Nuclear) H$\alpha$ EW corrected for continuum absorption according to GANDALF or according to Table \ref{Thalpha}; 
(11): (Nuclear) [OIII]/H$\beta$; 
(12): (Nuclear) [NII]/H$\alpha$;
(13): (Nuclear) [SII]/H$\alpha$;
      notice that the nuclear value of the  [OI]/H$\alpha$ ratio (available for only 101 galaxies) is not reported.  
(14): (Nuclear) classification using the BPT ([NII]/H$\alpha$ and [SII]/H$\alpha$) diagrams;  
(15): (Nuclear) classification using the WHAN diagram}
\label{Tclass}
\end{longtable}
\end{landscape}

\newpage
  \begin{figure*}
  \centering
\includegraphics[scale=0.46]{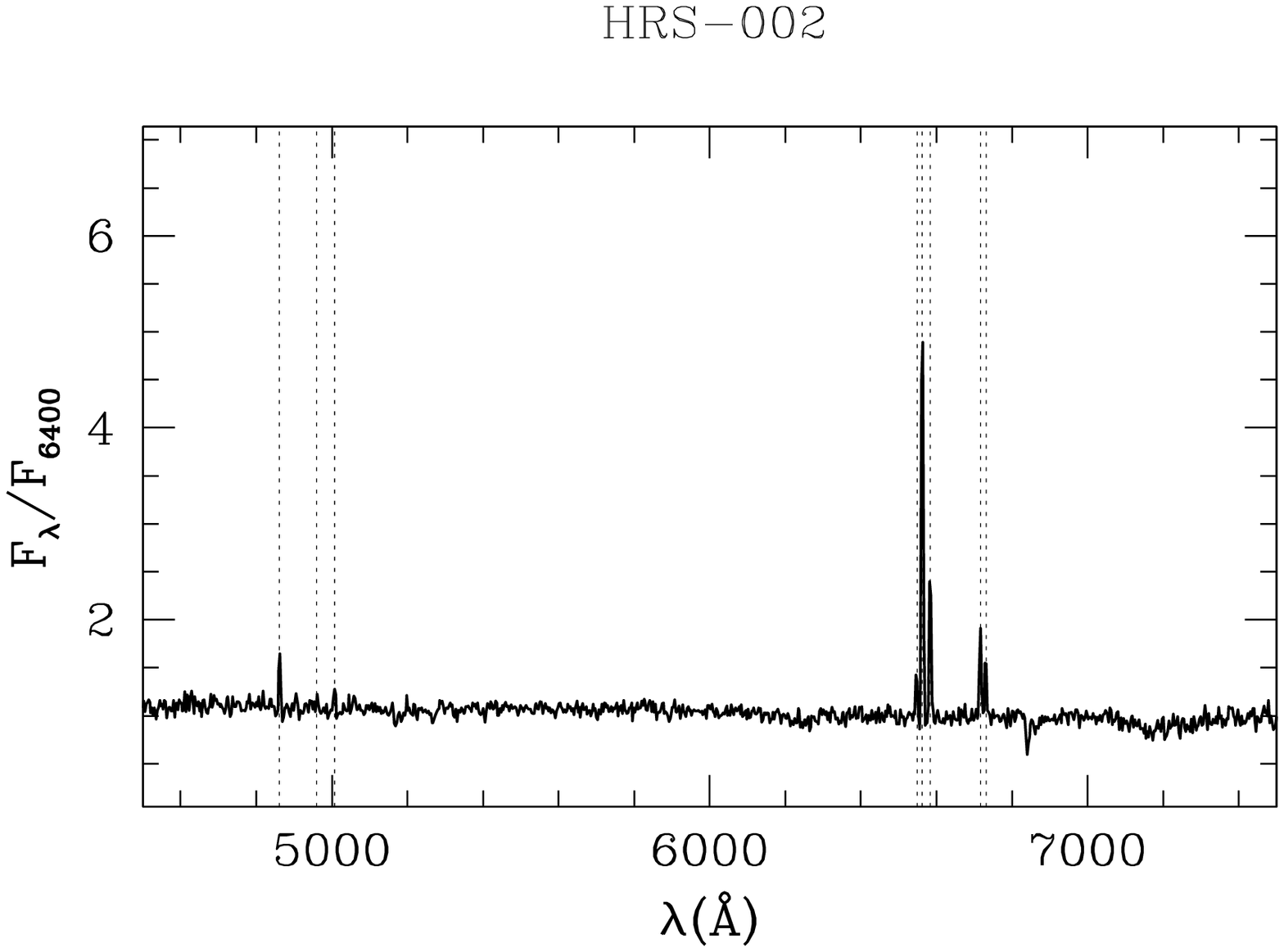}\includegraphics[scale=0.46]{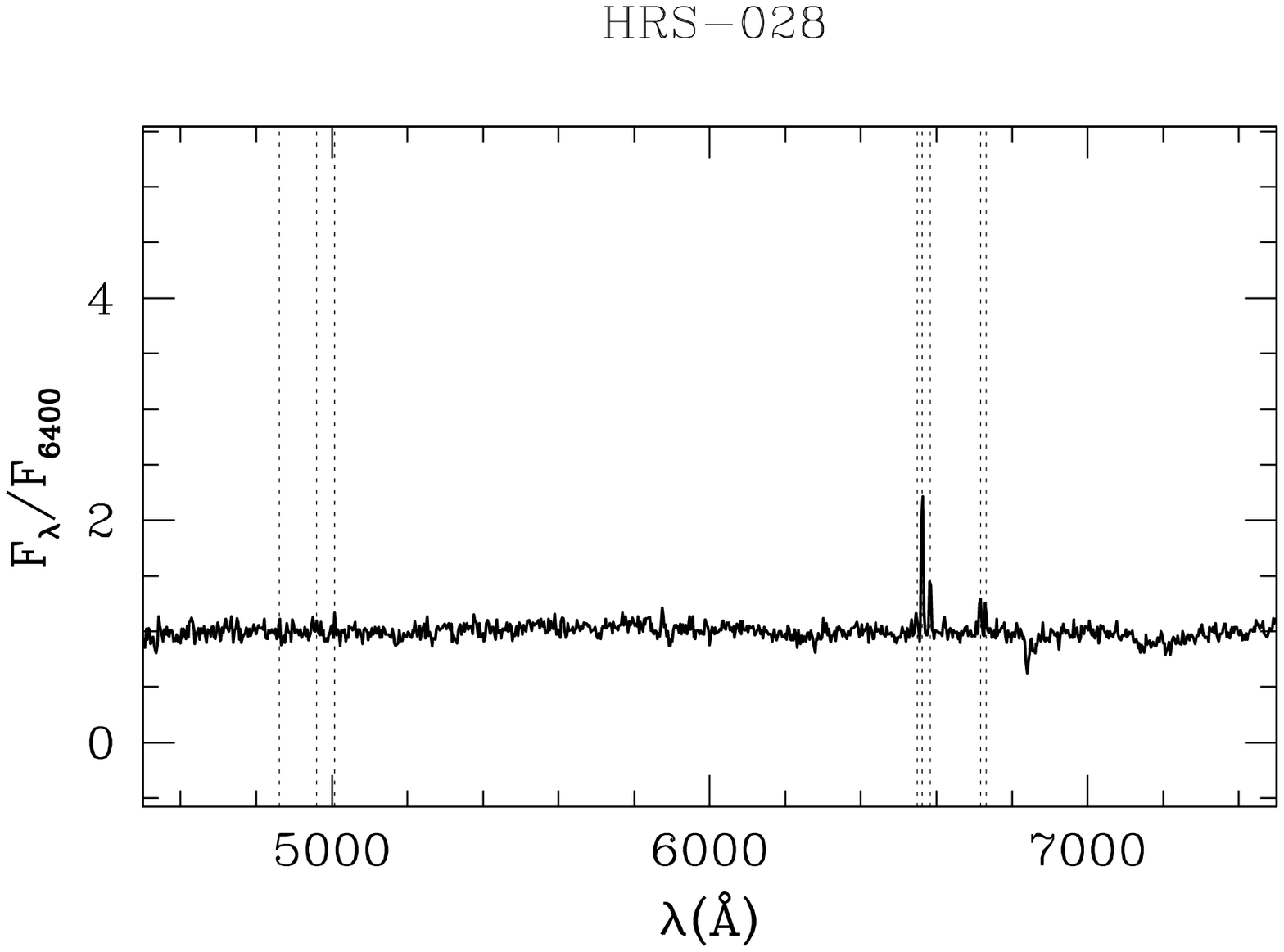}\\
\includegraphics[scale=0.46]{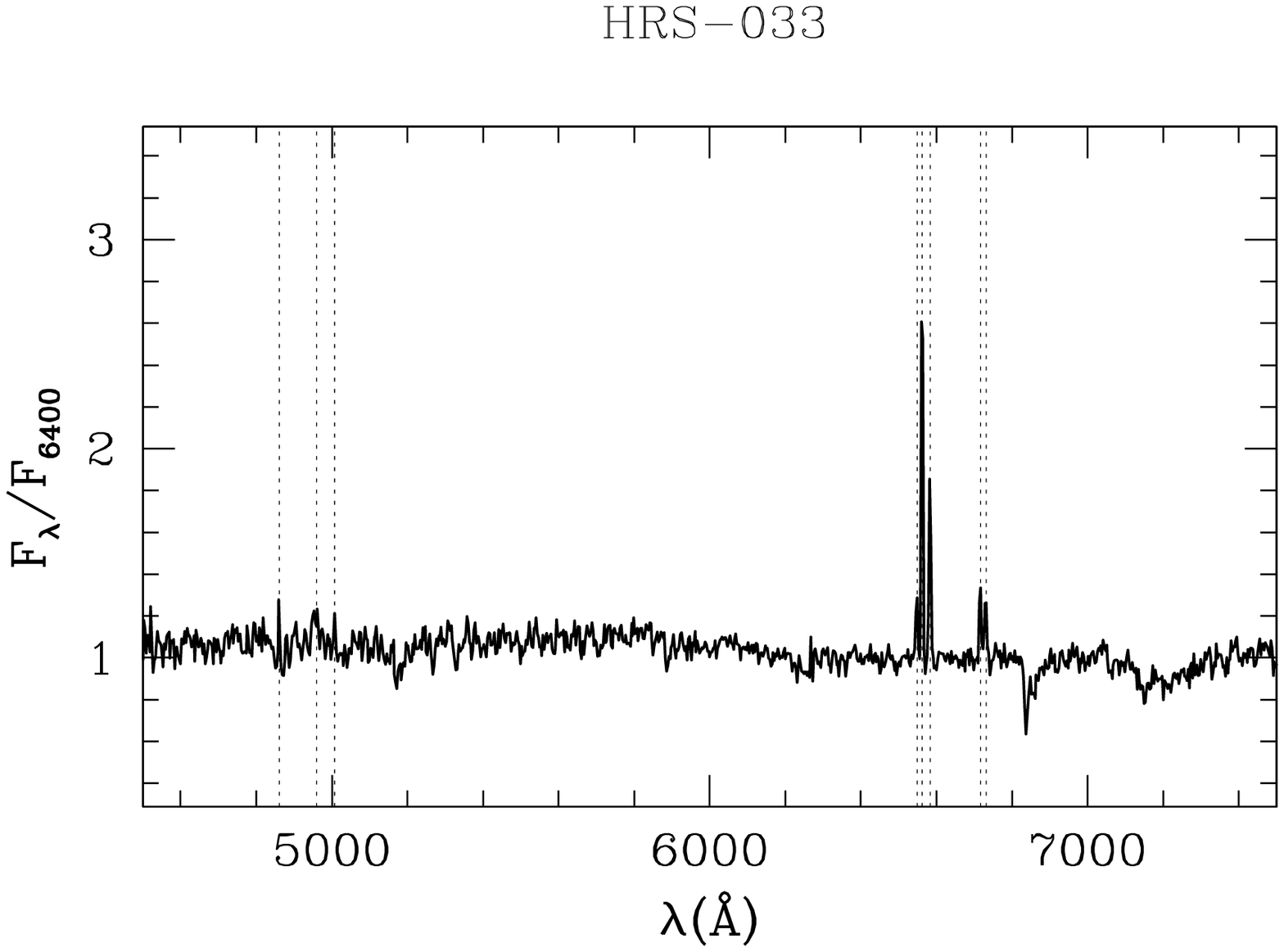}\includegraphics[scale=0.46]{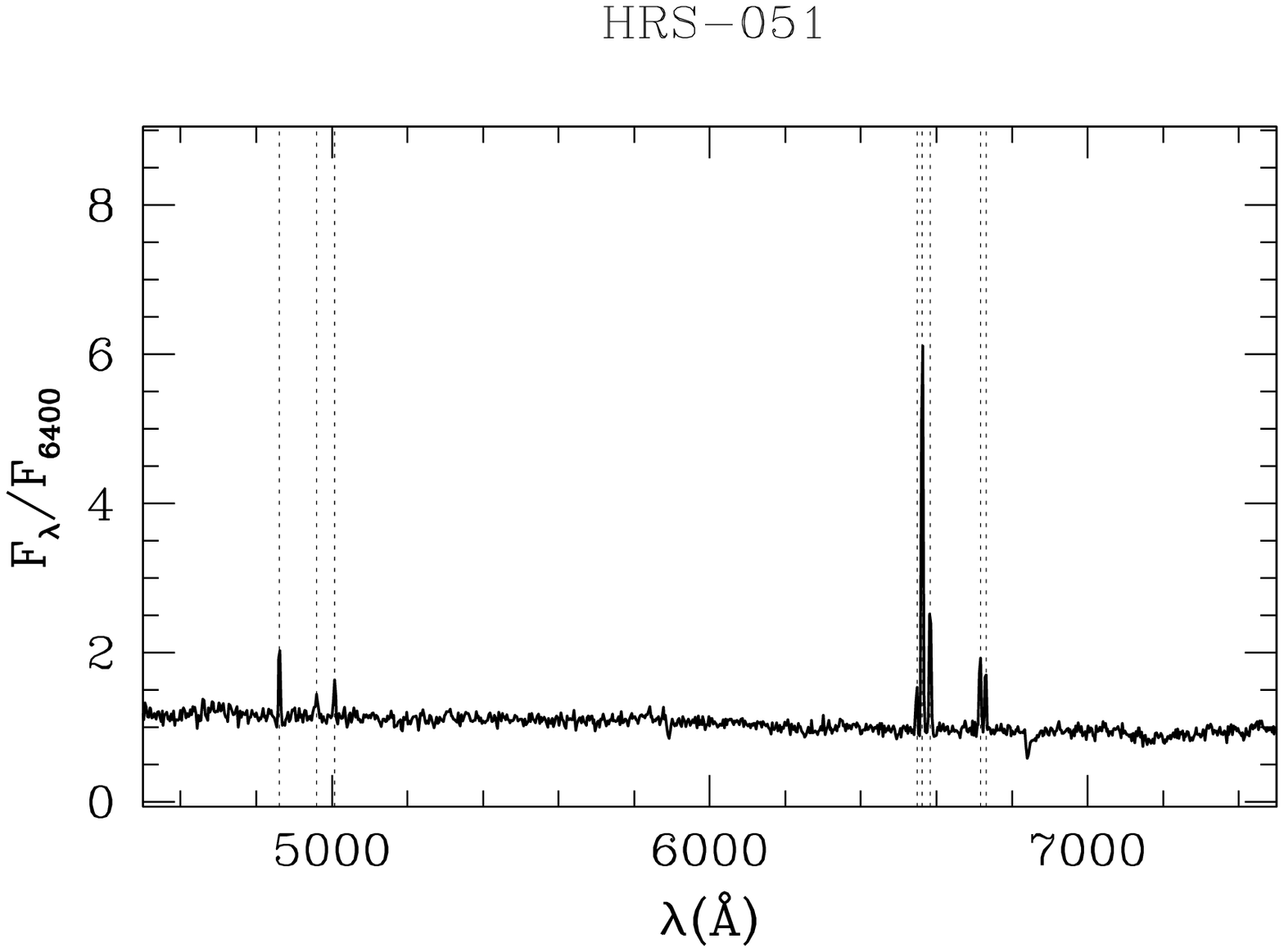}\\
\includegraphics[scale=0.46]{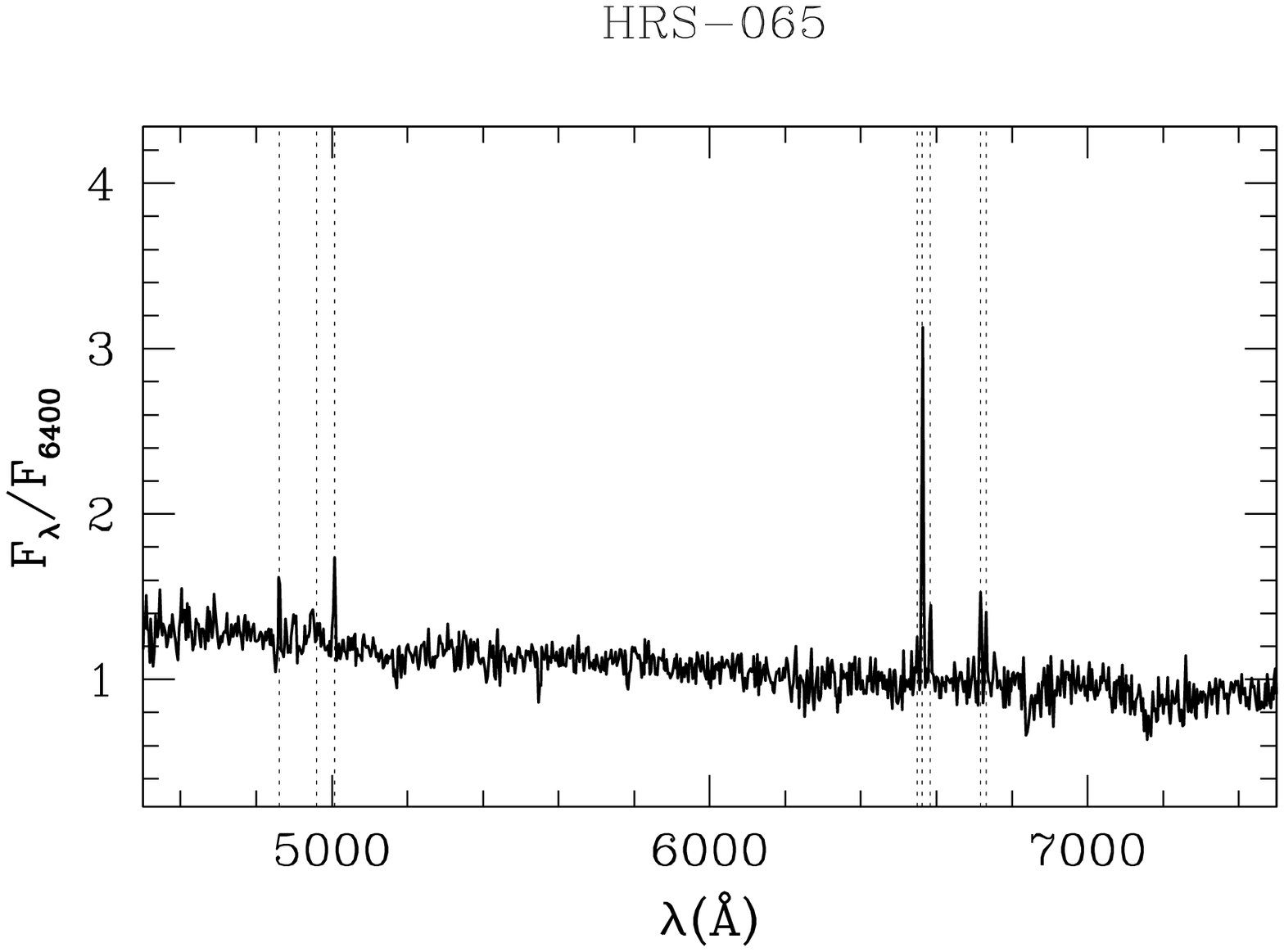}\includegraphics[scale=0.46]{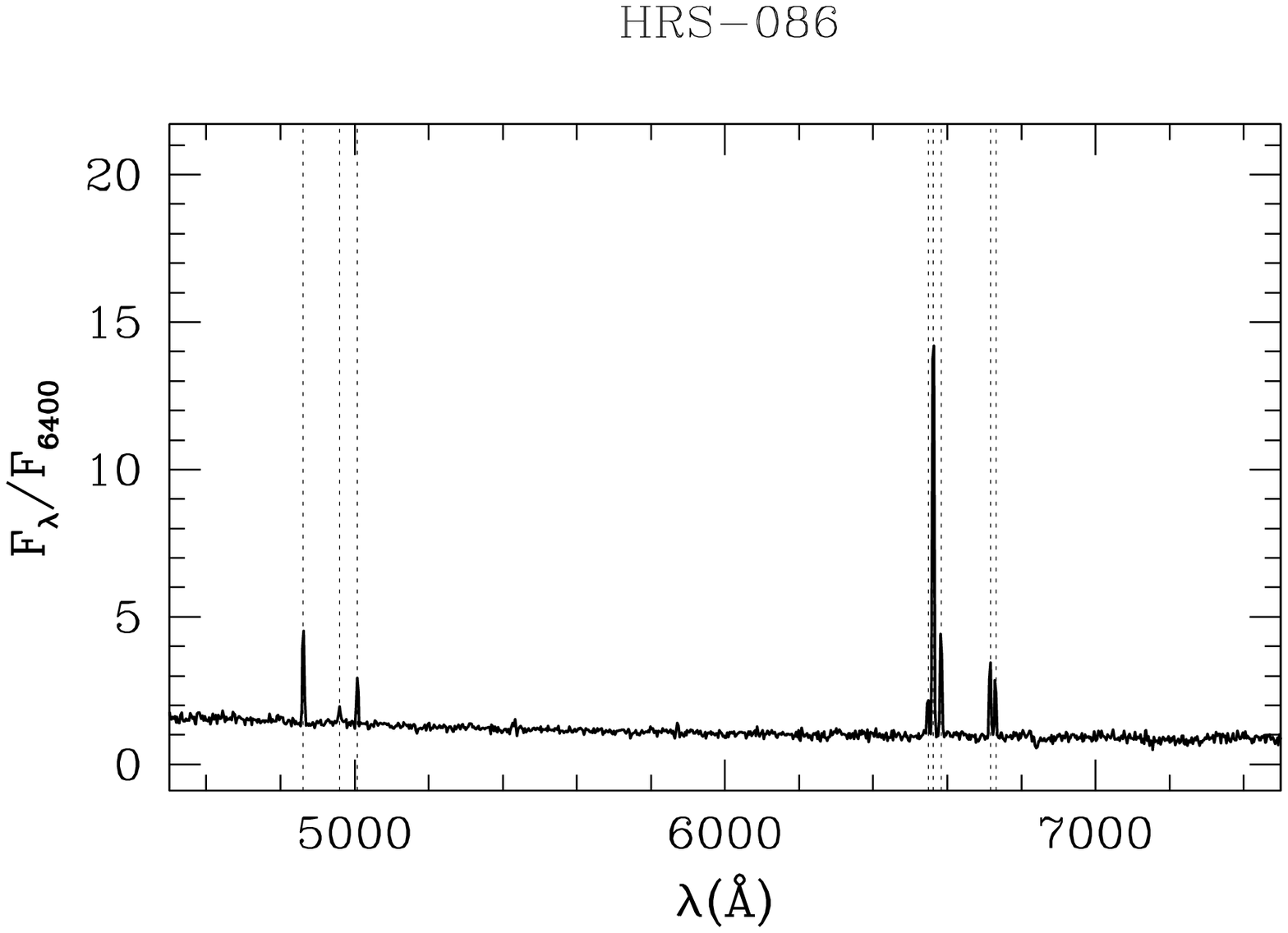}\\
\includegraphics[scale=0.46]{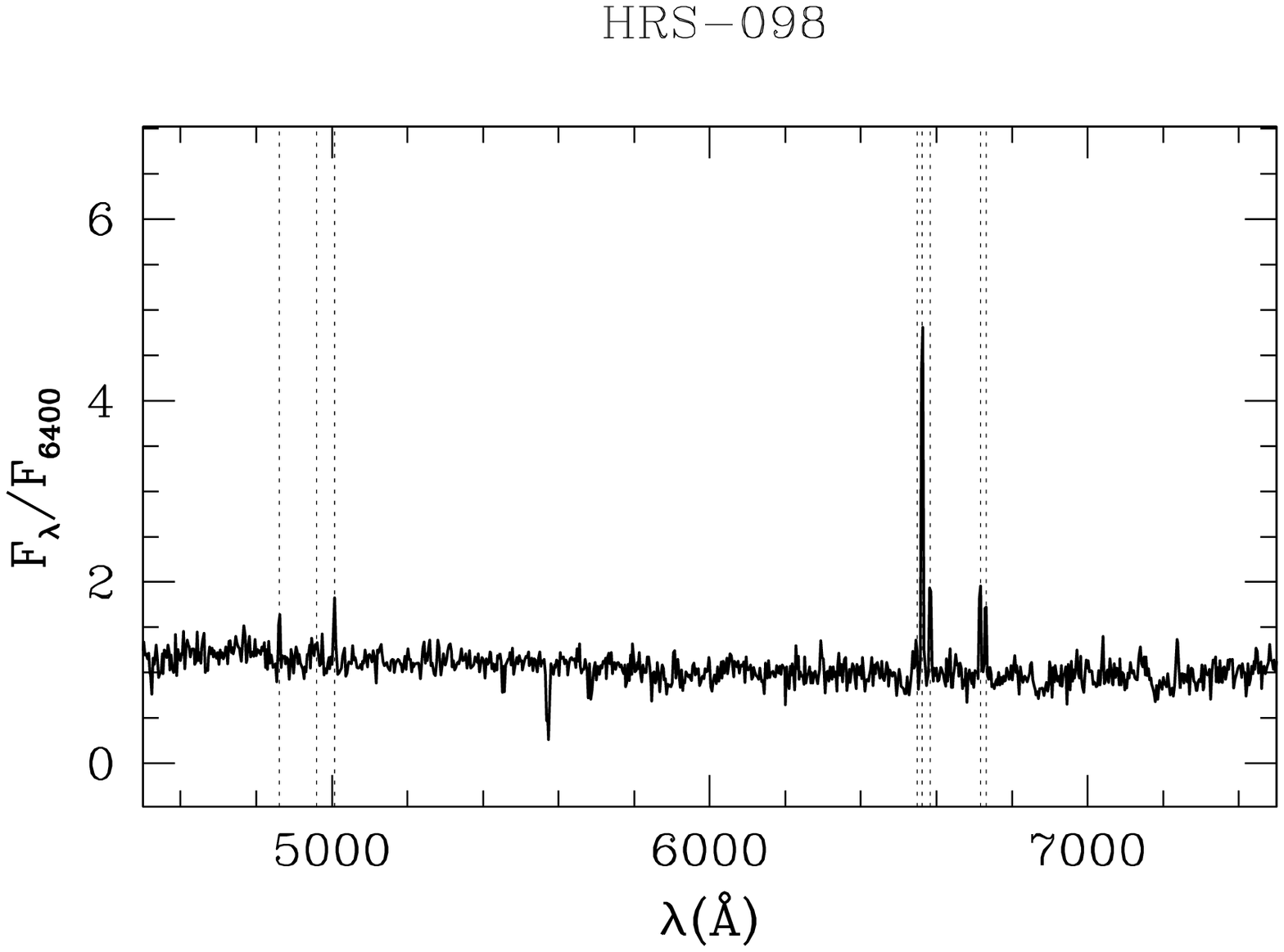}\includegraphics[scale=0.46]{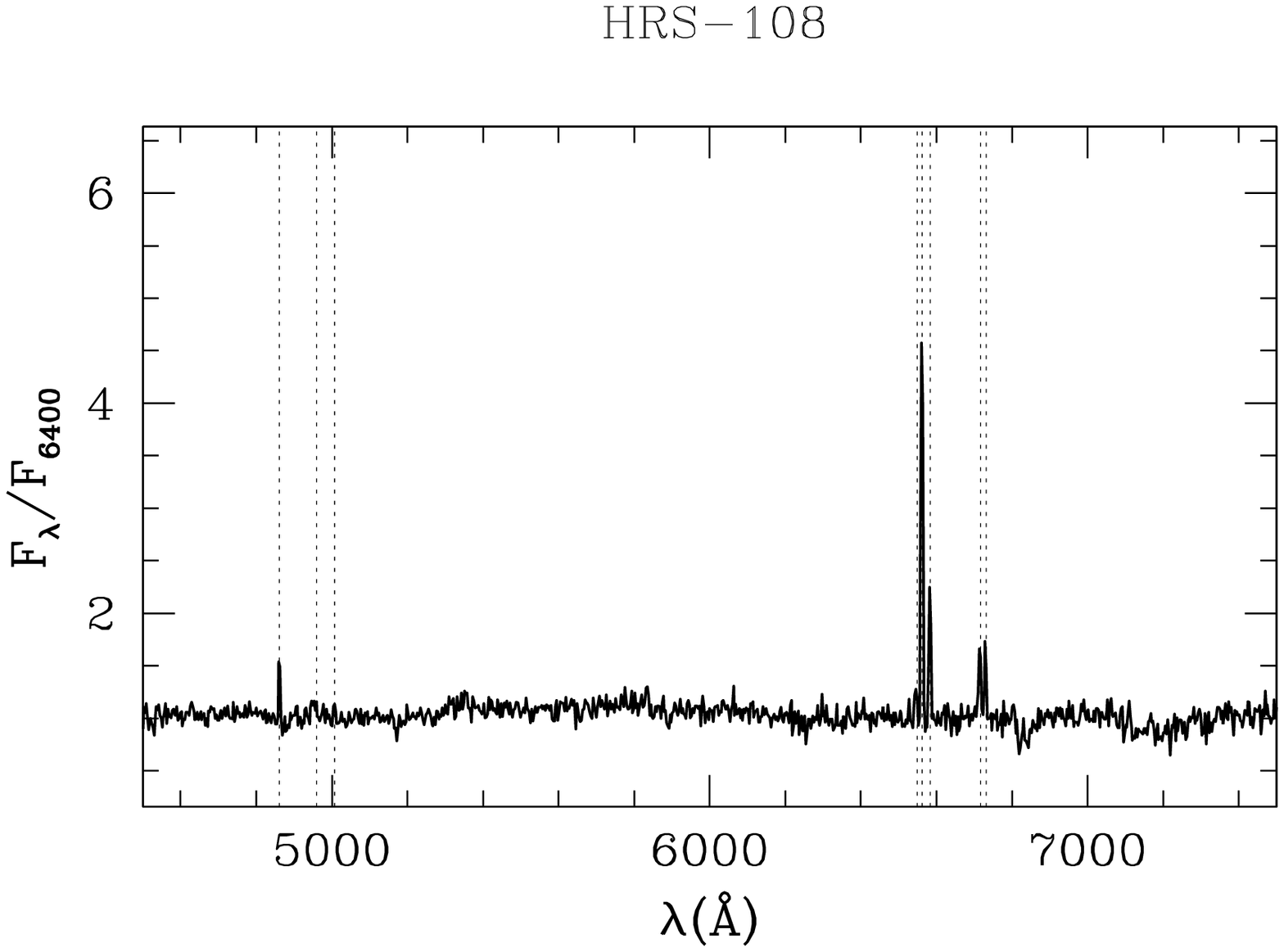}\\
  \end{figure*}
  \begin{figure*}
  \centering
\includegraphics[scale=0.46]{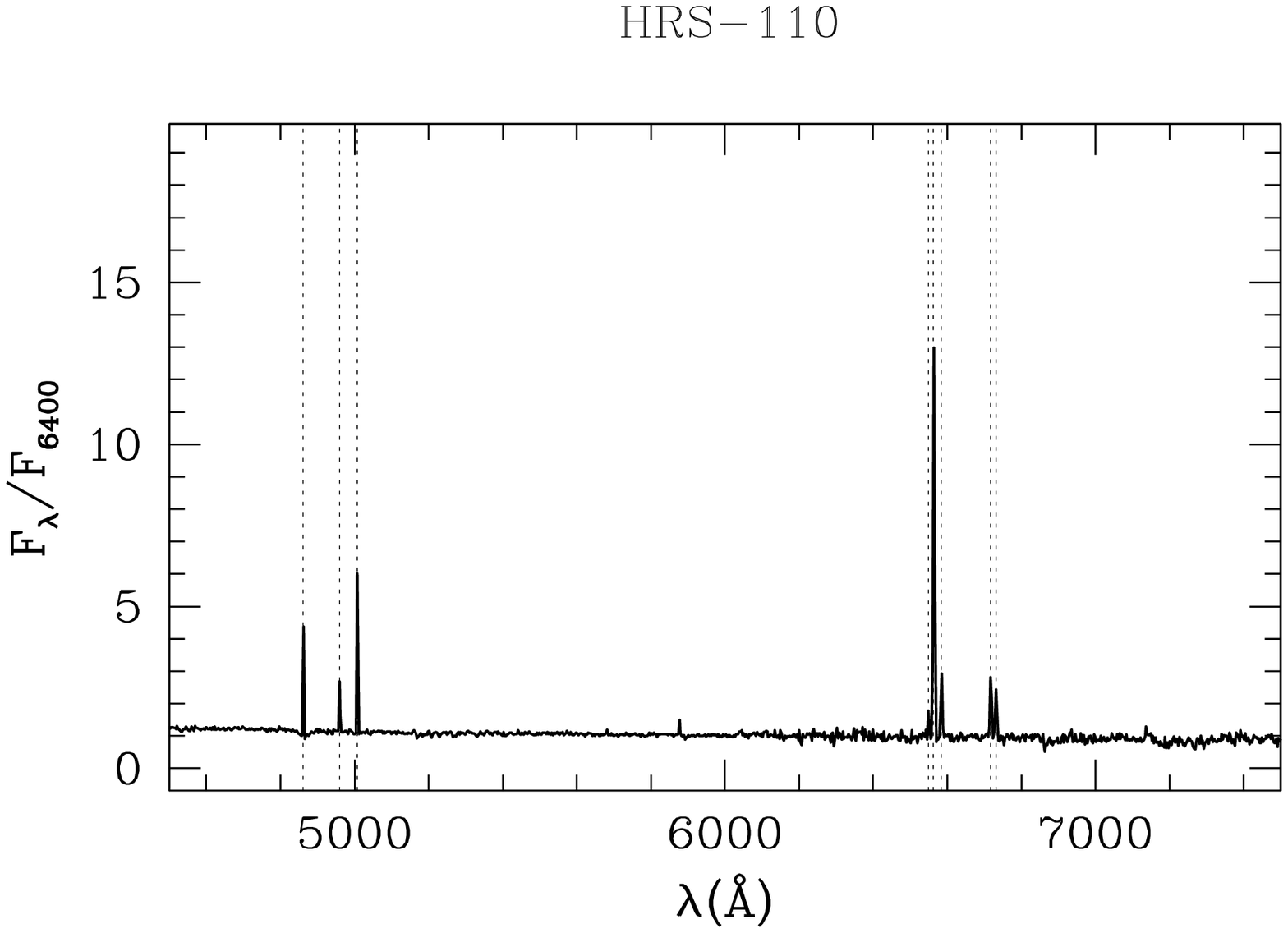}\includegraphics[scale=0.46]{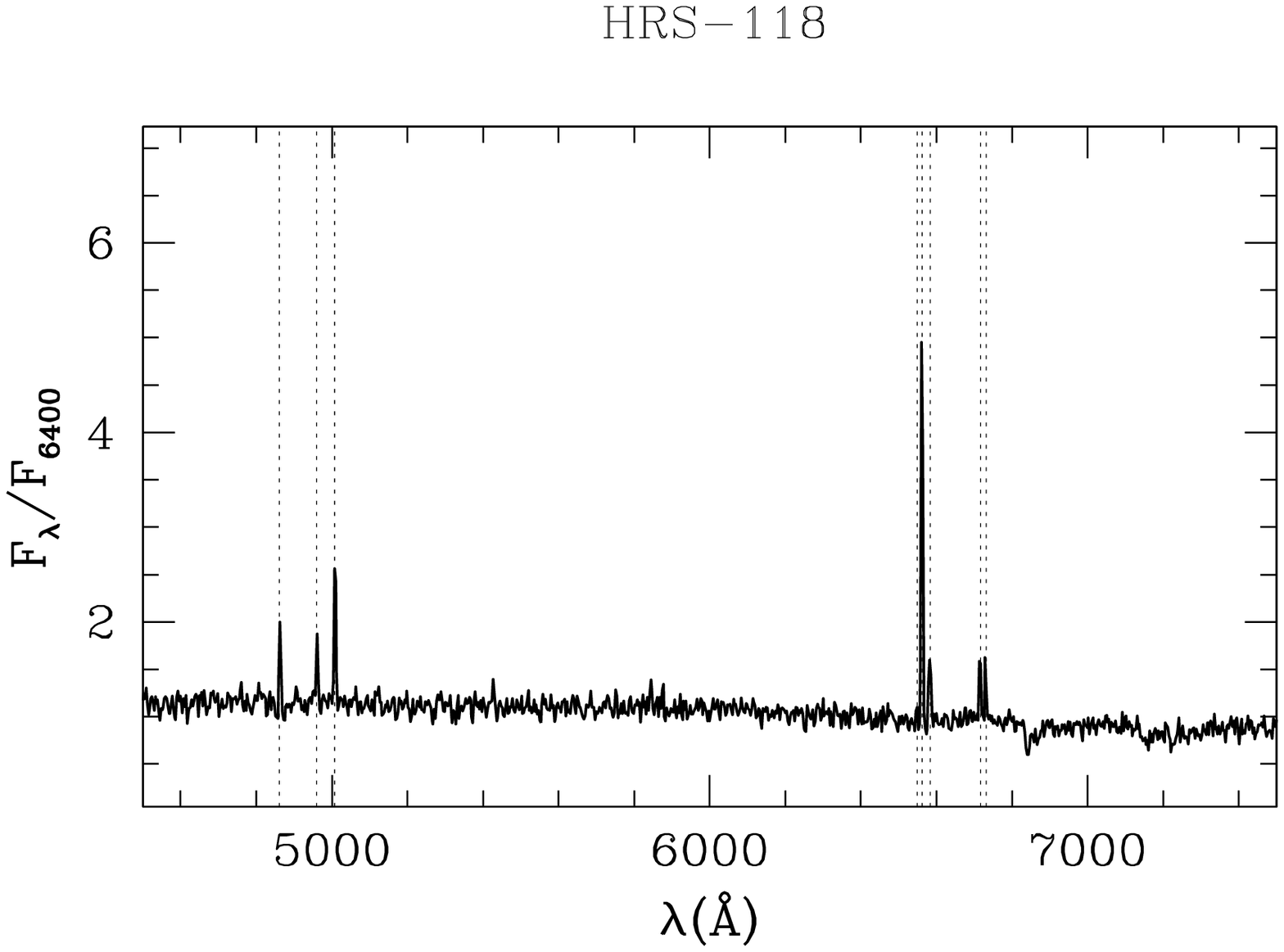}\\
\includegraphics[scale=0.46]{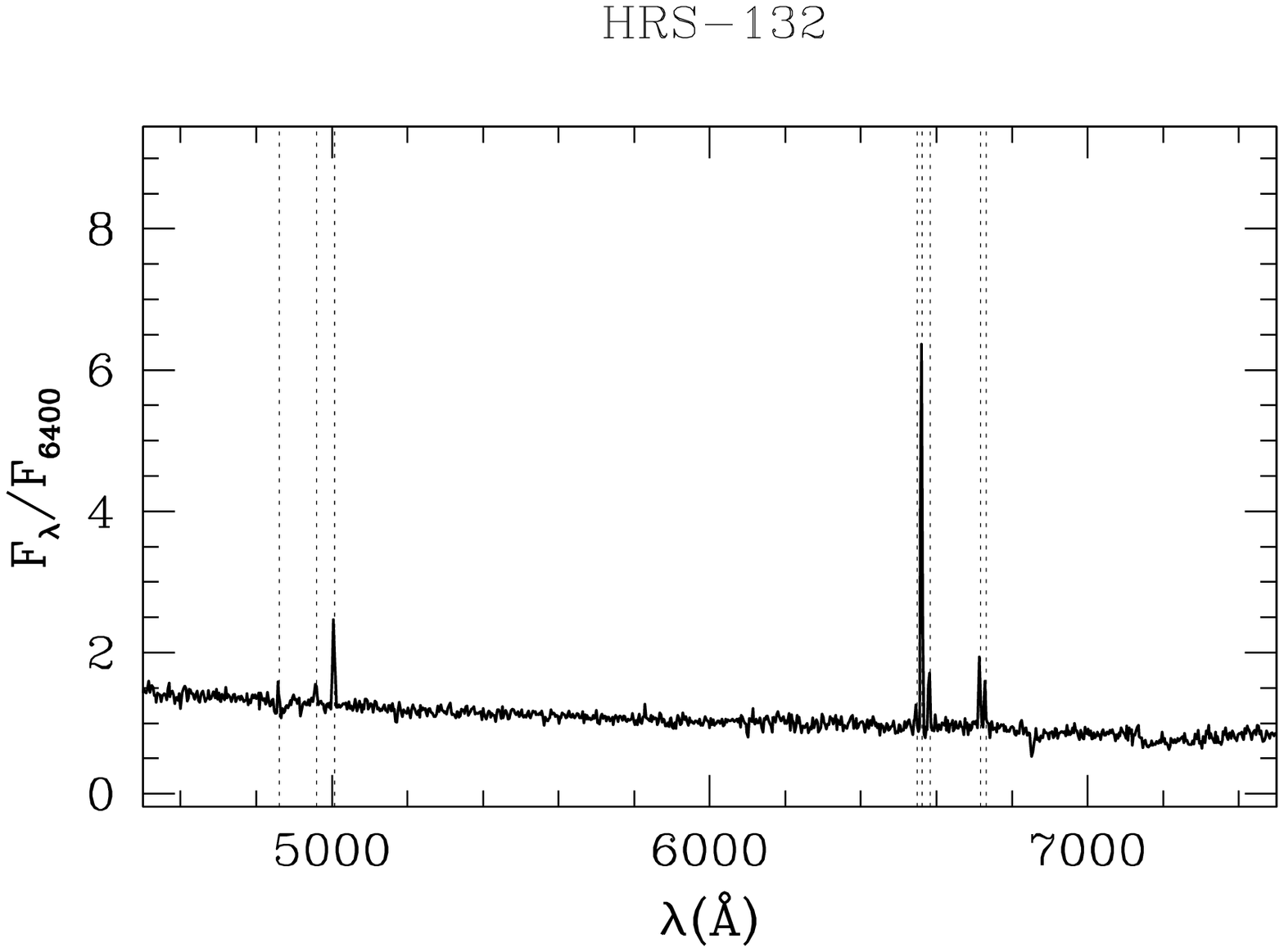}\includegraphics[scale=0.46]{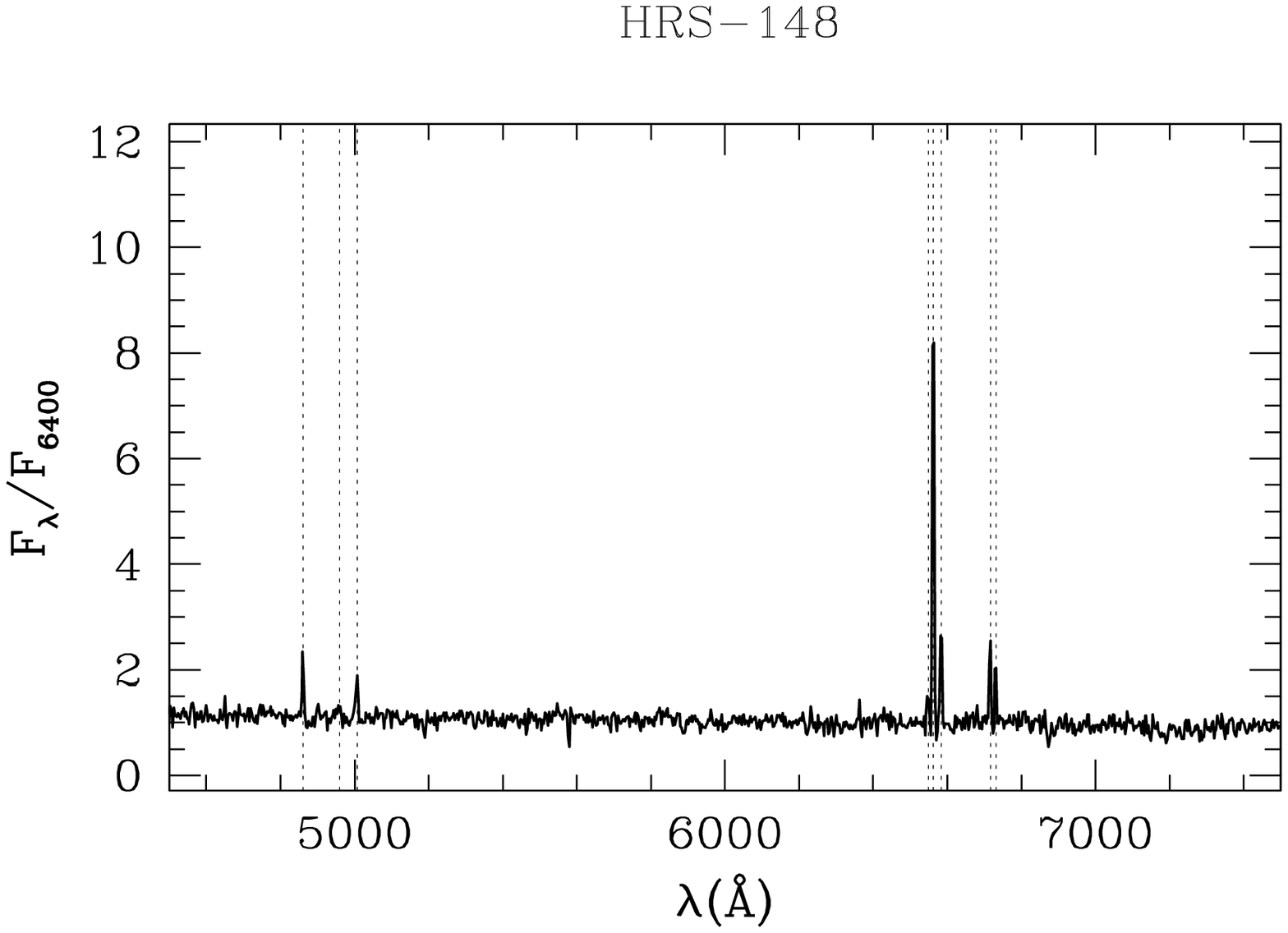}\\
\includegraphics[scale=0.46]{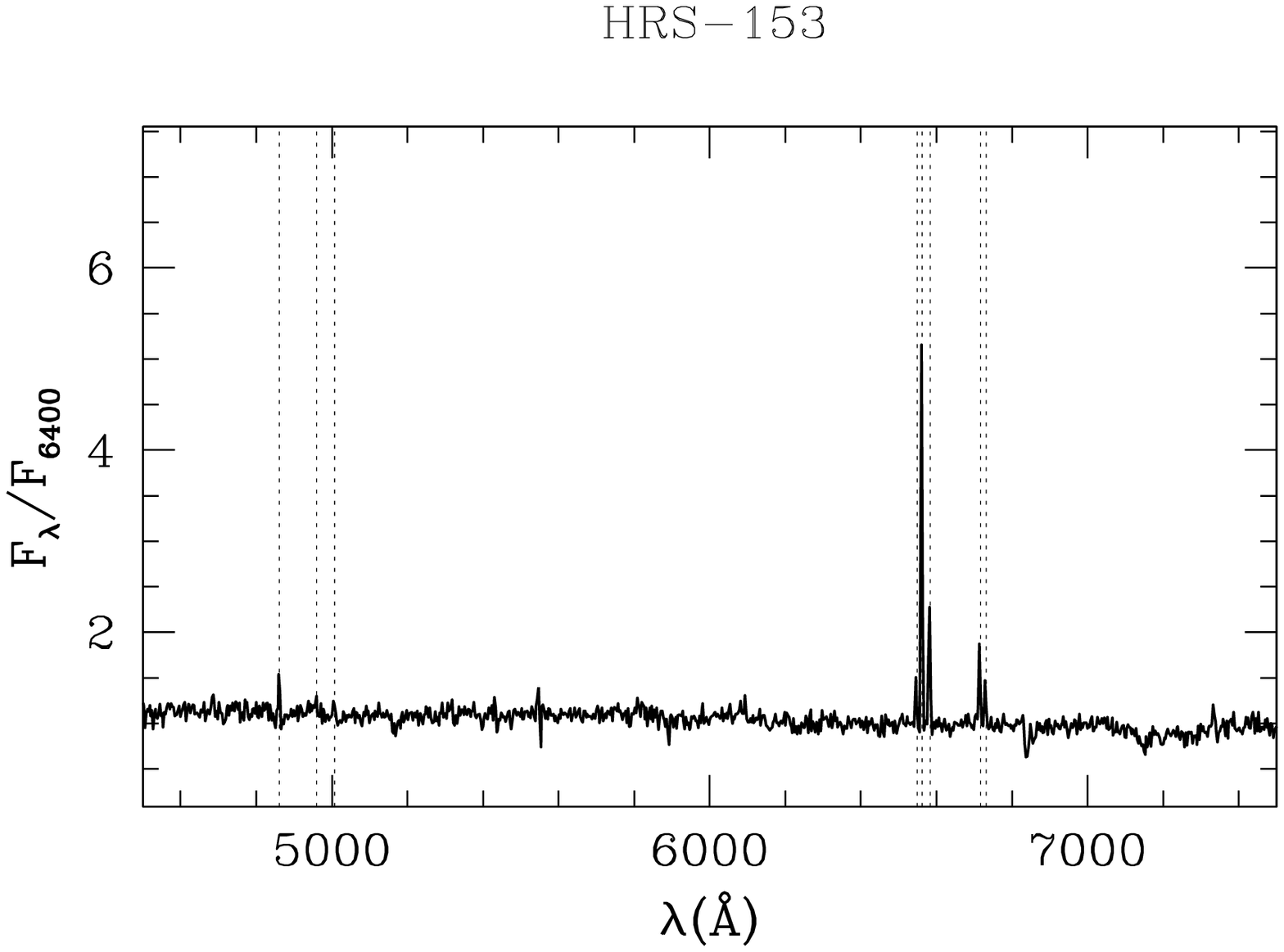}\includegraphics[scale=0.46]{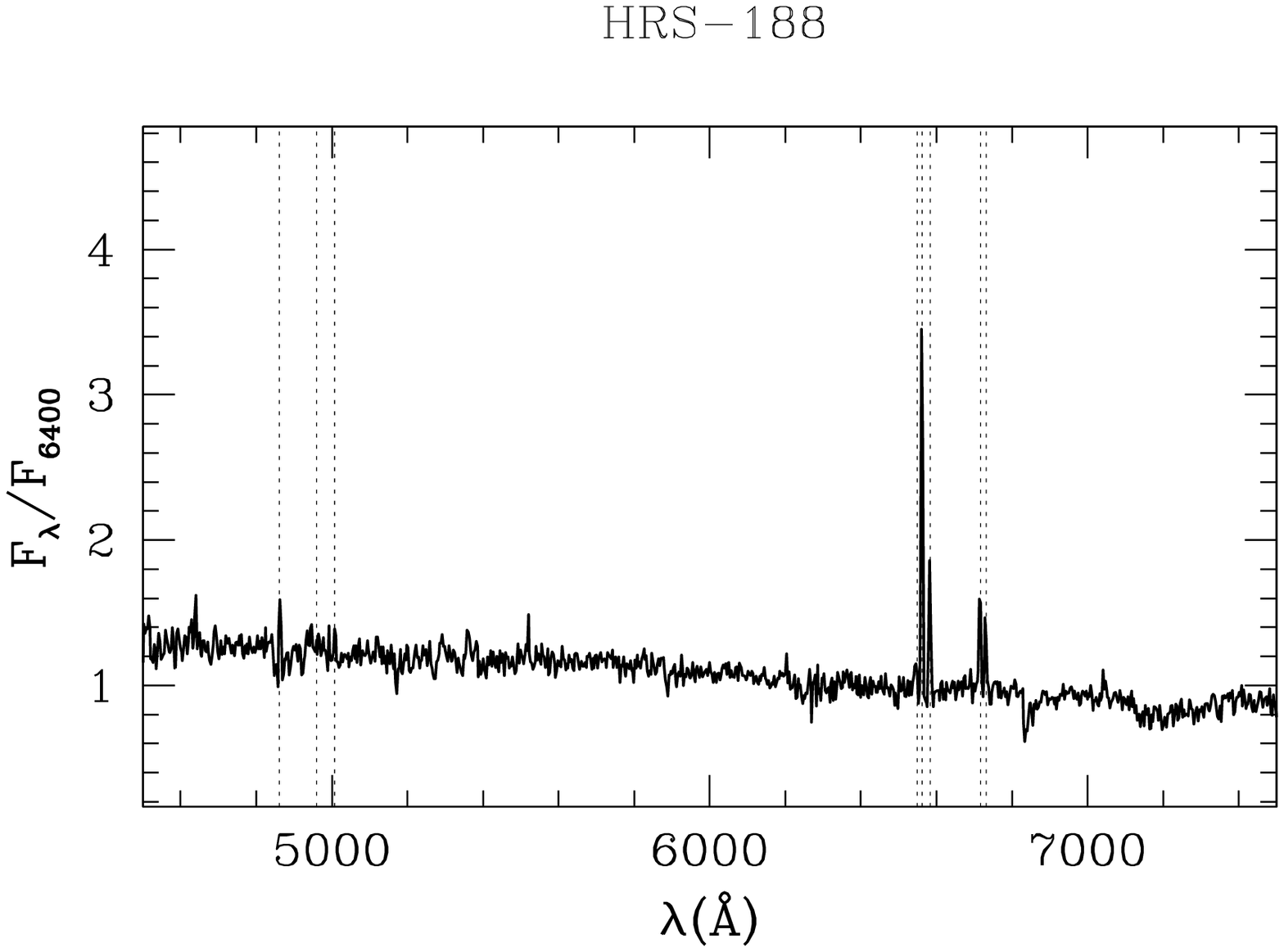}\\
\includegraphics[scale=0.46]{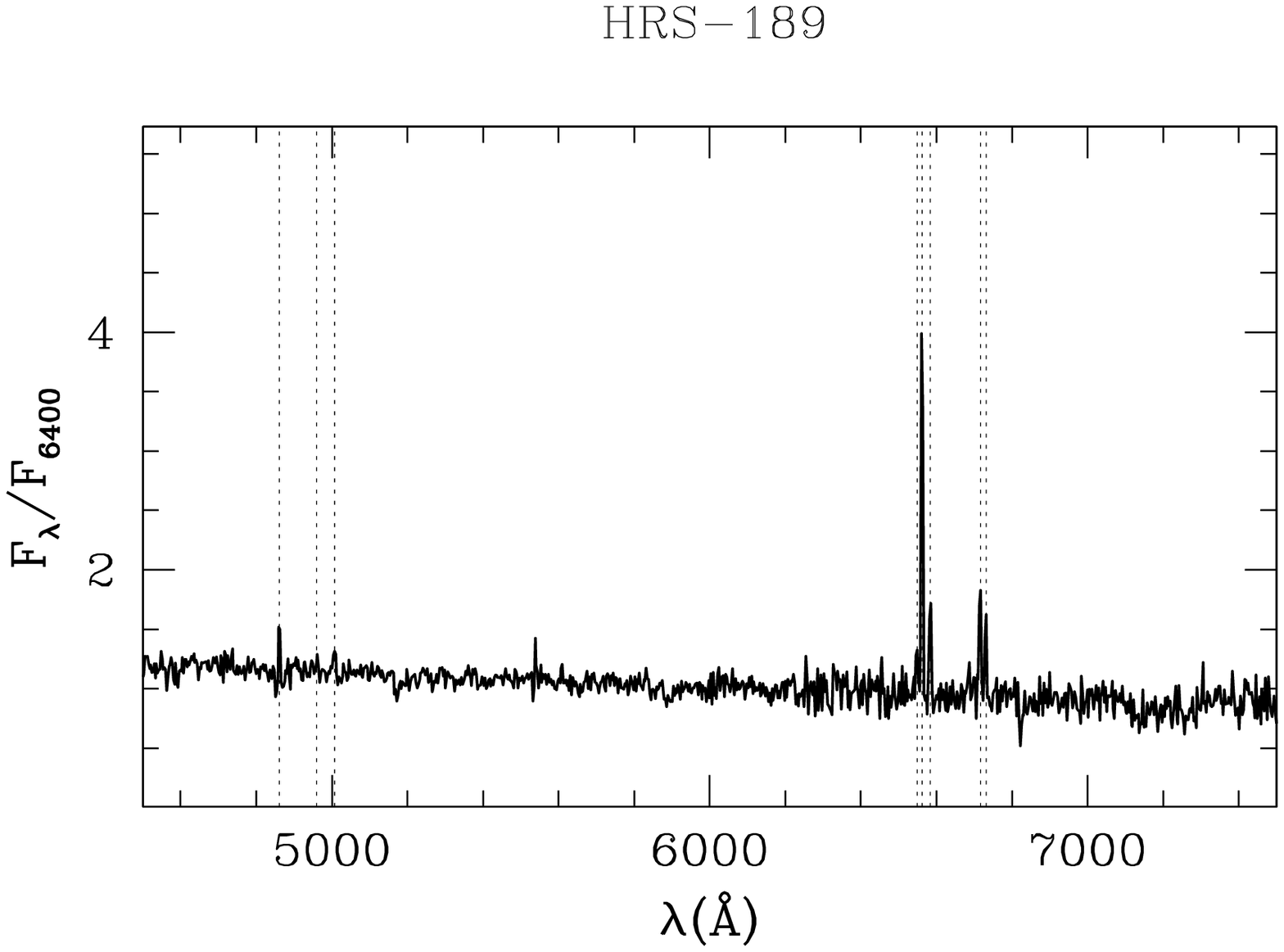}\includegraphics[scale=0.46]{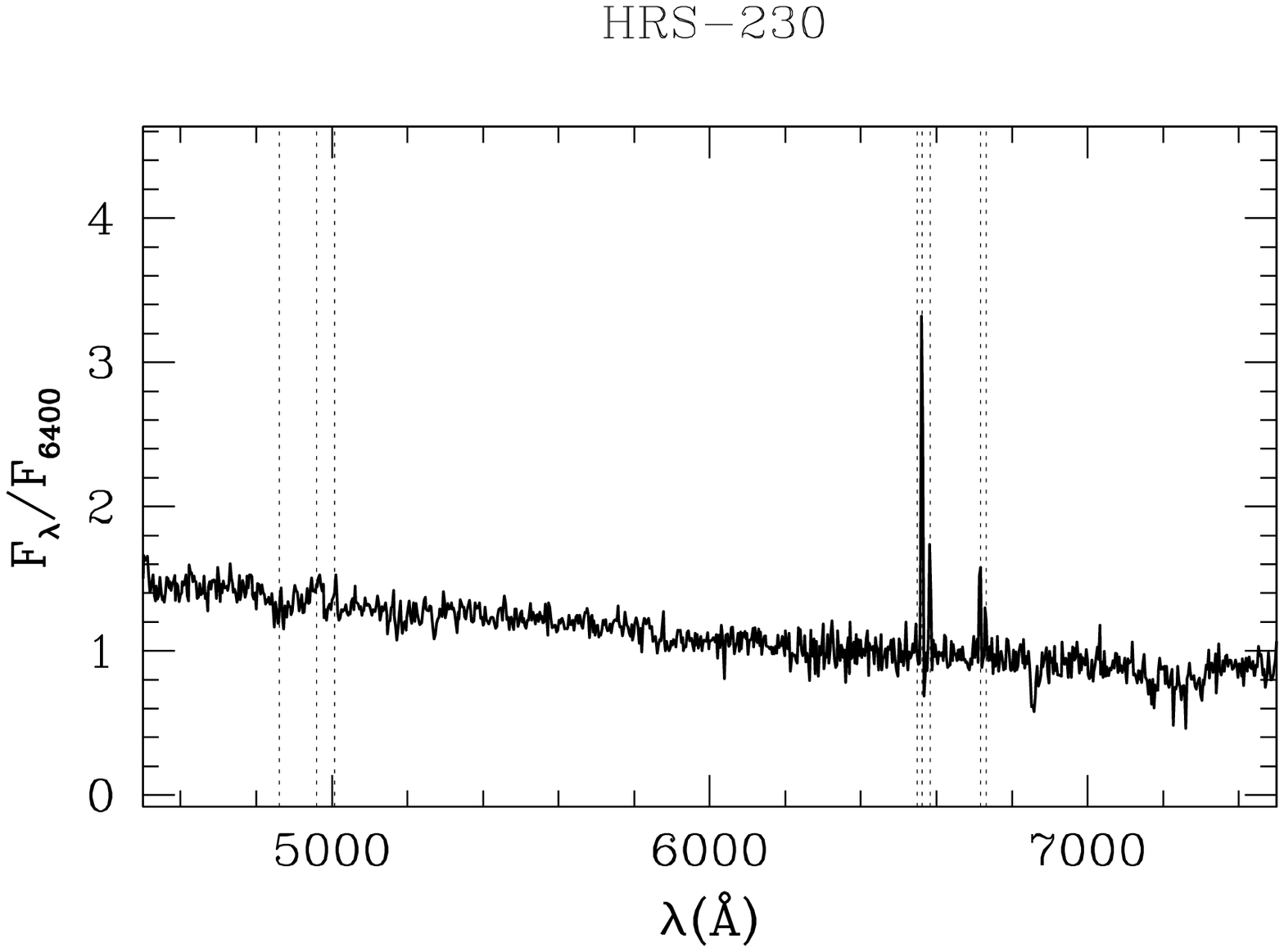}\\
  \end{figure*}
  \begin{figure*}
  \centering
\includegraphics[scale=0.46]{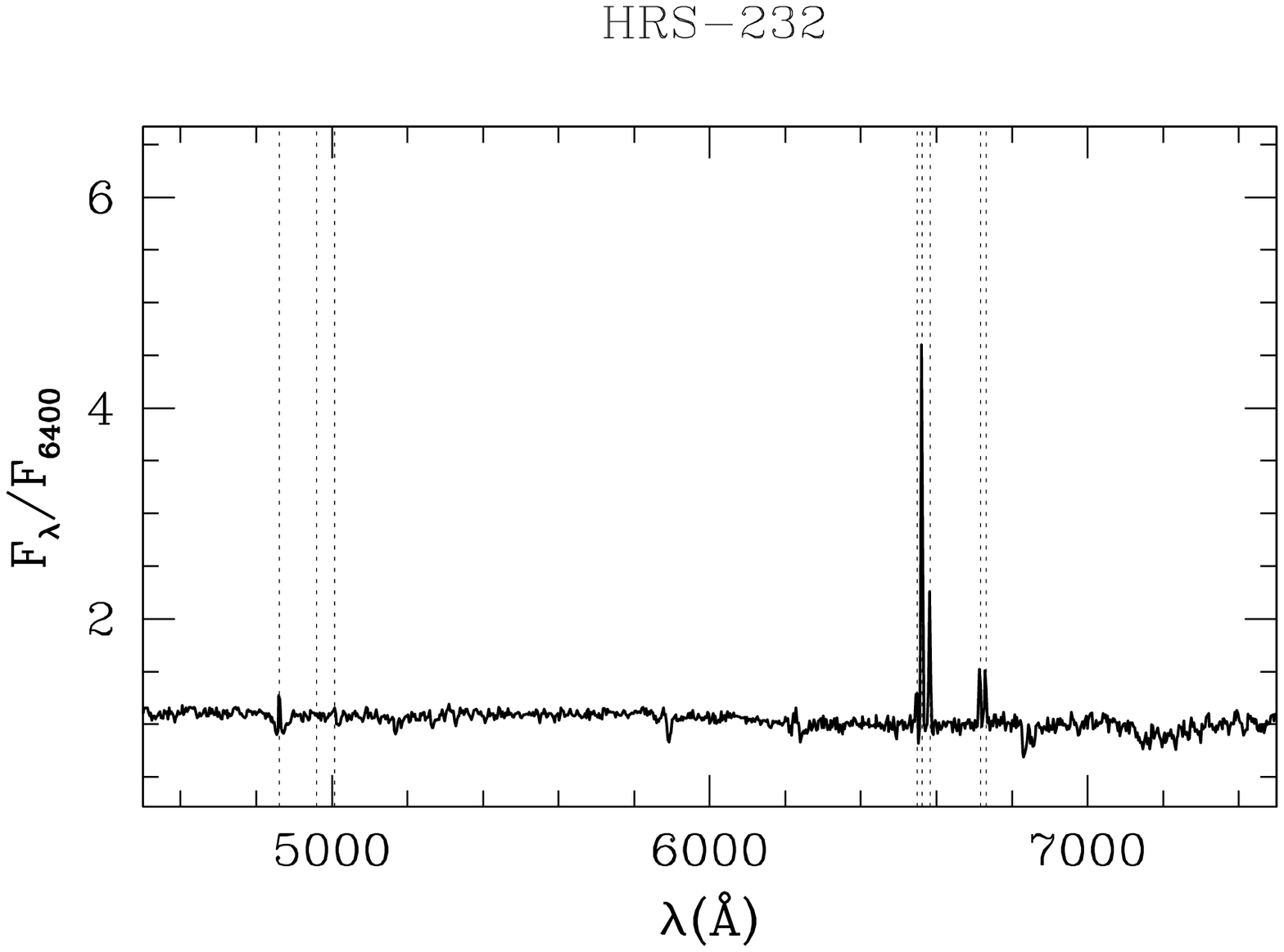}\includegraphics[scale=0.46]{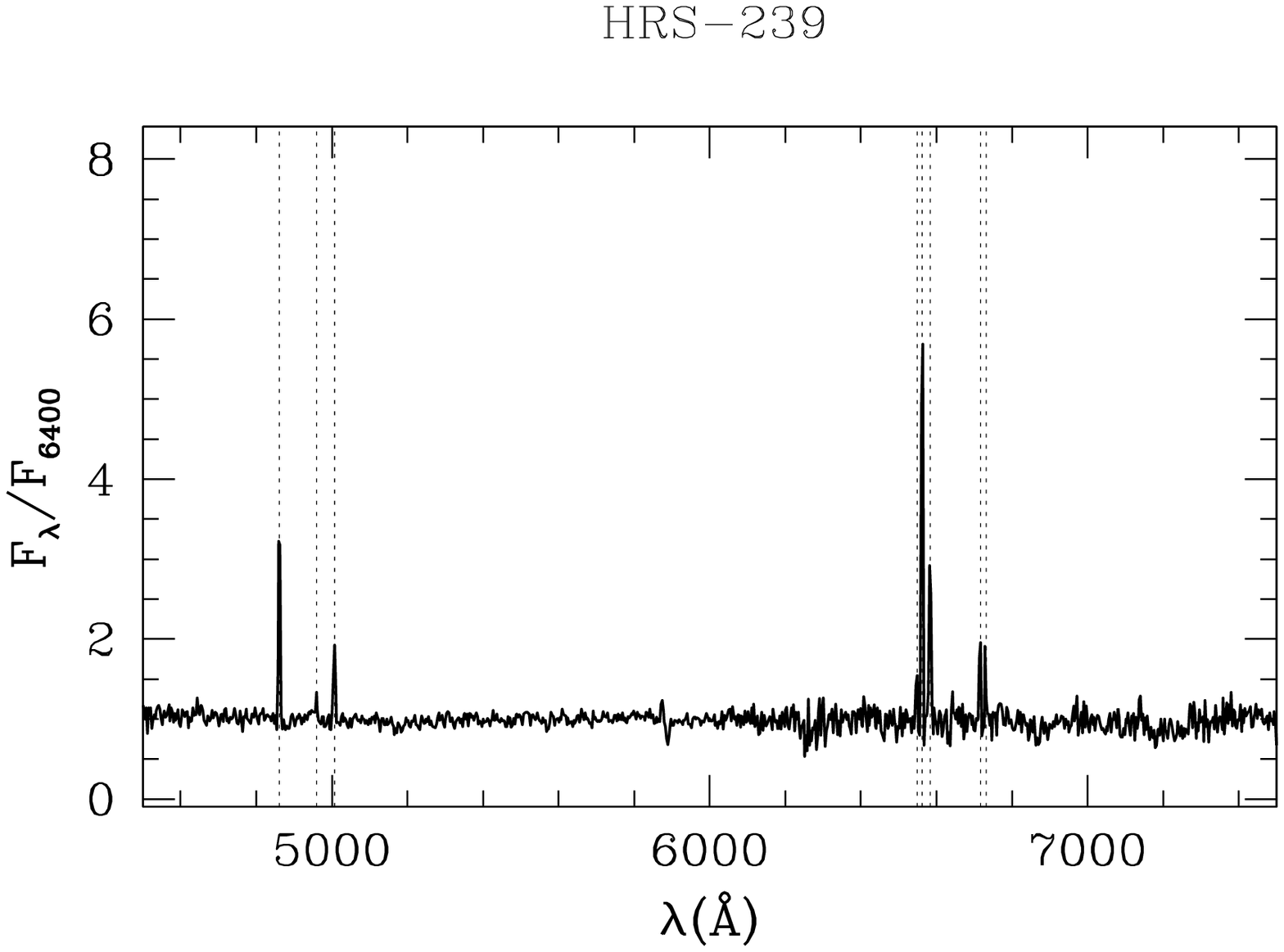}\\
\includegraphics[scale=0.46]{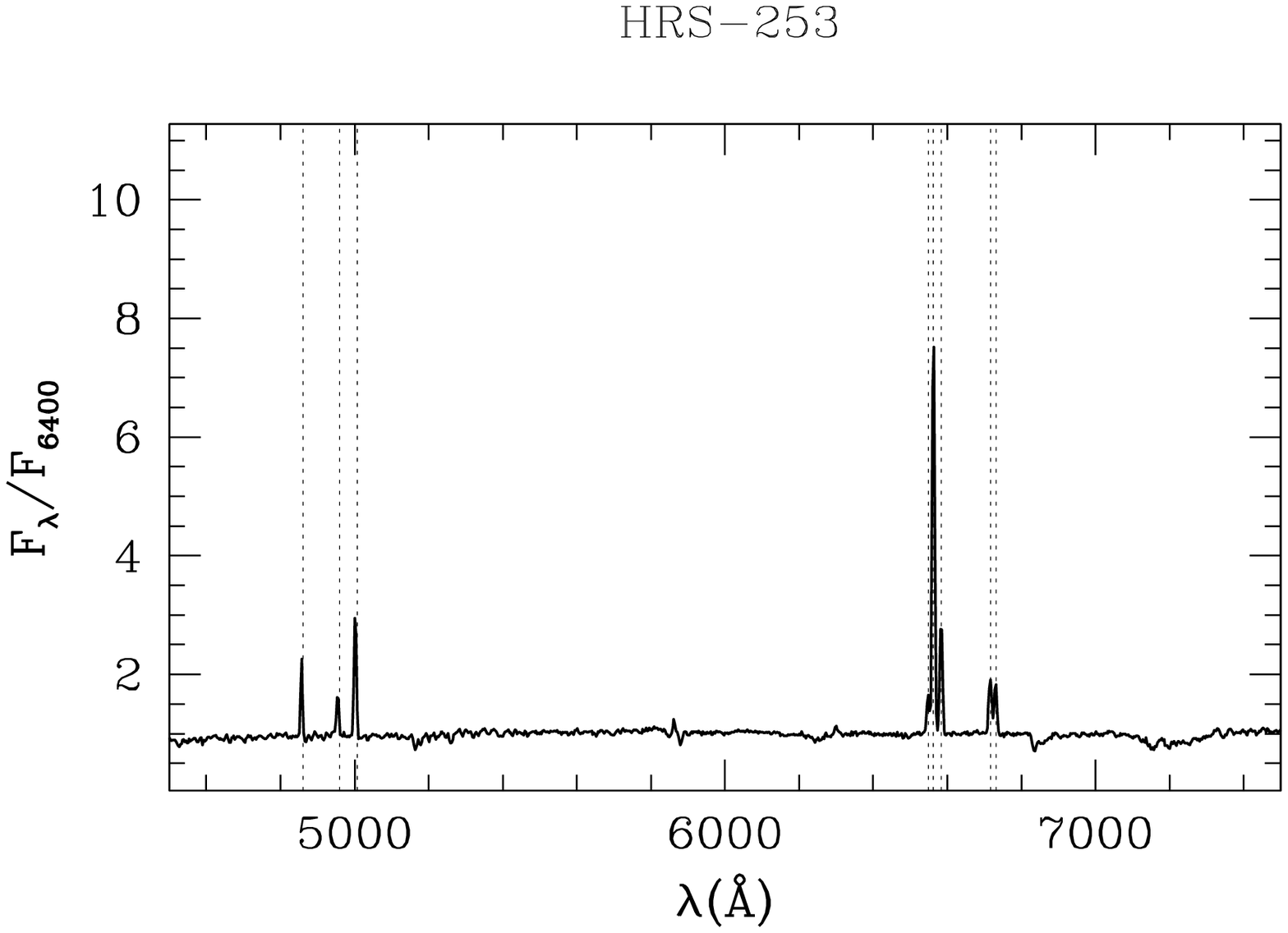}\includegraphics[scale=0.46]{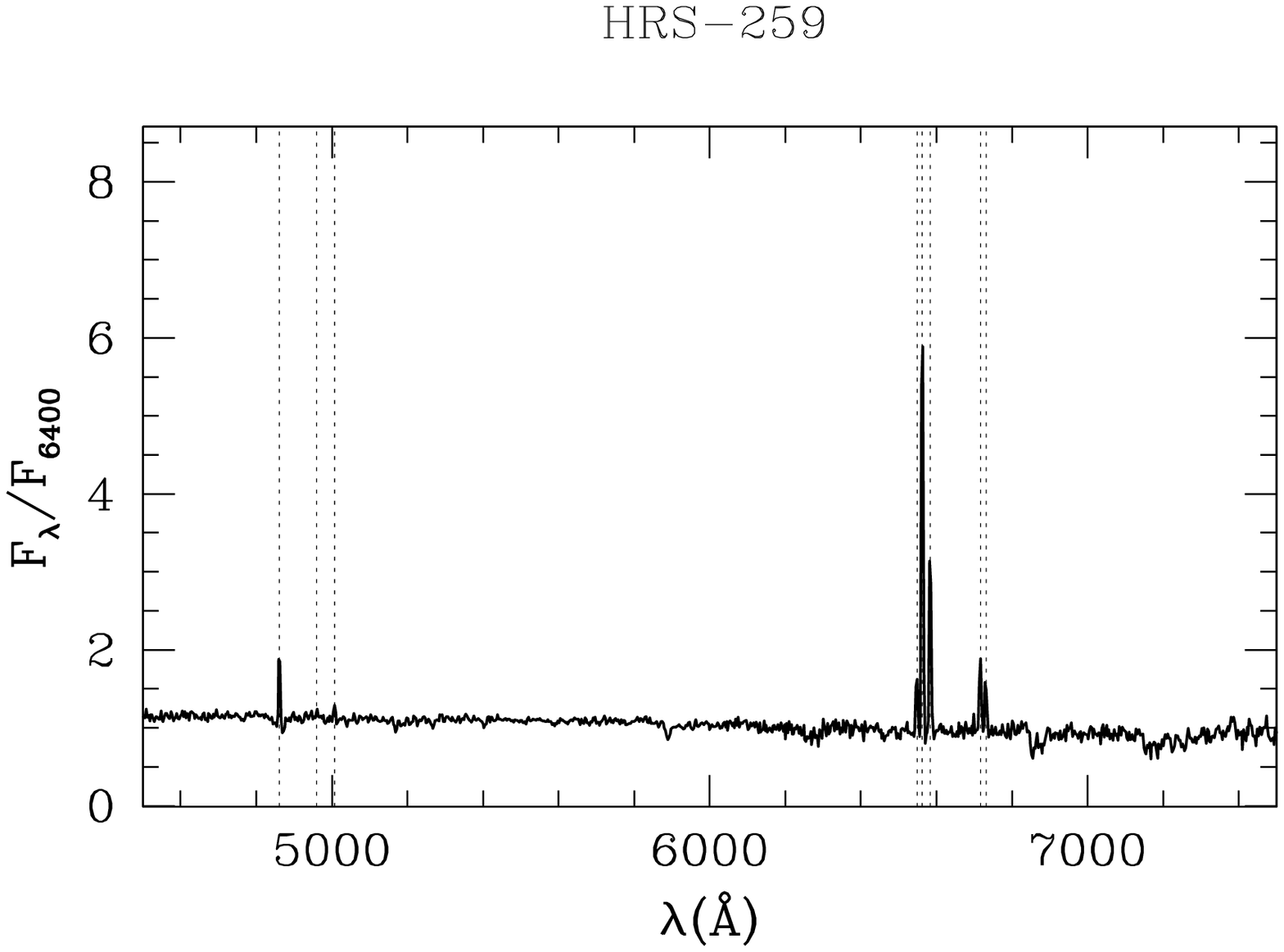}\\
\includegraphics[scale=0.46]{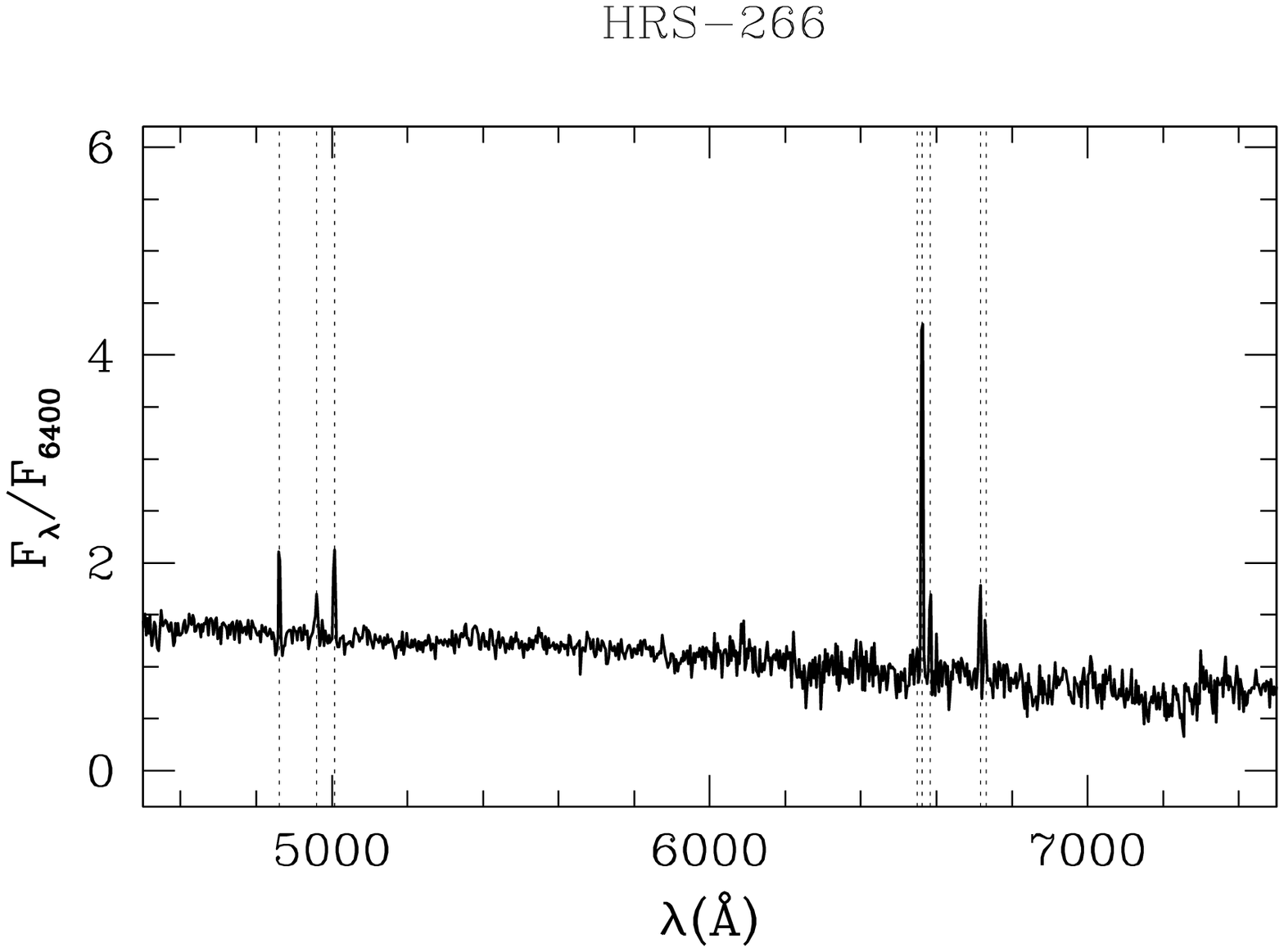}\includegraphics[scale=0.46]{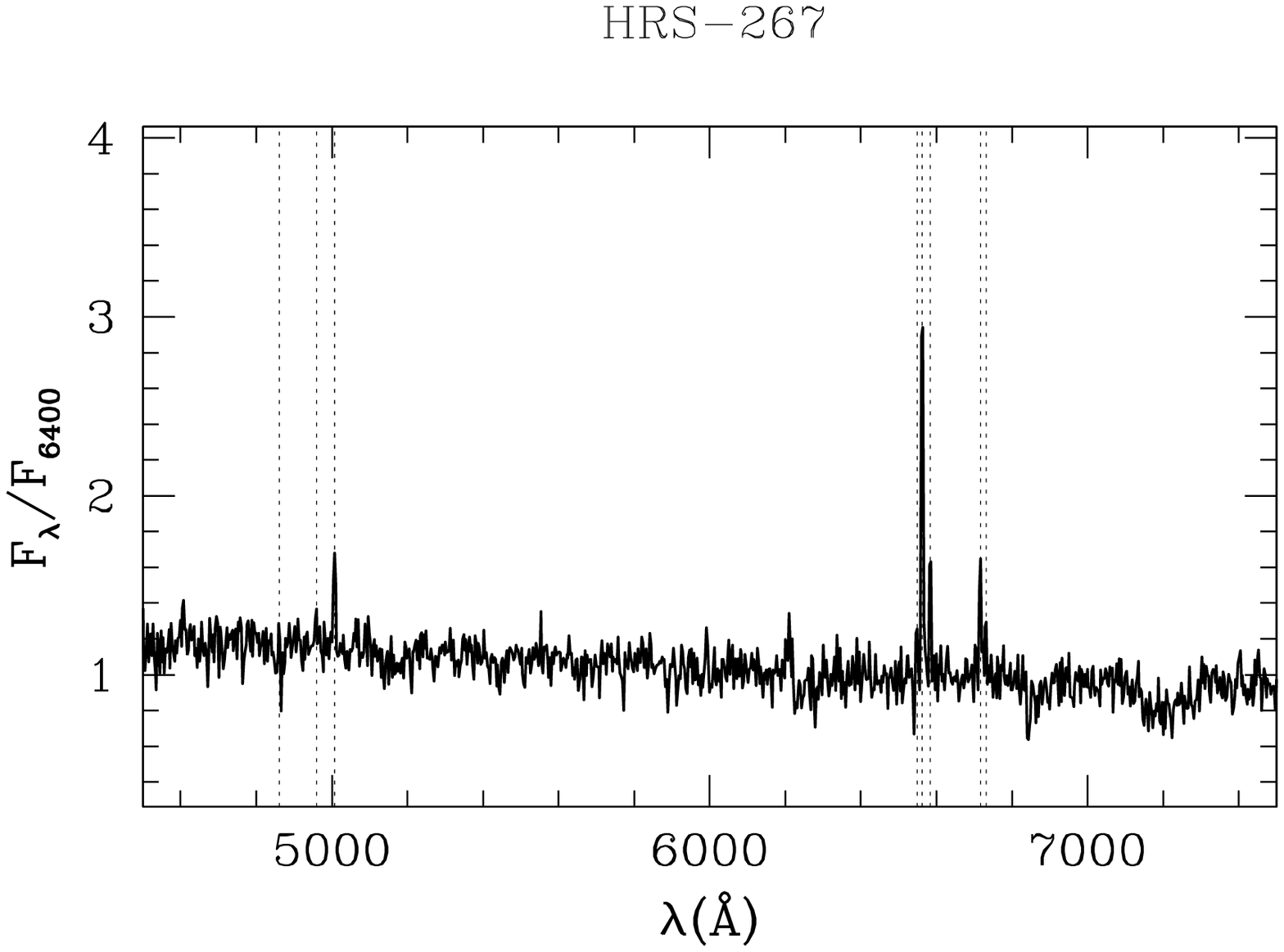}\\
\includegraphics[scale=0.46]{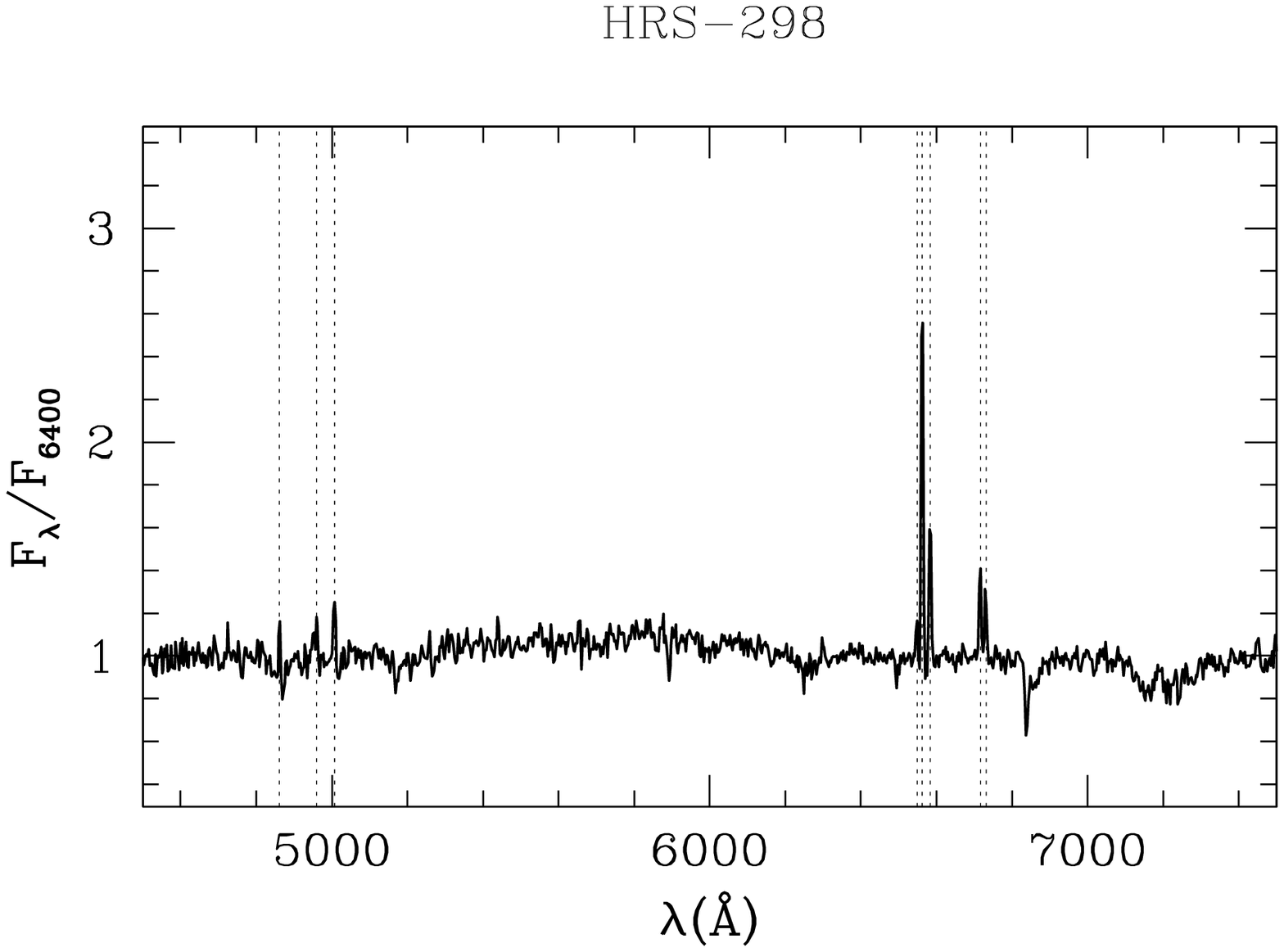}\includegraphics[scale=0.46]{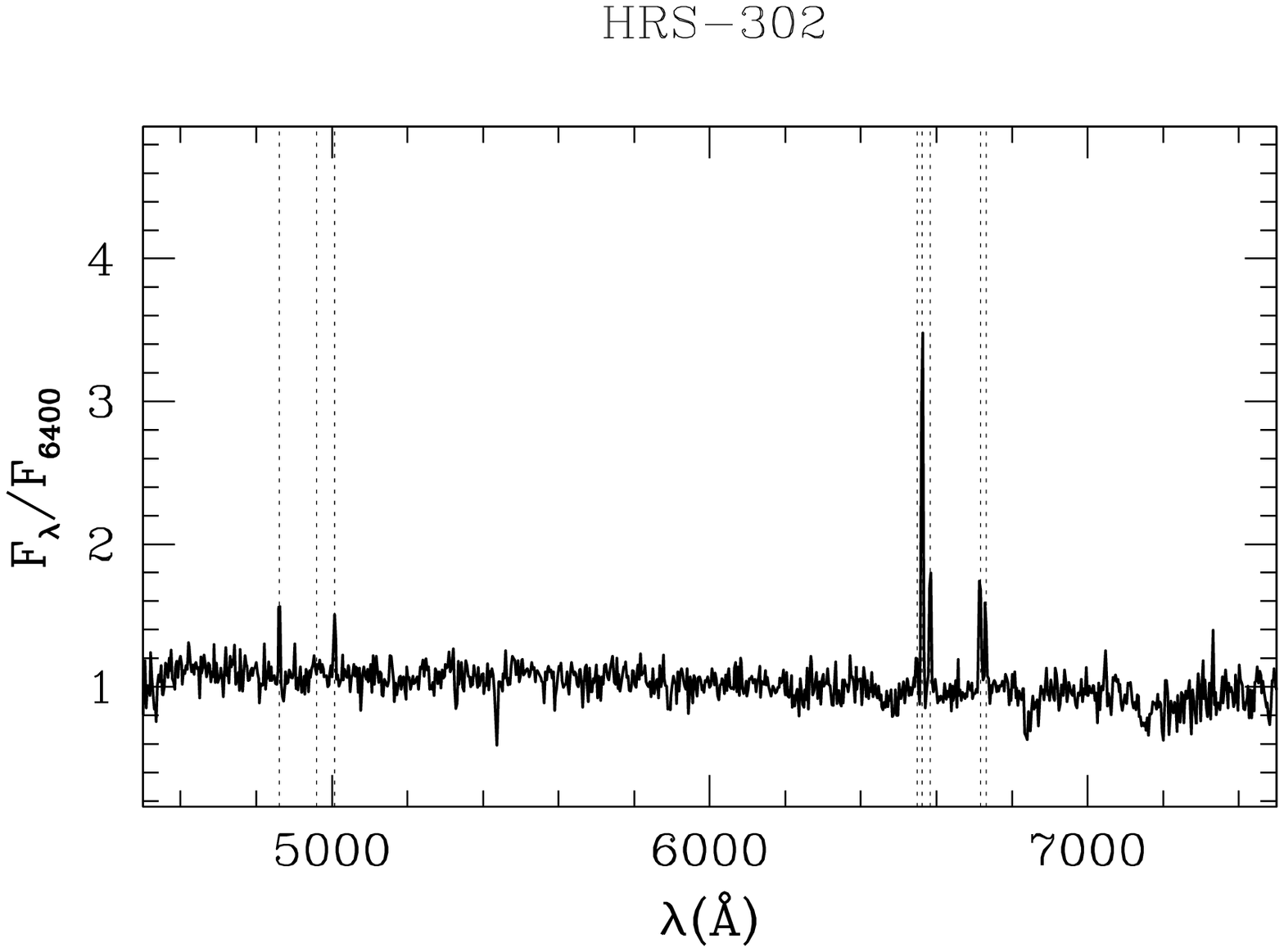}\\
  \end{figure*}           
  \begin{figure*}
  \centering
\includegraphics[scale=0.49]{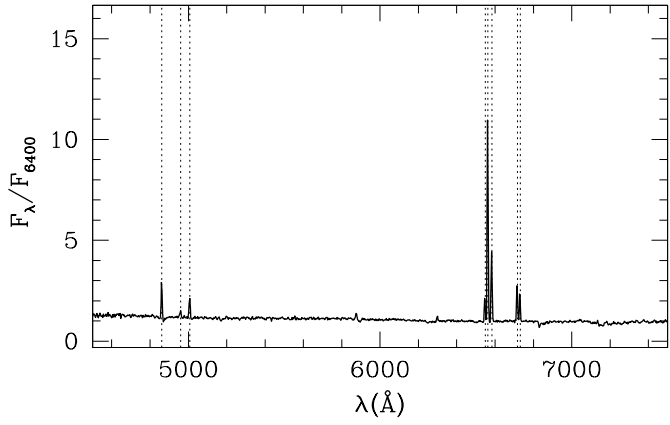}\includegraphics[scale=0.49]{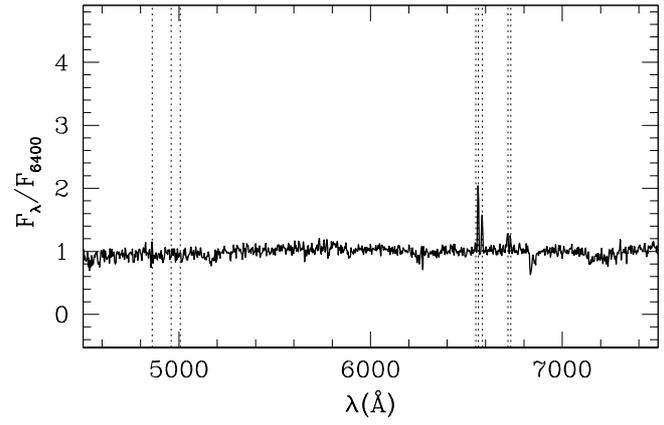}
  \caption{Spectra taken at Loiano with the blue and red grisms covering approximately from 4500 to 7500 \AA. The spectra have been Doppler shifted to rest frame and
  normalized to the flux in the interval 6400-6500 $\AA$. The vertical broken lines mark the rest-frame position of H$\beta~\lambda$4861; [OIII]~$\lambda$4958; [OIII]~$\lambda$5007; [NII]~$\lambda$ 6549; H$\alpha~\lambda$ 6563; [NII]~$\lambda$ 6584;
  [SII]$\sim$ $\lambda6717$; [SII]$\sim$ $\lambda6731$.}.
  \label{spectra}  
  \end{figure*}
  
  \begin{figure*}
  \centering
\includegraphics[scale=0.30]{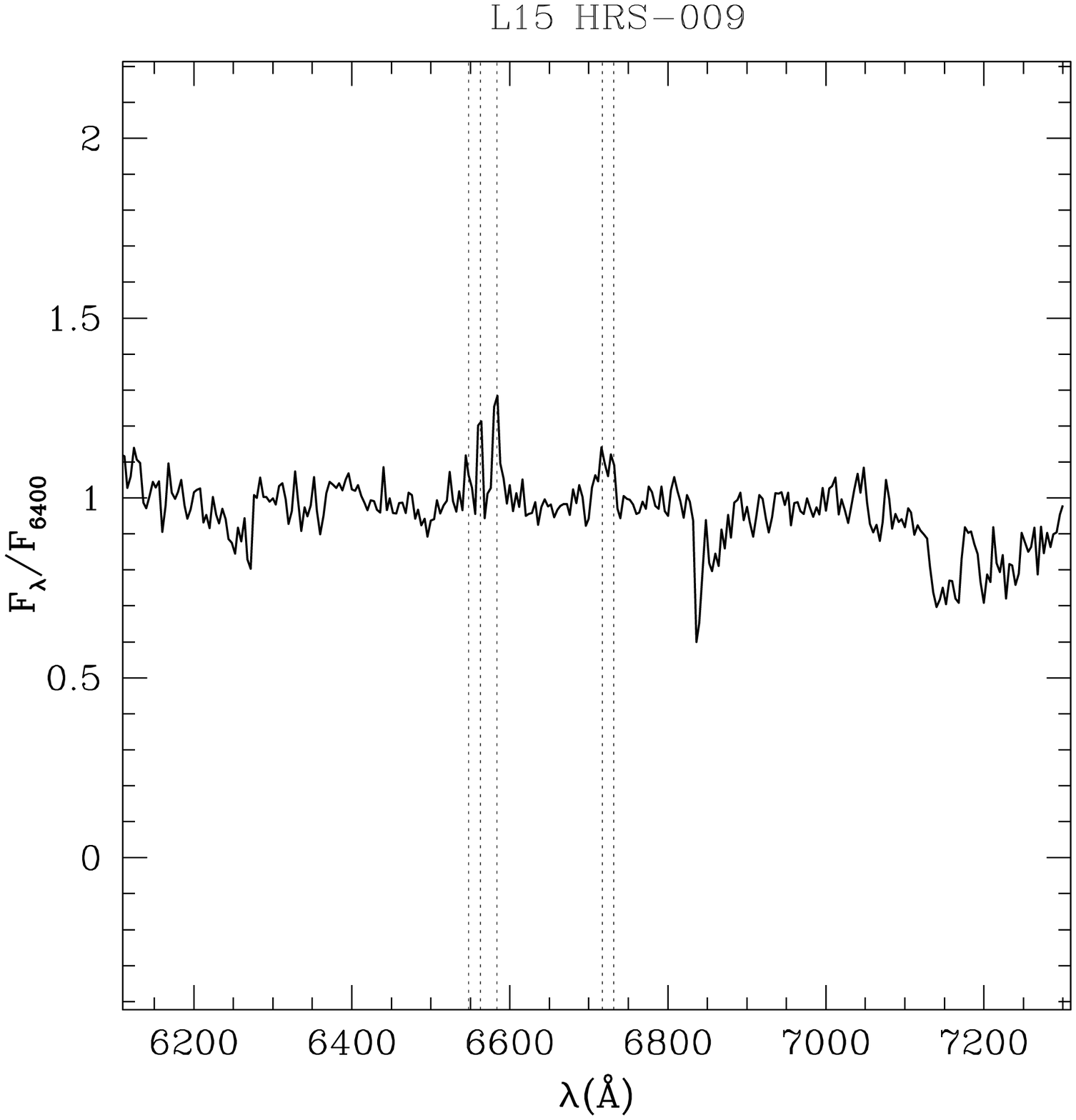}\includegraphics[scale=0.30]{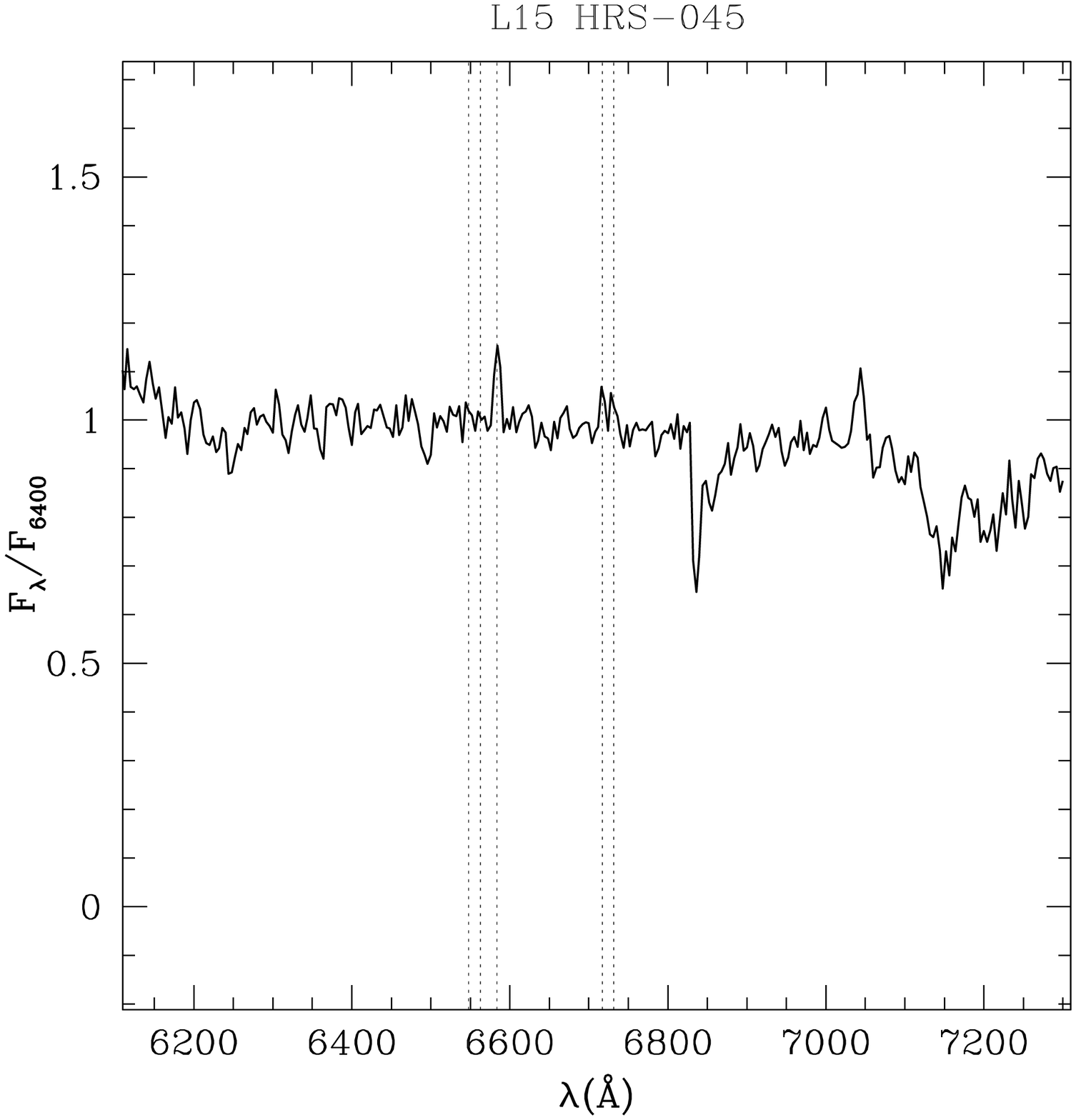}\includegraphics[scale=0.30]{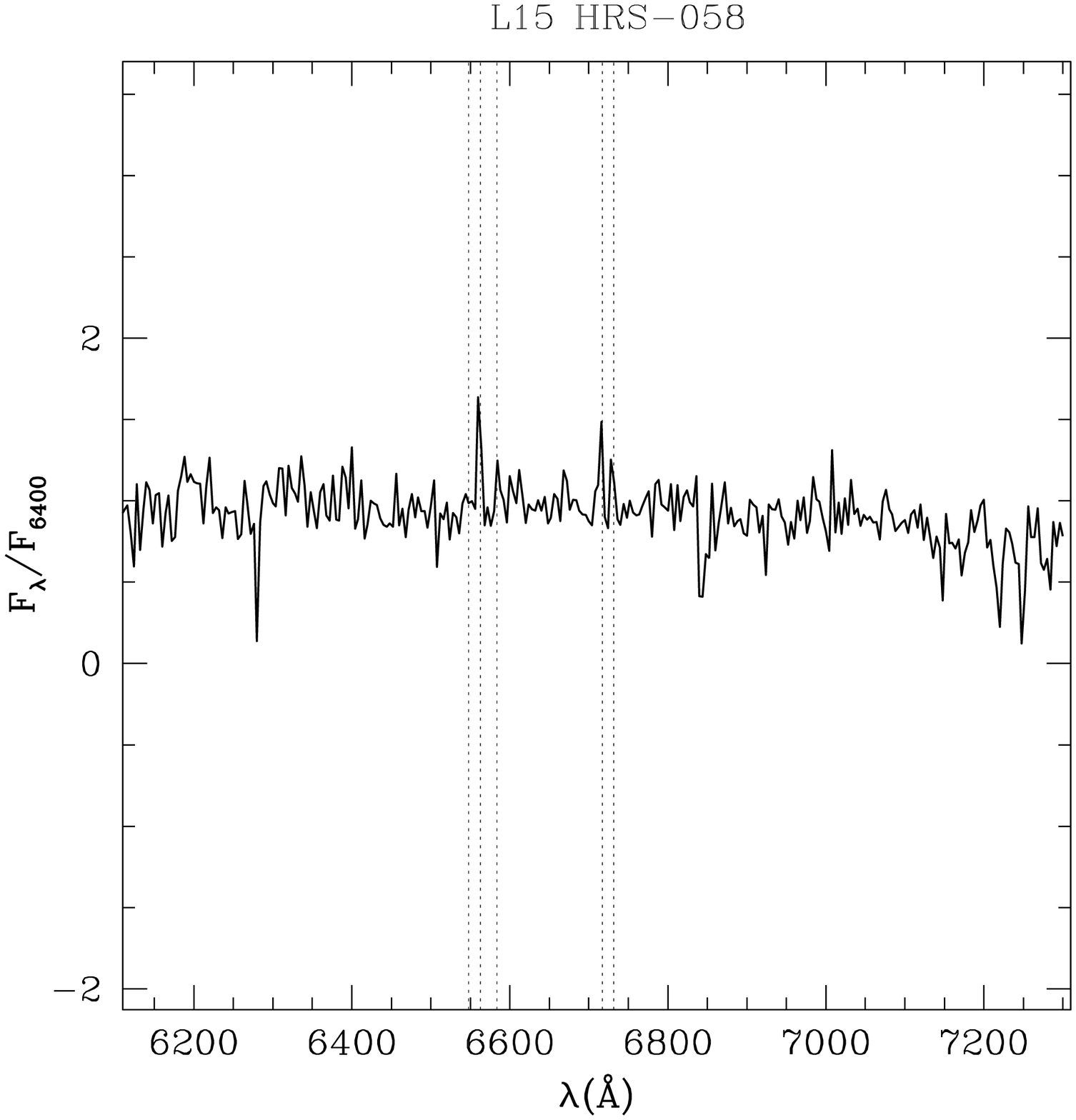}\\
\includegraphics[scale=0.30]{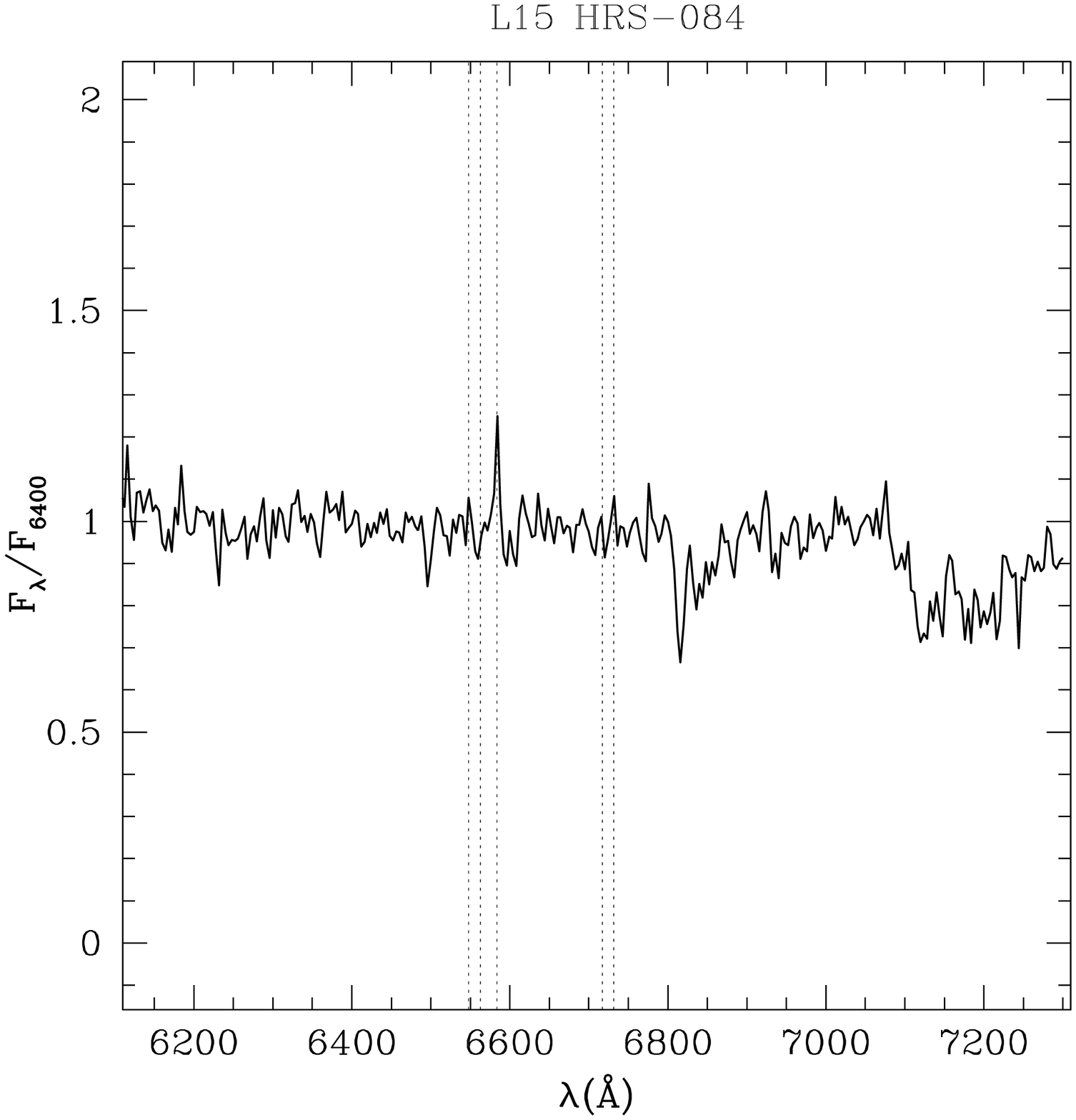}\includegraphics[scale=0.30]{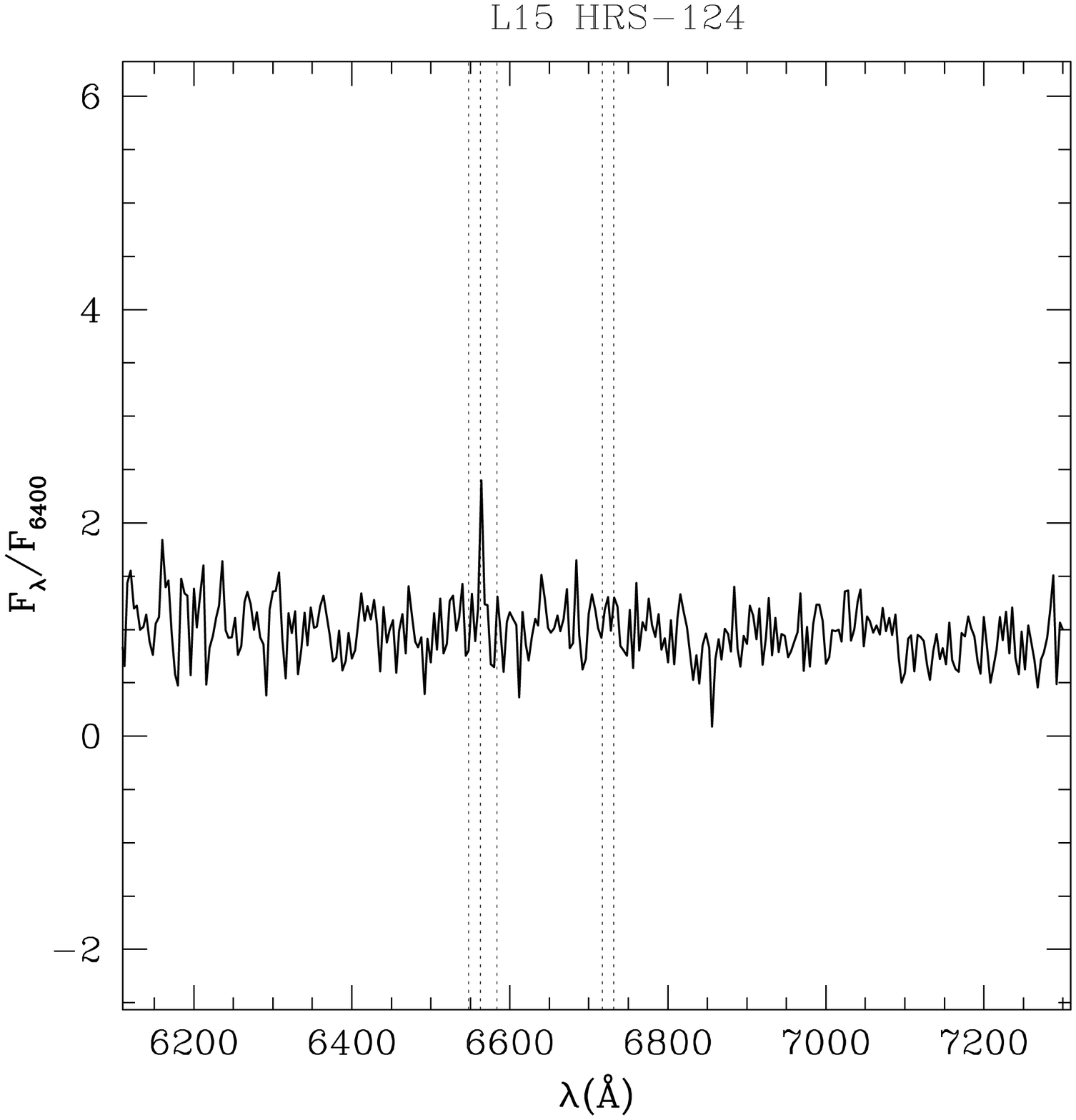}\includegraphics[scale=0.30]{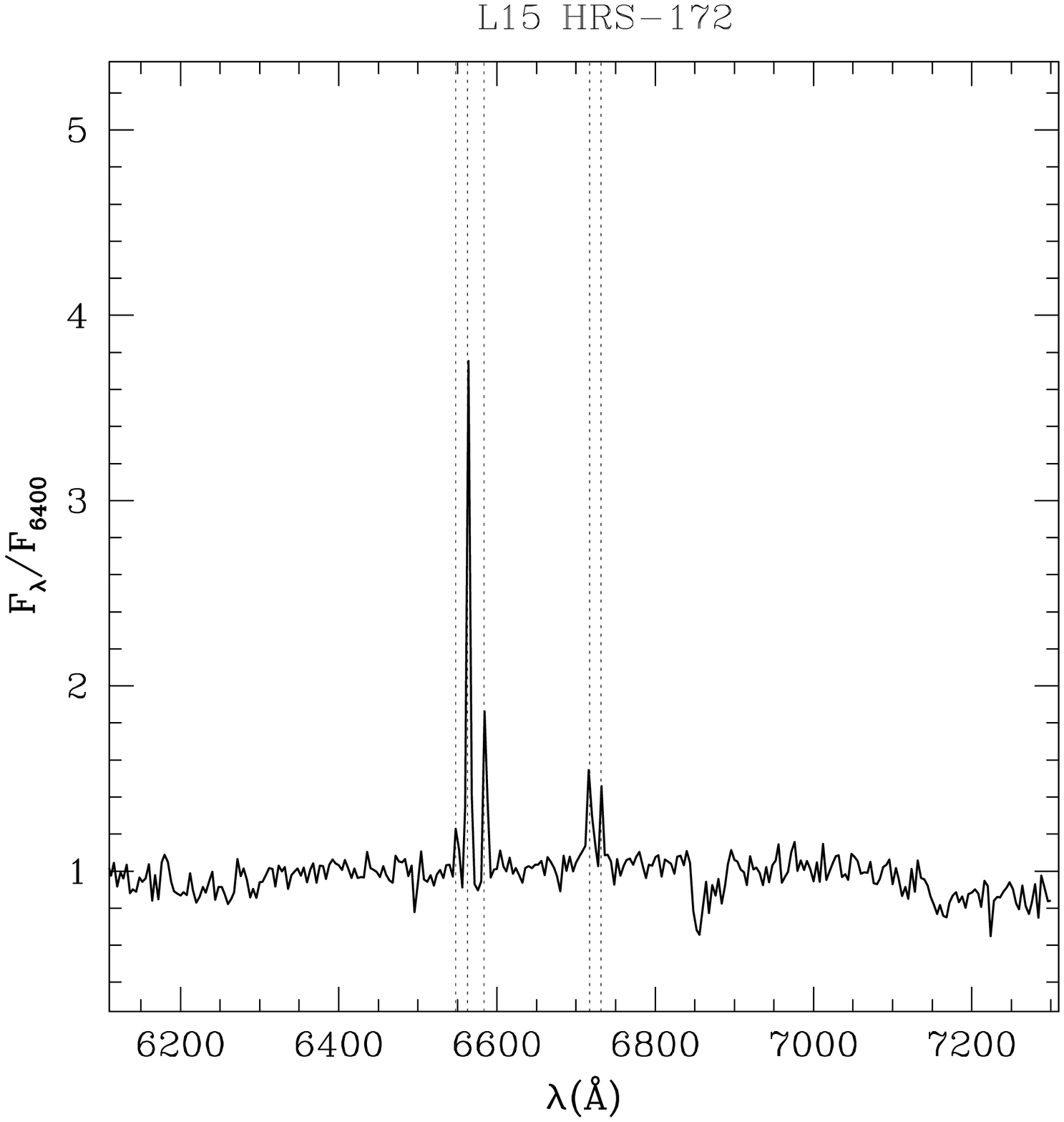}\\
\includegraphics[scale=0.30]{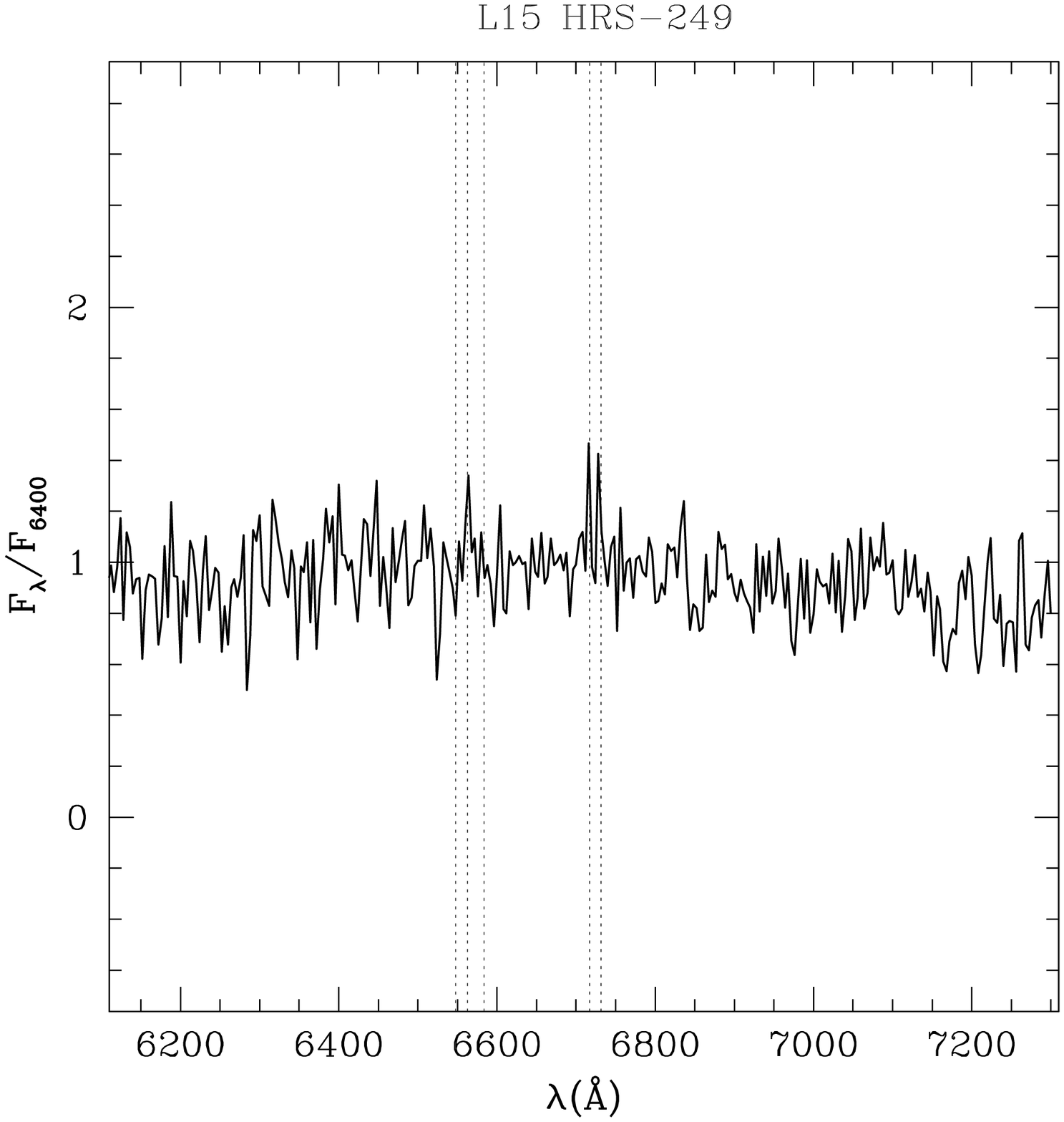}\includegraphics[scale=0.30]{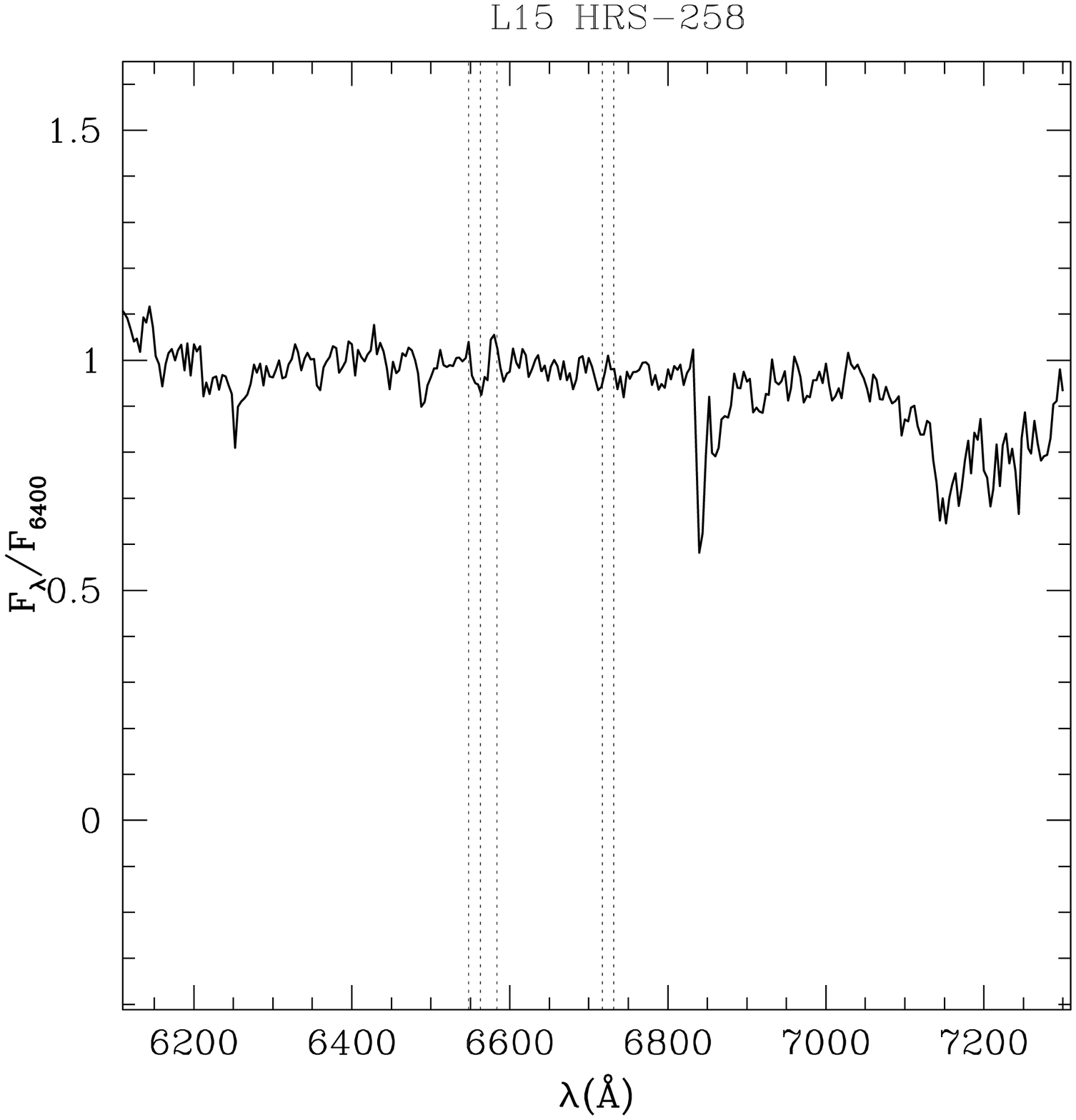}\includegraphics[scale=0.30]{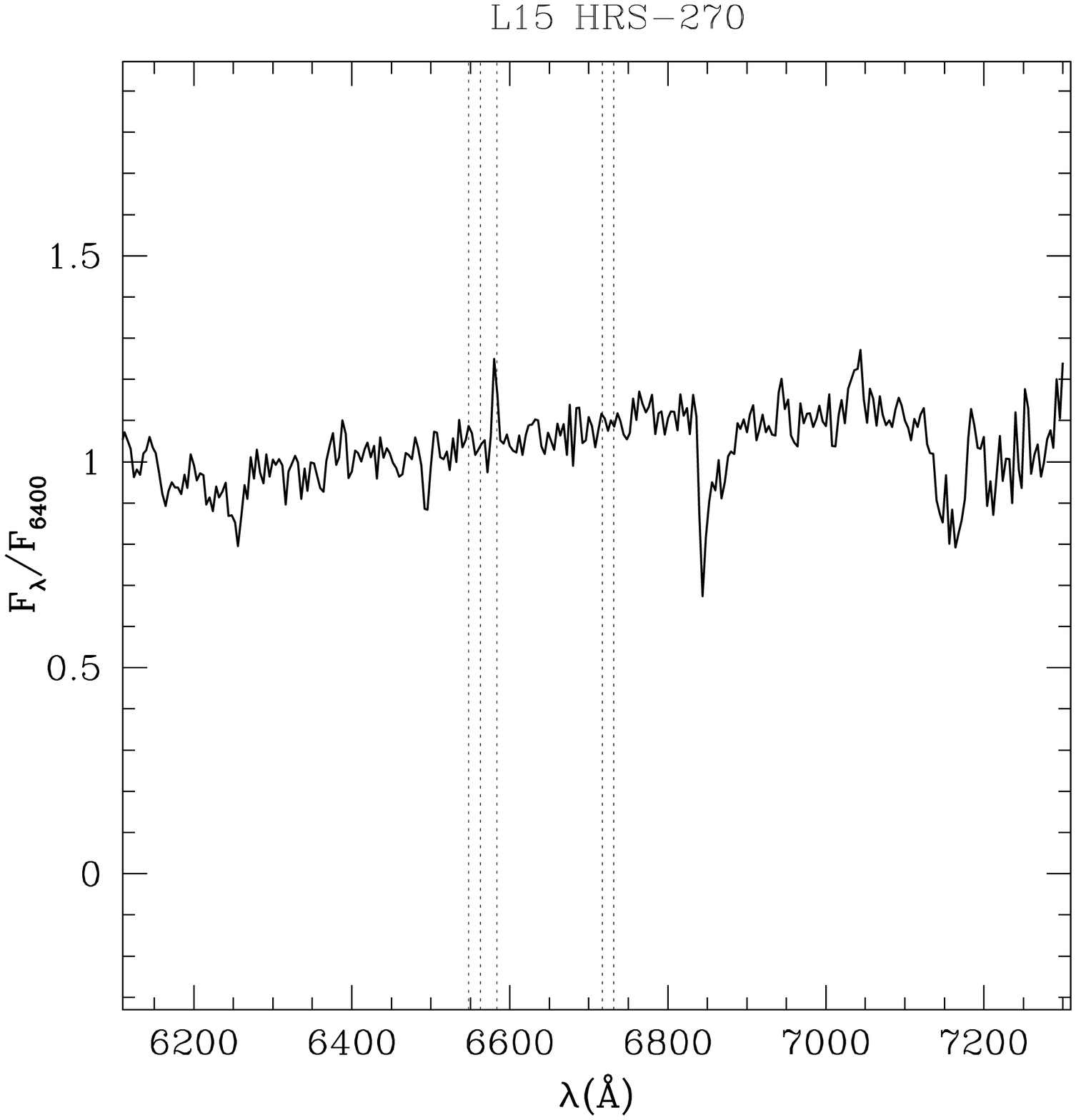}\\
  \end{figure*}
  \begin{figure*}
  \centering
\includegraphics[scale=0.30]{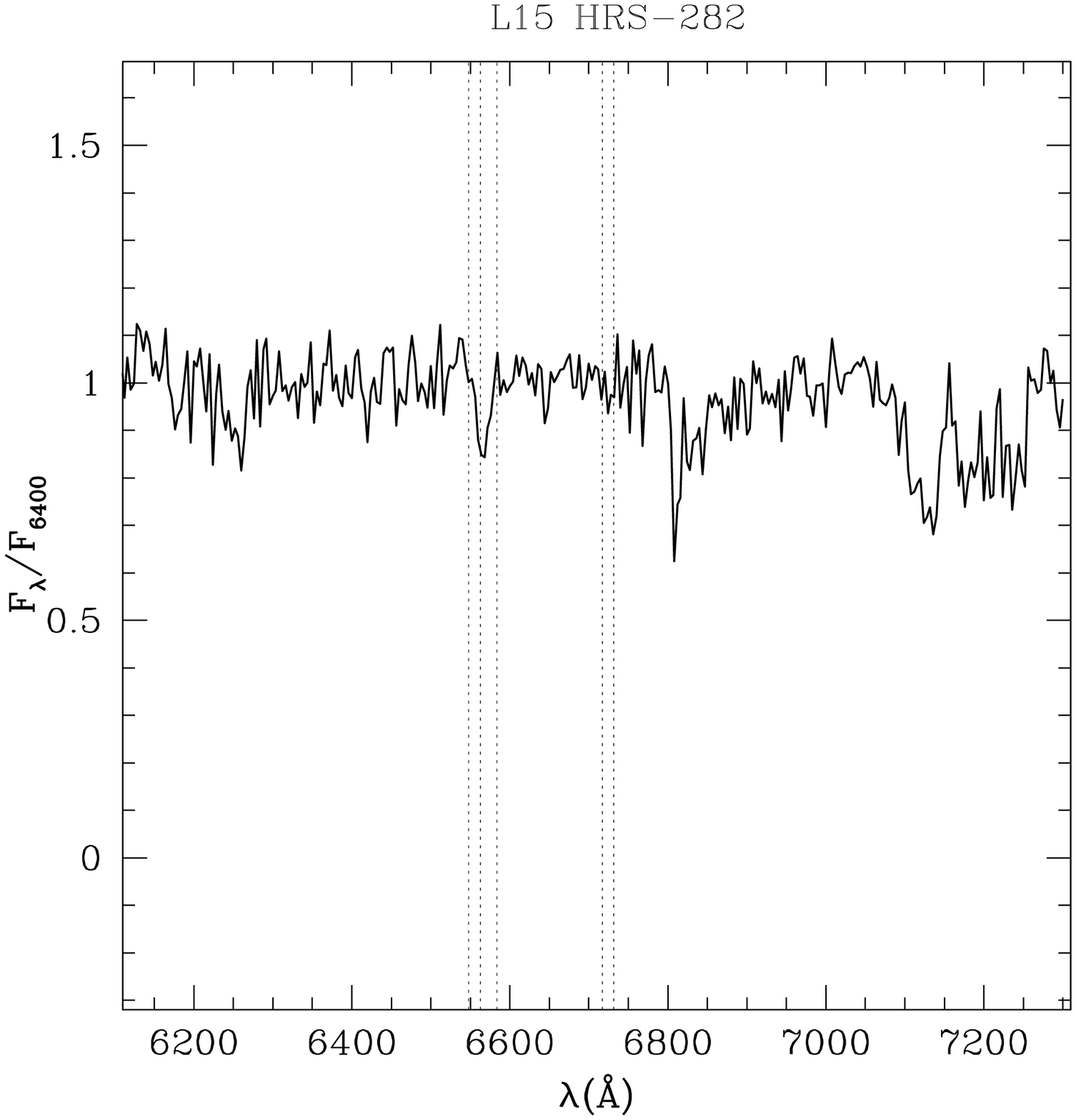}\includegraphics[scale=0.30]{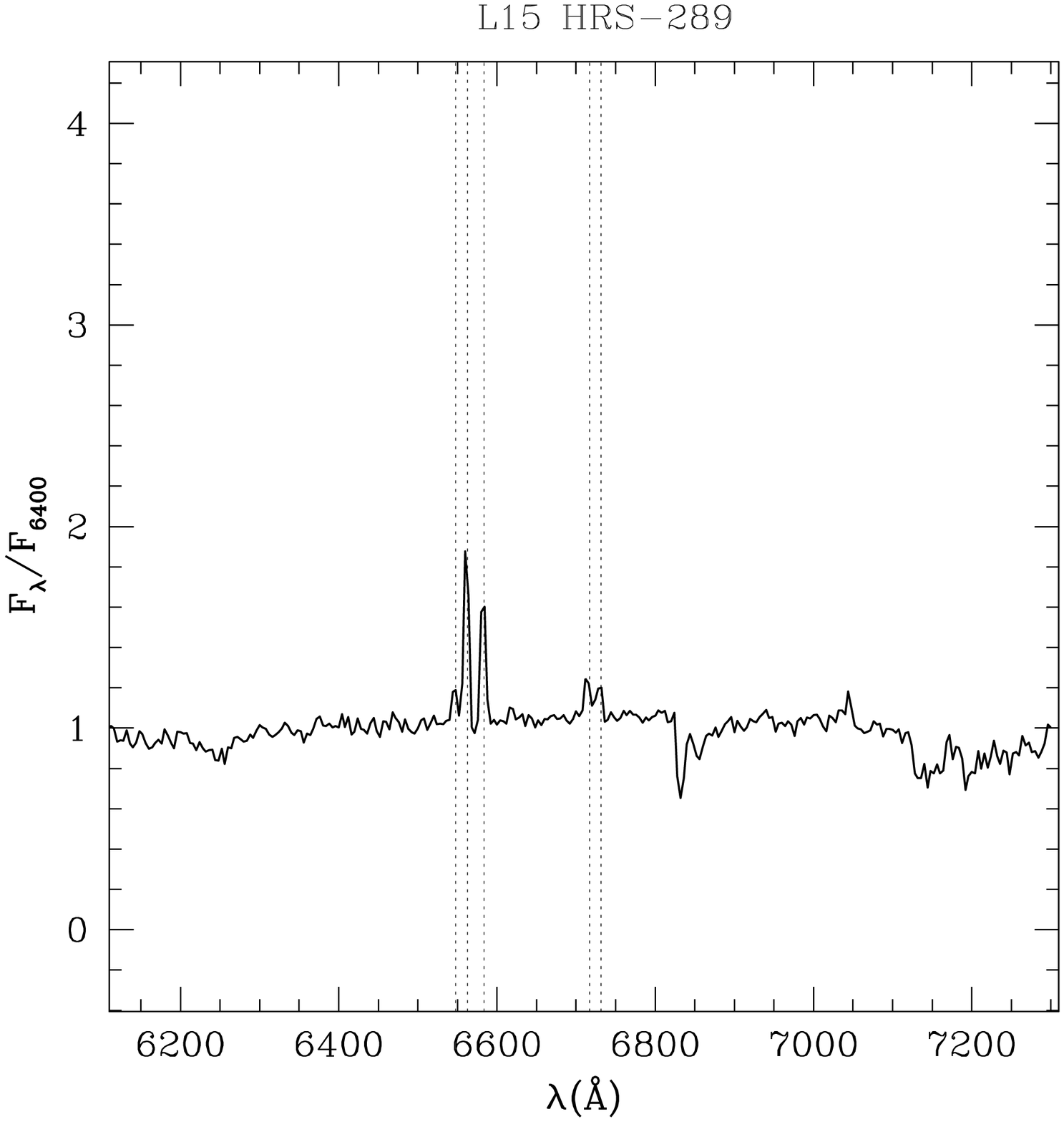}\\
\caption{Unpublished HRS spectra taken at Loiano in 2015 with the red grism. They cover approximately from 6200 to 7200 \AA. The spectra have been Doppler shifted to rest frame and
  normalized to the flux in the interval 6400-6500 $\AA$. The vertical broken lines mark the rest-frame position of  [NII]~$\lambda$ 6549; H$\alpha~\lambda$ 6563; [NII]~$\lambda$ 6584;
  [SII]$\sim$ $\lambda6717$; [SII]$\sim$ $\lambda6731$. (Similar red-grism spectra taken prior to 2015 are already published in Gavazzi et al. 2011 and 2013)}.
  \label{spectra1}  
  \end{figure*}
  \end{onecolumn}

\end{document}